\documentclass[eqsecnum, aps, pra, amsmath, amssymb, twocolumn, superscriptaddress, 10pt ]{revtex4-2}

\usepackage{graphicx}
\usepackage{dcolumn}
\usepackage[colorlinks=true,
            linkcolor=blue,
            urlcolor=blue,
            citecolor=blue]{hyperref}
\usepackage{bm}
\usepackage{enumerate}
\usepackage{lipsum}
\usepackage{threeparttable}
\usepackage{epstopdf}
\usepackage{float}
\usepackage{xcolor,colortbl}
\usepackage[caption = false]{subfig}
\usepackage{tabularx}
\usepackage{chngpage}
\usepackage{csquotes}

\usepackage{natbib}
\usepackage[maxfloats=256]{morefloats}

\begin{document}


\title{Double, triple, and quadruple magic wavelengths for cesium ground, excited, and Rydberg states}
\author{A.  Bhowmik}
\email{anal.bhowmik-1@ou.edu}
\affiliation{Homer L. Dodge Department of Physics and Astronomy, The University of Oklahoma, Norman, Oklahoma 73019,USA}
\affiliation{Center for Quantum Research and Technology, The University of Oklahoma, Norman, Oklahoma 73019, USA}
\author{M.  Gaudesius}
\affiliation{Homer L. Dodge Department of Physics and Astronomy, The University of Oklahoma, Norman, Oklahoma 73019,USA}
\affiliation{Center for Quantum Research and Technology, The University of Oklahoma, Norman, Oklahoma 73019, USA}
\author{G. Biedermann}
\affiliation{Homer L. Dodge Department of Physics and Astronomy, The University of Oklahoma, Norman, Oklahoma 73019,USA}
\affiliation{Center for Quantum Research and Technology, The University of Oklahoma, Norman, Oklahoma 73019, USA}
\author{D. Blume}
\email{doerte.blume-1@ou.edu}
\affiliation{Homer L. Dodge Department of Physics and Astronomy, The University of Oklahoma, Norman, Oklahoma 73019,USA}
\affiliation{Center for Quantum Research and Technology, The University of Oklahoma, Norman, Oklahoma 73019, USA}

\date{\today}





\begin{abstract}
    Dynamic polarizabilities of  cesium Rydberg states, explicitly  $nS_{1/2}$, $nP_{1/2}$, $nP_{3/2}$, $nD_{3/2}$, and $nD_{5/2}$,  where the principal quantum number $n$ is $40$ to $70$, are presented for linearly polarized light. The dynamic polarizability is calculated  using the sum-over-states approach.      We identify double magic wavelengths in the range of $1,000-2,000$~nm for simultaneous trapping of the ground state and a  Rydberg state, which   are, respectively, red-detuned and blue-detuned  with respect to a low-lying excited auxiliary state.   Based  on calculations of the radiative lifetime, blackbody radiation induced transitions, and  population transfer out of the Rydberg and auxiliary states (estimated within two-state as well as master equation models), we conclude that magic wavelength trapping is  particularly promising experimentally for the $nD_{J,|M_J|}$ Rydberg series with angular momentum $J=3/2$ and projection quantum numbers $M_J=\pm 1/2$  (auxiliary state $8P_{1/2}$) and $M_J=\pm 3/2$ (auxiliary state $8P_{3/2}$), using trap depths as large as $10$~$\mu$K.  Moreover,  by tuning the  angle between the quantization axis and the polarization vector of the light, we identify  triple and quadruple magic wavelengths,  for which the polarizabilities of the ground state, a Rydberg state,  and, respectively,  one and two low-lying excited states are equal. Our comprehensive theoretical study provides much needed guidance for on-going experimental efforts on cesium Rydberg-state based quantum simulations that operate on time scales up to several $\mu$s.
\end{abstract}

\maketitle
\section{Introduction}\label{sec_1}

An  excited atomic state with a high principal quantum number $n$ is referred to as a Rydberg state.  Rydberg states have intriguing  characteristics such as a large size,  large transition dipole moments,  strong controllable long-range interactions~\cite{Robicheaux2018}, and long lifetimes~\cite{Gallagher1994}. These properties are being leveraged in a range of applications across various  branches of  physics. Specifically, 
Rydberg atoms constitute an appealing platform for quantum logic devices~\cite{Saffman2010, Yuan2023, Buchemmavari2024},  quantum information protocols~\cite{ Adams2020},  quantum computing~\cite{Saffman2016}, quantum signal processing~\cite{Zeytinoglu2024}, quantum simulations~\cite{Malz2023}, investigations of fundamental few- and many-body physics questions~\cite{Lee2019, Zeiher2016, Bernien2017, Bharti2023}, nonlinear quantum optics~\cite{Firstenberg2016}, and so-called transmission imaging~\cite{Huo2022}.

The radiative lifetime of a Rydberg atom scales with the principal quantum number as $n^3$~\cite{Gallagher1994}. For $n \approx 50$, the lifetime is typically of the order of 100~$\mu$s, which is much longer than the radiative lifetime of, e.g., low-lying excited states of alkali atoms. The radiative lifetime of a Rydberg state may, however, be significantly reduced by blackbody radiation induced population transfer out of the Rydberg state~\cite{Beterov2009}, thereby  potentially placing strict limits on the length of experimental Rydberg-atom based  quantum information processing and quantum simulation protocols~\cite{Lukin2001, Xia2015, Endres2016}. 

Rydberg lifetime calculations are not only important from experimental and  quantum technology points of view, but also from a theory point of view. Since a state's lifetime is governed by a set of  matrix elements and transition wavelengths, accurate lifetime determinations can be used to benchmark  other quantities such as the atomic  polarizability~\cite{Laidig1990, Mitroy2010}. In this work, we calculate the spontaneous lifetime, blackbody radiation lifetime, and effective lifetime of several $^{133}$Cs Rydberg series, namely the $nS_{1/2}$, $nP_{1/2}$, $nP_{3/2}$, $nD_{3/2}$, and $nD_{5/2}$ series with  
$n \in [40,70]$.  These lifetimes play an important role in assessing the experimental feasibility of simultaneously trapping ground state and Rydberg states at  so-called magic wavelengths; their determination and interpretation are the main topic of this paper.

 In  most cold atom  experiments,  ground state atoms  are trapped by employing  a far-off-resonant red-detuned optical-dipole trap~\cite{GRIMM2000}. The reason is that the  polarizability of the ground state is positive when the  light is red detuned with respect to a low-lying transition line, thereby creating an attractive force for the ground state.   A  Rydberg atom cannot, in general, be held in place with this set-up  since the linearly polarized light creates a repulsive force for the Rydberg state, owing to the Rydberg states' negative "background polarizability"~\cite{Younge2010}.  In addition to resulting in anti-trapping,  excitation of an atom to a Rydberg state may induce  heating and decoherence due to entanglement of the spin and motional states~\cite{Keijzer2023}. In experiments that use a red-detuned optical dipole trap for the ground state, the heating is minimized by turning the trapping laser off while the atom resides in the Rydberg state~\cite{Saffman2010}.  For some applications, including quantum simulation studies, it is, however, desirable to keep the atom trapped while it resides in a Rydberg state.

 Several avenues for trapping Rydberg atoms have been pursued. $^{87}$Rb 
Rydberg atoms were trapped experimentally  in a one-dimensional ponderomotive optical lattice by repeatedly reversing the lattice potential on a time scale that is fast compared to the lifetime of the Rydberg state~\cite{Anderson2011}. Ensembles of $^{87}$Rb  Rydberg atoms have also been trapped  in a one-dimensional optical lattice \cite{Li2013}.  A ponderomotive bottle beam trap was used to demonstrate three-dimensional trapping of $^{87}$Rb  Rydberg atoms in the states $nS_{1/2}$, $nP_{1/2}$, and $nD_{3/2}$   with $60\leq n\leq 90$~\cite{Barredo2020}.   Since bottle beam traps are comparatively challenging to create experimentally~\cite{Xu2010,Barredo2020}, the quest for alternative Rydberg state trapping protocols is ongoing. 
Very recently, e.g., the ground state and Rydberg state of a neutral ytterbium  atom were trapped simultaneously  in a single frequency optical tweezer trap, which confines the ground state (standard red-detuned dipole trap),  by utilizing the polarizability of the  ionic core~\cite{Wilson2022}. This approach is restricted to huge Rydberg states since the outer electron must reside in the region where the light intensity is negligible so that the positive polarizability of the ionic core is not "overshadowed" by the negative polarizability of the valence electron.

Alternatively, it was proposed in 1999 that the ground state and a Rydberg state could be trapped simultaneously at magic wavelengths  for which the  differential ac-Stark shift of the two states is cancelled~\cite{Katori1999}. Even though  magic wavelengths are nowadays used routinely in cold atom experiments (see, e.g., Ref.~\cite{Ye2008} for a discussion of the use of magic wavelengths  in quantum metrology), their use in the context of Rydberg experiments is still comparatively rare~\cite{Saffman2005, Goldschmidt2015, Bai2020}.   

Building on Refs.~\cite{Yerokhin2016, Bai2020_1},  we carefully investigate a scheme for simultaneously trapping the cesium ground state and a Rydberg state in an attractive potential well at magic wavelengths for which the ground state is red-detuned and the Rydberg state is blue-detuned.  Our  calculations of the dynamic polarizability, which governs the trapping force,   assume a linearly polarized Gaussian-shaped light beam.  In this scheme, the Rydberg state is trapped thanks to the admixture of an auxiliary state. Light scattering and transitions out of the two-state manifold formed by the ground state and the Rydberg state, which can severely compromise the usefulness of the scheme, are analyzed.  A comparatively high light intensity, e.g., guarantees trapping (this is our aim) but also enhances the likelihood of unwanted transitions to the auxiliary state. The $nD_{3/2,|M_J|}$  series with projection quantum numbers $M_J=\pm 1/2$ and $\pm 3/2$ are found to be the most promising candidates for magic wavelength trapping.

Moreover, we identify several triple magic wavelengths at which not only the ground state and a Rydberg state are trapped (referred to as  double magic wavelengths) but also an additional low-lying excited state, which we refer to as an intermediate state.  Triple magic wavelengths were, e.g., also found for  ytterbium and cadmium~\cite{Topcu_2016,Zhang_2024}. Lastly, we also identify a number of quadruple magic wavelengths, at which the ground state, the  Rydberg state, and two intermediate states are trapped simultaneously. The identified triple and quadruple magic wavelengths provide exciting new prospects for Rydberg state physics.

The remainder of this paper is organized as follows. Section~\ref{sec_2} presents the radiative, blackbody, and effective lifetimes of Rydberg states. Section~\ref{sec_3} presents the static and dynamic polarizabilities for several cesium Rydberg series. From the dynamic polarizabilities, we extract double magic wavelengths as a function of $n$ for various auxiliary states. Section~\ref{sec_4} assesses the experimental realizability of the magic wavelength based trapping scheme  by analyzing  the Rabi frequency and the transition probability  between the Rydberg and auxiliary states. Section~\ref{sec_5} presents  triple and quadruple magic wavelengths and the corresponding polarizabilities.   The paper ends with conclusions in Sec.~\ref{sec_6}. Technical details are relegated to appendices. The supplemental material tabulates a large number of magic wavelengths and their characteristics as a function of  $n$.

\section{Lifetime of Rydberg states}\label{sec_2}
The spontaneous lifetime $\tau_{\text{SP}}$  of the $k$-th atomic state is calculated from the radiative transition probability  via~\cite{Johnson2007} 
\begin{equation}\label{eq_1}
\tau_{\text{SP}}=\dfrac{1}{\sum_iA_{ki}},
\end{equation}
where $A_{ki}$ is the transition probability from the $k$-th state to the $i$-th state.  The sum runs over all states $i$ with $E_i<E_k$, where $E_i$ and $E_k$ denote the energies of the $i$-th and the $k$-th states, respectively.  Throughout this work, the transition probabilities account only for the  strongest transitions, namely the dipole transitions \cite{Johnson2007}: 
\begin{equation}\label{eq_2}
A_{ki}=\dfrac{4\alpha\omega_{ki}^3}{3c^2(2J_k+1)}|\langle\psi_k||d||\psi_i\rangle|^2.
\end{equation}
Here, $\alpha$  denotes the fine structure constant, $c$ the speed of light in vacuum, $J_k$ the total electronic angular momentum of the $k$-th state, $\omega_{ki}$ the transition frequency, and  $|\langle\psi_k||d||\psi_i\rangle|$ the magnitude of the reduced electric dipole matrix element   between the unperturbed atomic states $\psi_k$ and $\psi_i$ ($d$ is equal to the $z$-component of the position vector of the electron). 

As the energy difference between two adjacent Rydberg states falls within  the microwave frequency range,   the lifetime of a Rydberg state can be drastically modified by the blackbody radiation (BBR)   photons of the microwave background field at room temperature~\cite{Gallagher1979}, leading to an increased effective decay rate.  The BBR lifetime $\tau_{\text{BBR}}$ of an atomic state can be accurately parameterized by introducing the dimensionless scaling coefficients $A$, $B$, $C$, and $D$~\cite{Beterov2009}:
\begin{equation}
\label{eq_3}
\tau_{\text{BBR}}=\dfrac{(n-\delta_{L,J})^D}{A}\dfrac{\exp\bigg(\dfrac{315780B}{(n-\delta_{L,J})^C (T/\mbox{K})}\bigg)-1}{2.14\times 10^{10}} \;\;\mbox{s}
\end{equation}
(here, the temperature needs to be provided in Kelvin and the equation yields the lifetime  in seconds); the values of $A$, $B$, $C$, and $D$ for cesium~\cite{Beterov2009}  are collected in Table~\ref{table_1} for completeness. In Eq.~(\ref{eq_3}),  $\delta_{L,J}$ denotes the quantum defect parameter, whose values are reported in Appendix~\ref{Appendix_A}. Since $\delta_{L,J}$  depends on the orbital angular momentum $L$ and the total angular momentum $J$ of the valence electron, $\tau_{\text{BBR}}$ depends on $n$, $L$, and $J$.  In what follows, we report $\tau_{\text{BBR}}$ at $T=300$~K.

\begin{table}[h]
\caption{Dimensionless scaling coefficients $A$,		$B$, 	$C$, and	$D$ for cesium. The values are taken from Ref.~\cite{Beterov2009}.} 
\centering 
\begin{tabular}{c |c|c | c| c} 
\hline\hline 
Rydberg series 	&	$A$	&	$B$	&	$C$	&	$D$	\\
\hline
$nS_{1/2}$	&	0.123   	&	0.231	&	2.517	&	4.375	\\
$nP_{1/2}$	&	0.041   	&	0.072	&	1.693	&	3.607	\\
$nP_{3/2}$	&	0.038   	&	0.056	&	1.552	&	3.505	\\
$nD_{3/2}$	&	0.038   	&	0.076	&	1.790	&	3.656	\\
$nD_{5/2}$	&	0.036   	&	0.073	&	1.770	&	3.636	\\
\hline 
\end{tabular}
\label{table_1} 
\end{table}

Combining $\tau_{\text{SP}}$ and $\tau_{\text{BBR}}$,  
the effective lifetime $\tau_{\text{eff}}$ of a Rydberg state becomes~\cite{Gallagher1979}
\begin{equation}\label{eq_4}
\dfrac{1}{\tau_{\text{eff}}}=\dfrac{1}{\tau_{\text{SP}}}+\dfrac{1}{\tau_{\text{BBR}}}.
\end{equation}
We note that the  aforementioned lifetimes depend on $n$, $L$, and $J$ but not on  $M_J$.  Correspondingly, the Rydberg series are labeled by $nL_J$ in this section. Columns 2-4 in Table~\ref{table_2} report the BBR, spontaneous, and effective lifetimes for selected  $n$ between $40$ and $70$ for the Rydberg series $nS_{1/2}$, $nP_{1/2}$, $nP_{3/2}$, $nD_{3/2}$, and $nD_{5/2}$.  The good agreement between our $\tau_{\text{eff}}$ and theoretical and experimental literature values (column 5 of Table~\ref{table_2}) validates our calculations.

\begin{figure}[!t]
{\includegraphics[ scale=.30]{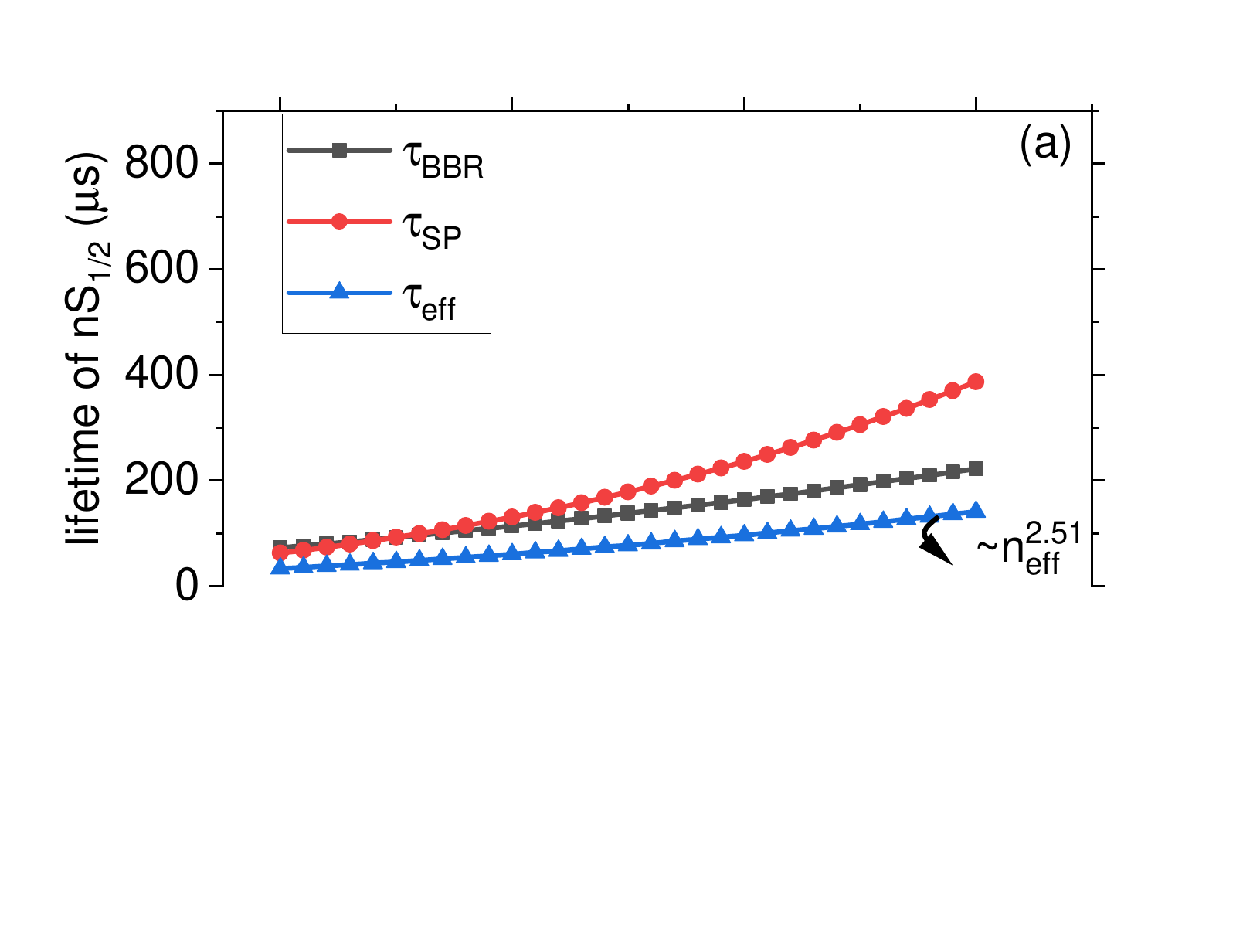}}\\
\vspace{-3.17cm}
{\includegraphics[ scale=.30]{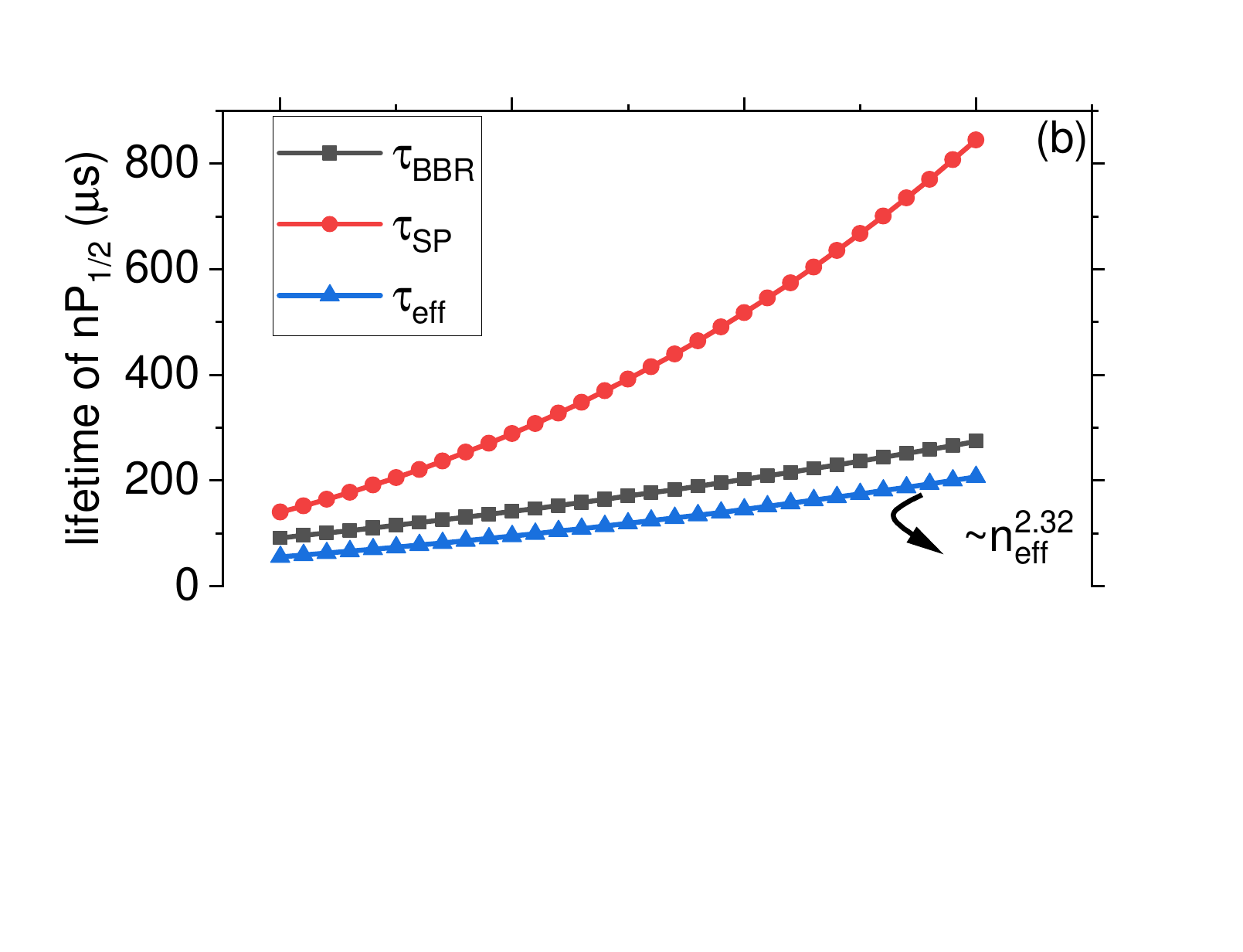}}\\
\vspace{-3.17cm}
{\includegraphics[ scale=.30]{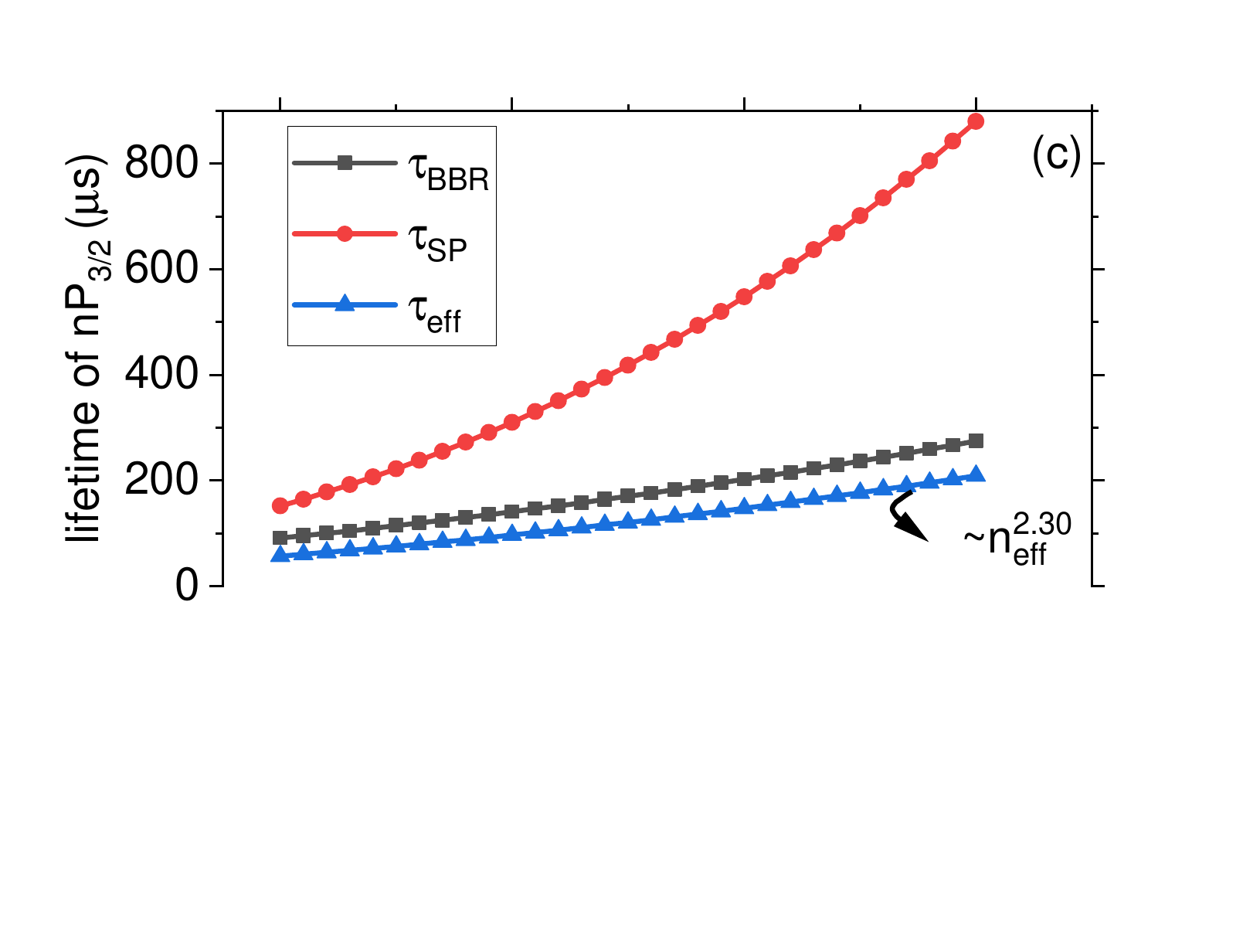}}\\
\vspace{-3.17cm}
{\includegraphics[ scale=.30]{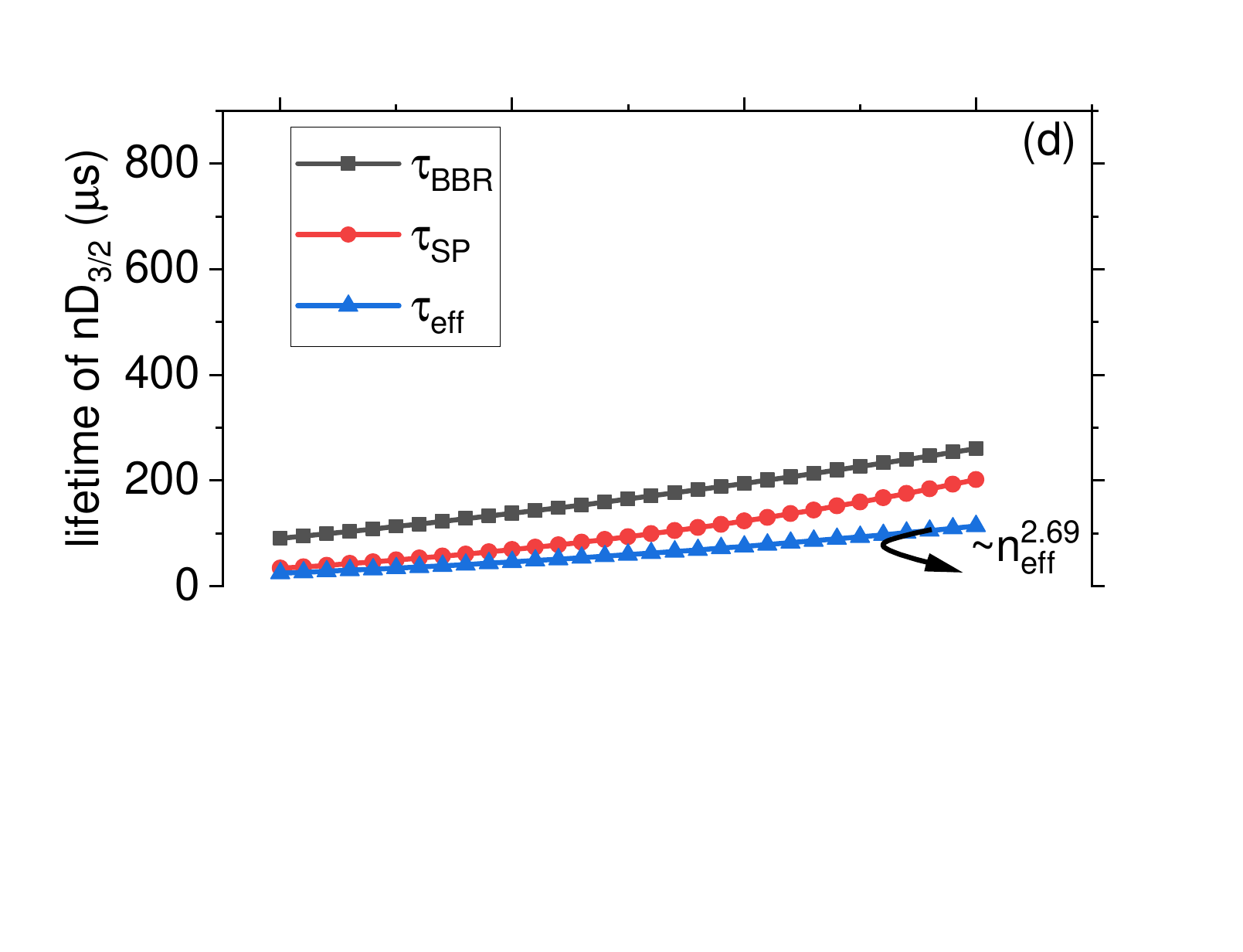}}\\
\vspace{-3.17cm}
{\includegraphics[ scale=.30]{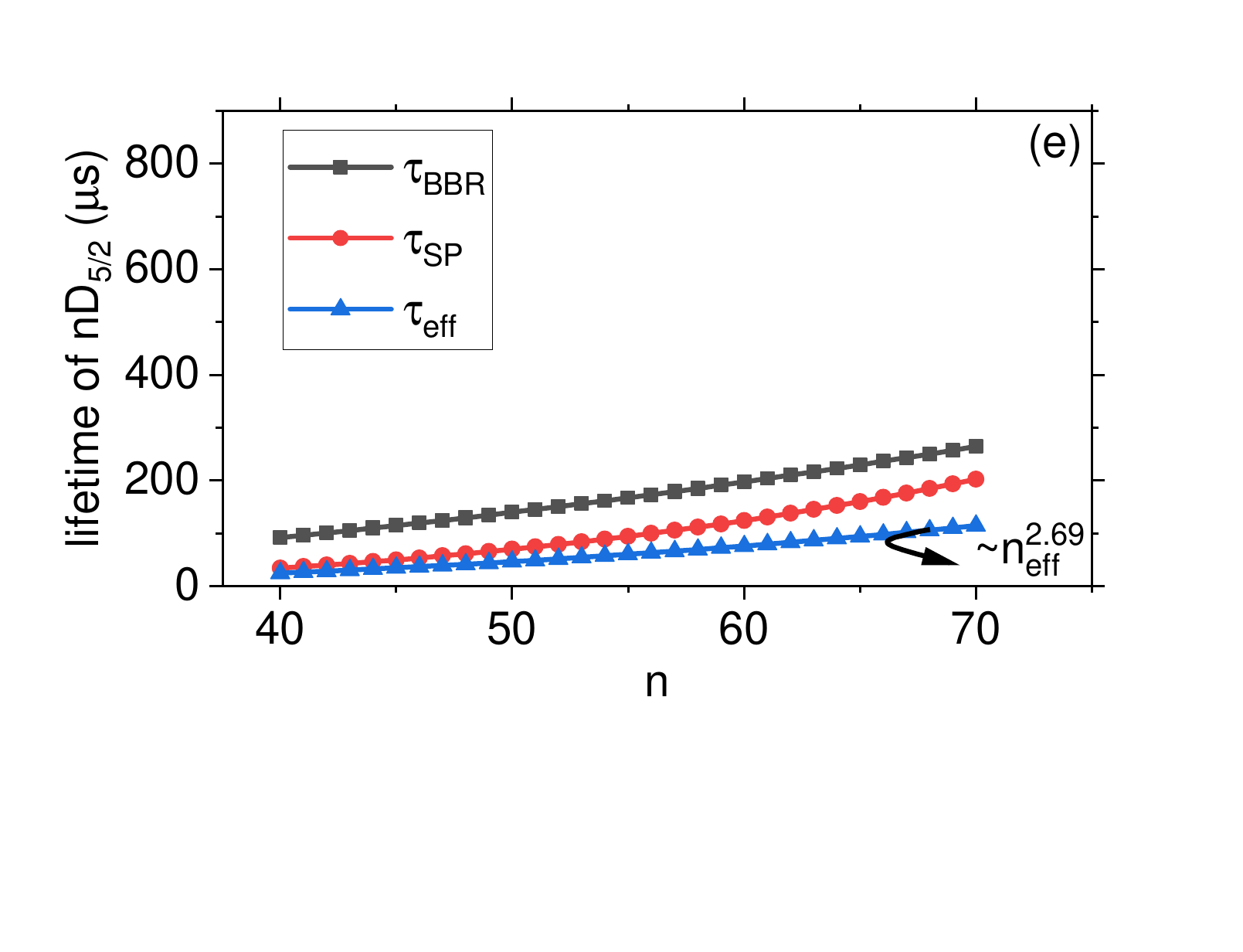}}\\
\vspace{-1.47cm}
\vspace*{-0.1in}
\caption{Blackbody radiation lifetime $\tau_{\text{BBR}}$, spontaneous lifetime $\tau_{\text{SP}}$, and effective lifetime $\tau_{\text{eff}}$  for the Rydberg series   (a) $nS_{1/2}$, (b) $nP_{1/2}$, (c) $nP_{3/2}$, (d) $nD_{3/2}$, and (e)  $nD_{5/2}$ (the Rydberg states are labeled by $nL_J$). The scaling of $\tau_{\text{eff}}$ with $n_{\text{eff}}$ is shown in each panel.  }
\label{Fig1}
\end{figure}

Squares, circles, and triangles in Fig.~\ref{Fig1} show the BBR, spontaneous, and effective lifetimes for the Rydberg series $nS_{1/2}$, $nP_{1/2}$, $nP_{3/2}$, $nD_{3/2}$, and $nD_{5/2}$ as a function of  $n$. All three lifetimes are observed to increase with increasing $n$. For the $n$ considered, $\tau_{\text{SP}}$ and  $\tau_{\text{BBR}}$ cross for the $nS_{1/2}$ series (the crossing occurs at $n=45$) but not for the other series. For the  $nP_{1/2}$ and $nP_{3/2}$ series, $\tau_{\text{SP}}$ is larger than $\tau_{\text{BBR}}$ for the range of $n$ considered. Conversely,  for the $nD_{3/2}$ and $nD_{5/2}$ series, $\tau_{\text{BBR}}$ is larger than $\tau_{\text{SP}}$ for the range of $n$ considered.  It is well-known that  the spontaneous lifetime of Rydberg states increases as $n^3$~\cite{Gallagher1994}. Since the quantum defect is small,  one finds $n^3 \approx (n_{\text{eff}})^3$, where the effective principal quantum number $n_{\text{eff}}$ is defined in Eq.~(\ref{eq_A2}).  The room temperature  BBR photons lead to a decrease of the Rydberg atom lifetime and a modification of the scaling law. Fitting our numerical results for $n \in [40, 70]$,  we find that the scaling laws for the $nS_{1/2}$, $nP_{1/2}$, $nP_{3/2}$, $nD_{3/2}$, and $nD_{5/2}$ series become $\tau_{\text{eff}} \propto (n_{\text{eff}})^{2.51}$, $(n_{\text{eff}})^{2.32}$, $(n_{\text{eff}})^{2.30}$, $(n_{\text{eff}})^{2.69}$, and $(n_{\text{eff}})^{2.69}$, respectively. This shows that the $nP_{1/2}$ and $nP_{3/2}$ states exhibit, among the Rydberg series  considered, the largest scaling law changes due to the BBR photons.

\begin{widetext}

\begin{table}[h]
\caption{Blackbody radiation lifetime $\tau_{\text{BBR}}$, spontaneous lifetime $\tau_{\text{SP}}$, and effective lifetime $\tau_{\text{eff}}$ for 
selected 
Rydberg states.  The effective lifetimes $\tau_{\text{eff}}$ are compared to  literature data (see column~5).  Entries marked by the superscript $a$ are taken from Ref.~\cite{Beterov2009} while those marked by $b$ ($c$) are theoretical 
(experimental) data 
taken from Ref.~\cite{Song2022}.} 
\centering 
\begin{tabular}{c |c|c | c| c} 
\hline\hline 
state	&	$\tau_{\text{BBR}}$ ($\mu$s)	&	$\tau_{\text{SP}}$	($\mu$s)&	$\tau_{\text{eff}}$ ($\mu$s)	&	$\tau_{\text{eff}}$	($\mu$s) (other)\\
	\hline
$40S_{1/2}$	&	72.859	&	62.679	&	33.693	&	33.032$^a$	\\
$50S_{1/2}$	&	114.170	&	130.885	&	60.979	&	60.414$^a$	\\
$60S_{1/2}$	&	164.091	&	236.301	&	96.842	&	97.082$^a$, 95.2$^b$, $95.4\pm 2.7^c$	\\
$70S_{1/2}$	&	222.360	&	387.026	&	141.223	&	143.28$^a$, 139.1$^b$,  $138.6\pm 2.4^c$	\\
	&		&		&		&		\\
$40P_{1/2}$	&	91.728	&	140.467	&	55.505	&	55.491$^a$	\\
$50P_{1/2}$	&	141.683	&	288.971	&	95.130	&	95.070$^a$	\\
$60P_{1/2}$	&	202.501	&	517.278	&	144.830	&	145.53$^a$	\\
$70P_{1/2}$	&	273.989	&	844.637	&	205.970	&	206.88$^a$	\\
	&		&		&		&		\\
$40P_{3/2}$	&	91.031	&	152.193	&	57.260	&	56.961$^a$	\\
$50P_{3/2}$	&	141.012	&	310.208	&	97.152	&	96.944$^a$	\\
$60P_{3/2}$	&	202.219	&	547.844	&	147.230	&	147.70$^a$	\\
$70P_{3/2}$	&	274.543	&	879.678	&	208.860	&	209.24$^a$	\\
	&		&		&		&		\\
$40D_{3/2}$	&	90.622	&	34.063	&	24.823	&	24.757$^a$	\\
$50D_{3/2}$	&	137.857	&	69.288	&	46.222	&	46.112$^a$	\\
$60D_{3/2}$	&	194.589	&	123.603	&	75.604	&	75.589$^a$	\\
$70D_{3/2}$	&	260.565	&	201.756	&	113.853	&	113.71$^a$	\\
	&		&		&		&		\\
$40D_{5/2}$	&	92.107	&	34.402	&	25.011	&	25.047$^a$	\\
$50D_{5/2}$	&	140.032	&	69.856	&	46.595	&	46.606$^a$	\\
$60D_{5/2}$	&	197.583	&	124.407	&	76.252	&	76.34$^a$,  76.3$^b$, $75.7\pm 0.7^c$	\\
$70D_{5/2}$	&	264.503	&	202.710	&	114.650	&	114.76$^a$, 114.1$^b$,  $115.4\pm 2.2^c$	\\
\hline 
\end{tabular}
\label{table_2} 
\end{table}

\end{widetext}

\section{Polarizability  and  double magic wavelengths for linearly polarized light}\label{sec_3}

The dynamic electric-dipole polarizability $\alpha_v(\omega)$ (or dynamic polarizability in what follows) describes the response of the single-valence  state $v$  to an  external electric field of magnitude $E$ and angular frequency $\omega$.  Assuming a Gaussian shaped laser beam, a positive dynamic polarizability indicates an attractive force (i.e., trapping) while a negative  dynamic polarizability indicates a repulsive force (i.e., anti-trapping). Throughout this section, we are interested in determining so-called double magic wavelengths $\lambda^{(d)}$ at which the cesium ground state and a Rydberg state experience the same negative Stark shifts, i.e., the same positive dynamic  polarizabilities.

\begin{figure}
\vspace*{-1in}
\hspace*{0.2in}
{\includegraphics[ scale=.70]{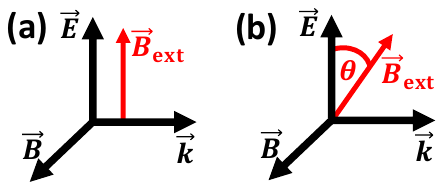}}
\vspace*{-4in}
\caption{Definition of the angle $\theta$. In (a), the external magnetic field vector $\vec{B}_{\text{ext}}$ lies along the electric field vector $\vec{E}$ (polarization vector); this corresponds to $\theta=0$. In (b), the  angle $\theta$ between $\vec{B}_{\text{ext}}$ and $\vec{E}$ is finite. Section~\ref{sec_5} discusses polarizability control through variation of  $\theta$.  }   
\label{Fig_schematic}
\end{figure}

To determine $\alpha_v(\omega)$, we decompose it into the core polarizability $\alpha_v^C(\omega)$, the valence-core polarizability $\alpha_v^{VC}(\omega)$, and the valence polarizability $\alpha_v^V(\omega)$,
\begin{eqnarray}\label{eq_4}
  \alpha_v(\omega)=  \alpha_v^C(\omega)+\alpha_v^{VC}(\omega)+\alpha_v^V(\omega).
\end{eqnarray}
We find that $\alpha_v^C(\omega)$ is essentially frequency-independent 
[$\alpha_v^C(\omega) \approx 16.12$~a.u.] while 
$\alpha_v^{VC}(\omega)$ is, to a good approximation, zero for all Rydberg states considered in this work; see Appendix~\ref{Appendix_B} for details.  The valence polarizability $\alpha_v^V(\omega)$ for linearly polarized light consists of the scalar component  $\alpha_{v,J}^{(0)}{(\omega)}$ and the tensor component $\alpha_{v,J}^{(2)}(\omega)$.  Note that $\alpha_{v,J}^{(2)}(\omega)$  vanishes for the $nL_{1/2}$ Rydberg series.  In this section,  the external magnetic field $\vec{B}_{\text{ext}}$, which sets the quantization axis, and the polarization direction  lie along the $z$-axis [see Fig.~\ref{Fig_schematic}(a)]. This implies that the angle $\theta$, which enters into the tensor polarizability [see Eq.~(\ref{eq_A11})], is zero; the ability to vary  $\theta$ will be exploited in Sec.~\ref{sec_5} to identify triple and quadruple magic wavelengths.   The theoretical framework, implementation, and computational details to determine the energies, wavefunctions, dipole matrix elements, and polarizabilities are presented in Appendices~\ref{Appendix_A}-\ref{Appendix_B}.

To benchmark our results, Table~\ref{table_3} compares the static (i.e., $\omega=0$) scalar and tensor polarizabilities, which are independent of  $M_J$, for selected $nL_J$ Rydberg states.  Our results agree well with theoretical results from Ref.~\cite{Yerokhin2016}, which calculated the polarizabilities---as we (see Appendix~\ref{Appendix_A})---using the Coulomb approximation method.  Table~\ref{table_3} also presents a comparison with experimental data from three different references.
The static scalar polarizability of $1.84\pm 0.02 \times 10^{12}$~a.u. for  the $70S_{1/2}$ Rydberg state from a very recent paper~\cite{Bai2020} agrees within three sigma with our result of $1.91 \times 10^{12}$~a.u..
Of the combined six experimental data from 1995~\cite{Lei1995} and 2014~\cite{Zhao2011}, one agrees with our theoretical prediction within error bars while the others deviate by between two and six sigma. Given that the polarizabilities reported in Table~\ref{table_3} vary over three orders of magnitude,  we interpret the rough agreement as a validation of our theoretical predictions despite the fact that the theory-experiment agreement is not perfect.

\begin{figure}[!h]
{\includegraphics[ scale=.30]{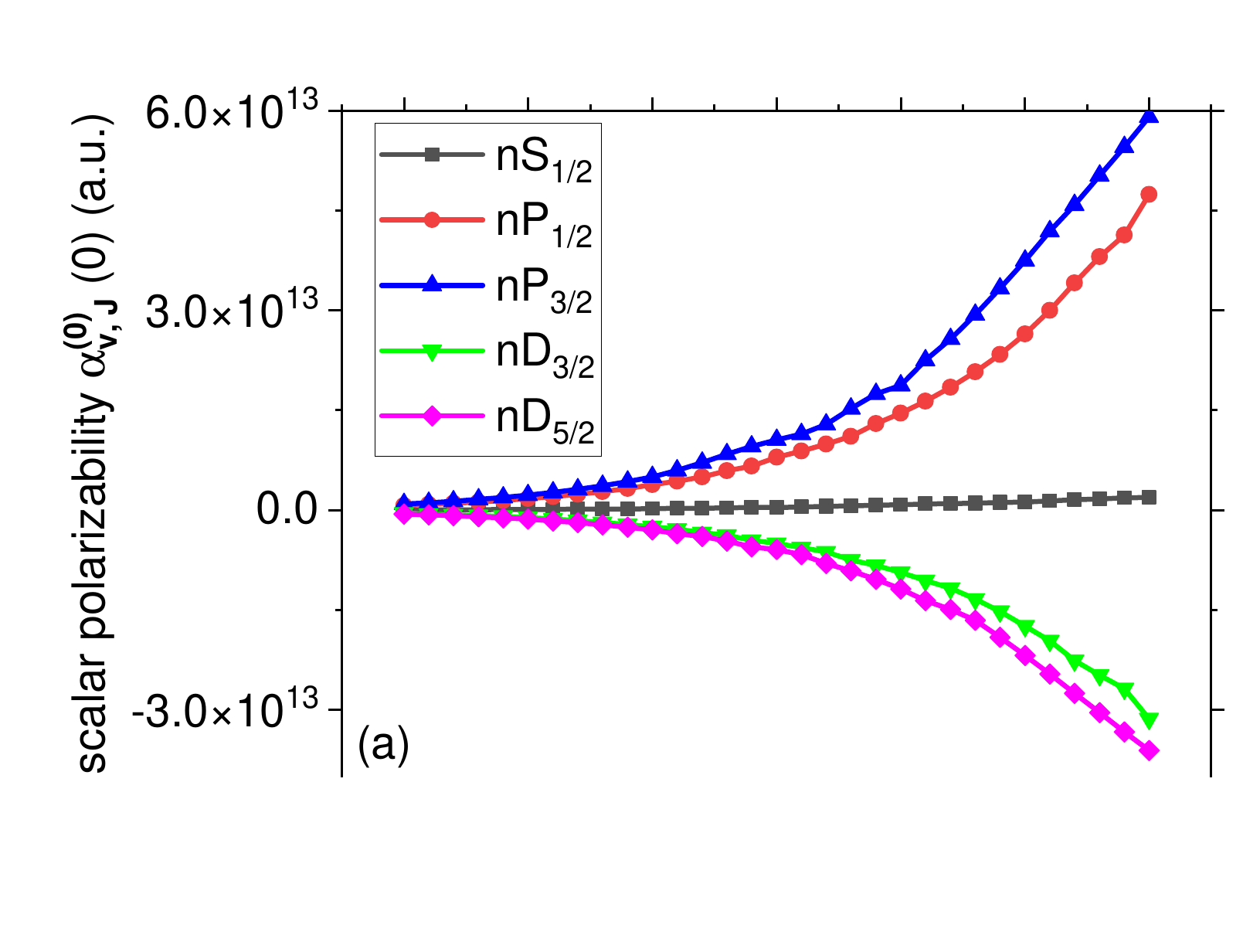}}\\
\vspace{-1.875cm}
{\includegraphics[ scale=.30]{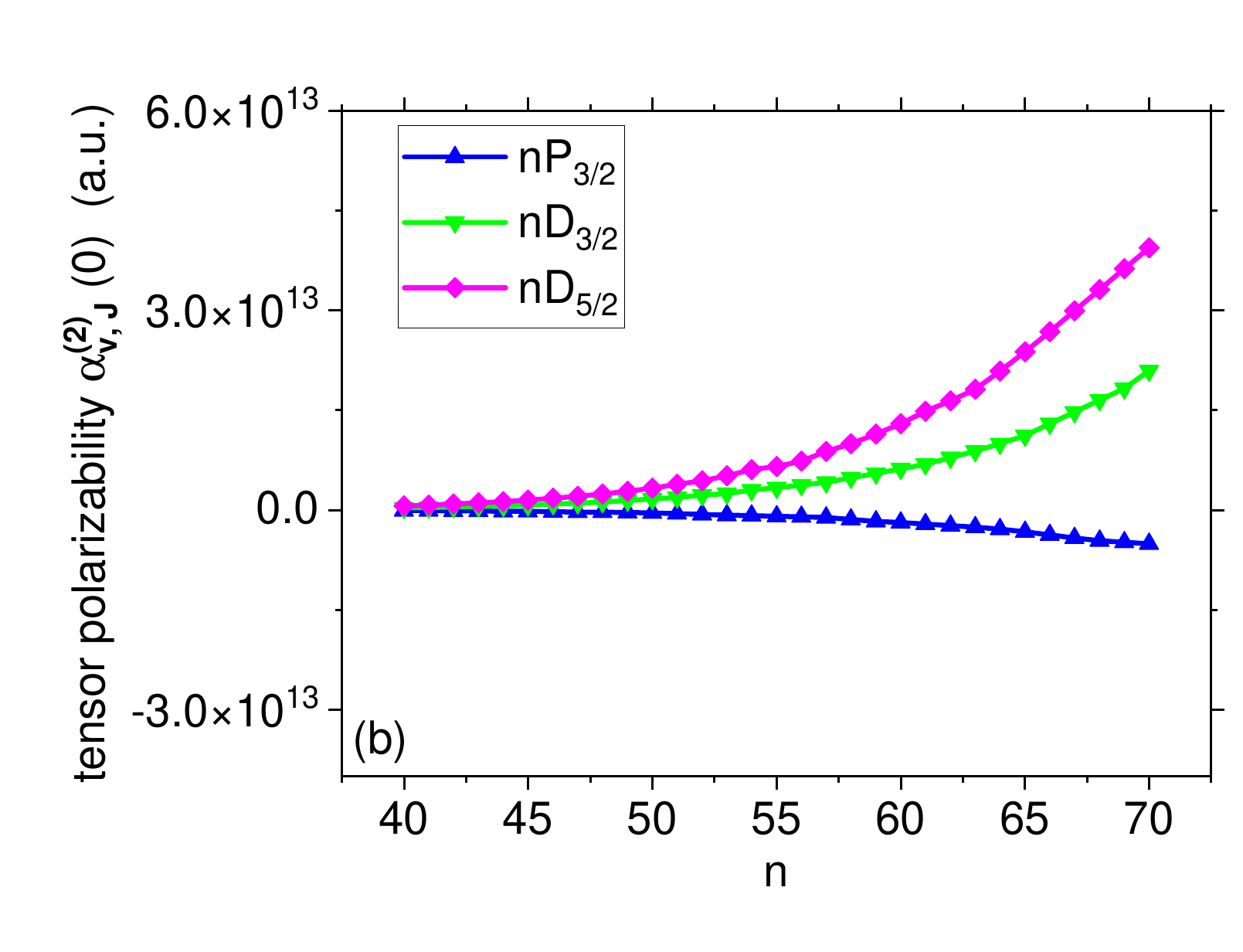}}\\
\vspace*{-0.15in}
\caption{(a) Static scalar polarizability $\alpha_{v,J}^{(0)}(0)$ for the   $nS_{1/2}$, $nP_{1/2}$, $nP_{3/2}$,  $nD_{3/2}$, and  $nD_{5/2}$ series and (b) static tensor polarizability $\alpha_{v,J}^{(2)}(0)$ for the  $nP_{3/2}$,  $nD_{3/2}$, and  $nD_{5/2}$ series as a function of $n$. } 
\label{Fig2}
\end{figure}

Figure~\ref{Fig2} shows the static scalar and tensor polarizabilities as a function of $n$.    The static scalar polarizabilities for the  $nS_{1/2}$, $nP_{1/2}$, and $nP_{3/2}$ series are positive while those for the $nD_{3/2}$ and  $nD_{5/2}$ series are negative.  The static  scalar polarizability for the $nS_{1/2}$ series   varies weakly with $n$   
compared to the other Rydberg series  
considered. Similar behavior  was  observed in Ref.~\cite{Yerokhin2016}.  
The static scalar and tensor polarizabilities of the 
$nP$- and $nD$-series  scale approximately as $n^7$.   The static tensor polarizabilities are  negative for the $nP_{3/2}$ series and positive for the $nD_{3/2}$ and  $nD_{5/2}$ series; recall that the  tensor polarizability is zero for the series with $J=1/2$.

\begin{widetext}

\begin{figure}[!h]
\vspace*{-0.8in}
{\includegraphics[ scale=.50]{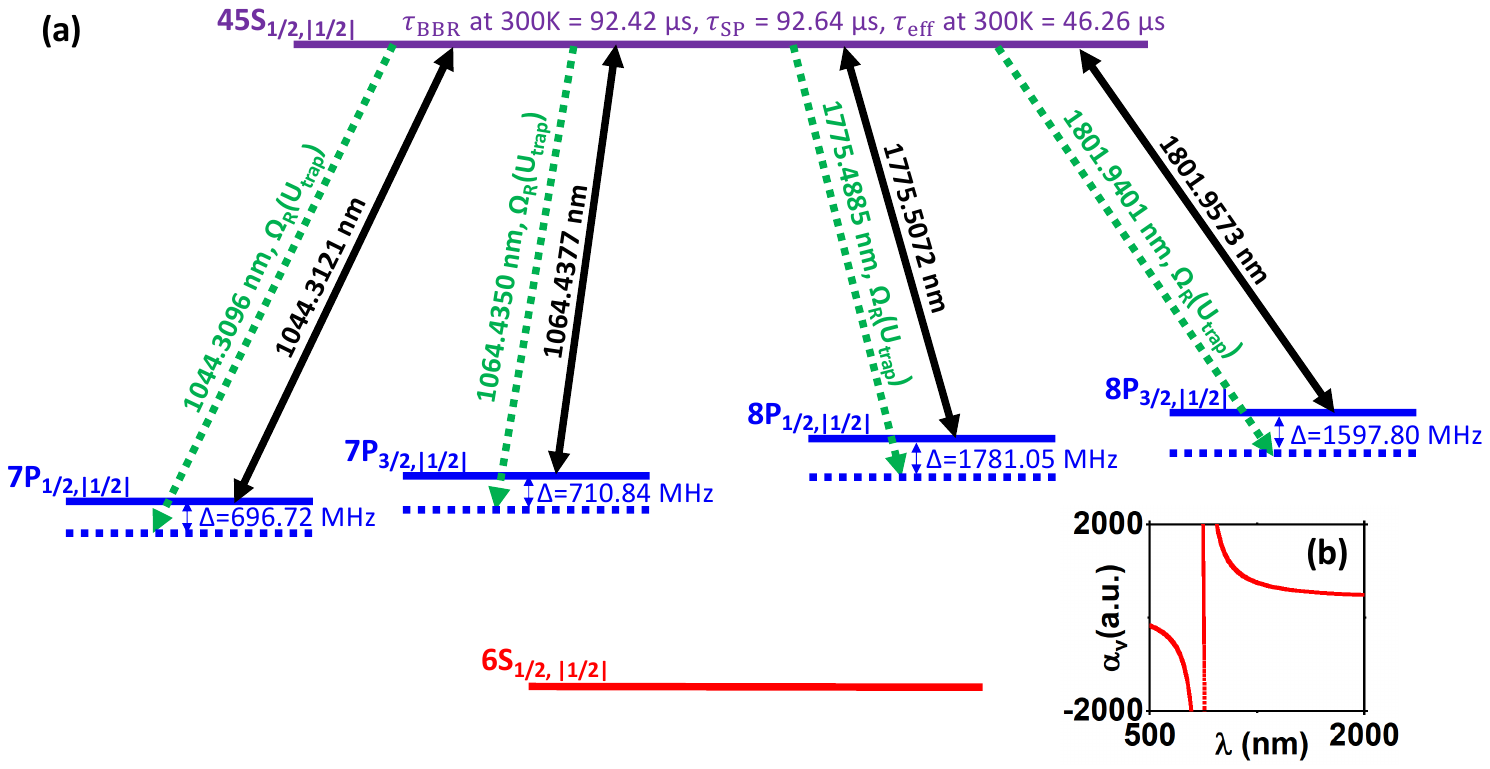}}
\vspace*{-0.8in}
\caption{(a) Schematic energy level diagram of the cesium $6S_{1/2,|1/2|}$ ground state, the $45S_{1/2, |1/2|}$ Rydberg state, and four auxiliary states ($7P_{1/2,|1/2|}$, $7P_{3/2,|1/2|}$, $8P_{1/2,|1/2|}$, and $8P_{3/2,|1/2|}$).  The blackbody radiation lifetime $\tau_{\text{BBR}}$, spontaneous lifetime $\tau_{\text{SP}}$, and effective lifetime  $\tau_{\text{eff}}$  of the  $45S_{1/2, |1/2|}$ state in the absence of any trapping light are noted. Black solid and green dotted arrows represent the resonance wavelength $\lambda^{(\text{res})}$ and the magic wavelength $\lambda^{(d)}$ of the trapping light, respectively. 
 The  detuning $\Delta$ to the  auxiliary state is indicated. The Rabi frequency $\Omega_R$ for the Rydberg state--auxiliary state transition depends on the   trap depth $|U_{\text{trap}}|$ (see discussion in Sec.~\ref{sec_4}). (b) Dynamic polarizability $\alpha_v(\omega)$ of the  ground state as a function of the wavelength $\lambda$.
 The large slope of $\alpha_v(\omega)$ near $870$~nm originates in the 
  $6S_{1/2}-6P_{1/2}$ resonance line at $894.5930$~nm and the $6S_{1/2}-6P_{3/2}$ resonance line at  
 $852.3473$~nm.}   
\label{Fig3}
\end{figure}

\end{widetext}

We now turn to the dynamic (i.e., $\omega \ne 0$)  polarizabilities for the $nS_{1/2,|M_J|}$, $nP_{1/2,|M_J|}$, $nP_{3/2,|M_J|}$,  $nD_{3/2,|M_J|}$, and  $nD_{5/2,|M_J|}$ Rydberg series. Note that since the dynamic polarizabilities depend on $|M_J|$, the labeling of the Rydberg series includes the $|M_J|$ subscript.  The background polarizability of a Rydberg state, i.e., the polarizability away from resonant transition frequencies (see below), is approximately $-e^2/(m_e \omega^2)$~\cite{Younge2010}, where $e$ and $m_e$ are the charge and mass of the electron, respectively. Therefore, in general, the Rydberg atom is pushed toward the low-intensity region of the Gaussian beam. This implies that a conventional trap, which operates with far-off-resonant red-detuned light for the ground state, is seen as an attractive potential well for the ground state but as a repulsive potential barrier for the Rydberg state. Thus, simultaneous trapping  of the ground and Rydberg states by the conventional approach is impossible. A possible solution is to  create a  red-detuned trap for the ground state with a wavelength that is blue-detuned and  very close to the resonance transition between the Rydberg state and an  auxiliary state~\cite{Bai2020_1}. Since we are, motivated by experimental considerations, interested in trapping light with wavelengths between around $1,000$ and $2,000$~nm, the number of auxiliary states is restricted to low-lying excited states. The  auxiliary states considered in this work (see Table~\ref{table_4}) are connected to the ground state and a Rydberg state via dipole transitions for linearly polarized light ($|\Delta L| =1$; $|\Delta J|=0,1$; and $\Delta M_J=0$).  While there exist lasers that operate in the $500-1,000$~nm wavelength regime, this wavelength window  is excluded from our analysis   since the magic wavelengths found in this wavelength regime  possess negative polarizabilities [see also Fig.~\ref{Fig3}(b)].

\begin{widetext}

\begin{table}[h]
\caption {Static scalar polarizability $\alpha_{v,J}^{(0)}(0)$ and static tensor polarizability $\alpha_{v,J}^{(2)}(0)$   are compared with theoretical and experimental results from the literature for  $n=40,50,60$, and $70$.  The polarizabilities are reported in atomic units 
 (a.u.) in the format  $x[y]$, which stands for $x\times 10^y$. Entries marked by the superscripts $a$, $b$, $c$, and $d$ are taken from Ref.~\cite{Yerokhin2016}  (theoretical data),  Ref.~\cite{Zhao2011}  (experimental data),  Ref.~\cite{Lei1995}  (experimental data), and Ref.~\cite{Bai2020} (experimental data), respectively.} 
\centering 
\begin{tabular}{c |c|c | c| c|c||c|c|c} 
  \hline
\hline
&\multicolumn{5}{c||}{scalar polarizability $\alpha_{v,J}^{(0)}(0)$}&\multicolumn{3}{c}{tensor polarizability $\alpha_{v,J}^{(2)}(0)$}\\
\hline
$n$	&	 $S_{1/2}$	&	 $P_{1/2}$	&	 $P_{3/2}$	&	 $D_{3/2}$	&	 $D_{5/2}$	&	 $P_{3/2}$	&	 $D_{3/2}$	&	 $D_{5/2}$	\\
\hline
40	&	$4.27[10]$	&	$6.96[11]$	&	$9.14[11]$	&	$-4.71[11]$	&	$-5.67[11]$	&	$-8.06[10]$	&	$3.15[11]$	&	$6.30[11]$	\\
	&	$4.27[10]^a$	&	$6.96[11]^a$	&	$9.14[11]^a$	&	$-4.71[11]^a$	&	$-5.67[11]^a$	&	$-8.06[10]^a$	&	$3.15[11]^a$	&	$6.30[11]^a$	\\
50	&	$2.15[11]$	&	$3.79[12]$	&	$4.98[12]$	&	$-2.49[12]$	&	$-3.00[12]$	&	$-4.36[11]$	&	$1.66[12]$	&	$3.31[12]$	\\
	&	$2.15[11]^a$	&	$3.79[12]^a$	&	$4.98[12]^a$	&	$-2.49[12]^a$	&	$-3.00[12]^a$	&	$-4.36[11]^a$	&	$1.66[12]^a$	&	$3.31[12]^a$	\\
	&		&		&		&	$-2.06\pm 0.06[12]^b$	&	$-2.79\pm 0.10 [12]^b$	&	$-4.36[11]^a$	&	$1.67\pm 0.07[12]^b$	&	$2.67\pm 0.23[12]^b$	\\
	&		&		&		&		&		&		&	$2.07\pm 0.06 [12]^c$	&	$3.77\pm 0.08[12]^c$	\\
	
60	&	$8.22[11]$	&	$1.46[13]$	&	$1.87[13]$	&	$-9.36[12]$	&	$-1.19[13]$	&	$-1.90[12]$	&	$6.17[12]$	&	$1.30[13]$	\\
70	&	$1.91[12]$	&	$4.74[13]$	&	$5.91[13]$	&	$-3.14[13]$	&	$-3.61[13]$	&	$-5.05[12]$	&	$2.10[13]$	&	$3.94[13]$	\\
&	$1.84\pm 0.02[12]^d$	&		&		&		&		&		&		&		\\
\hline 
\end{tabular}
\label{table_3} 
\end{table}

 \end{widetext}

\begin{figure}[!h]
{\includegraphics[ scale=.30]{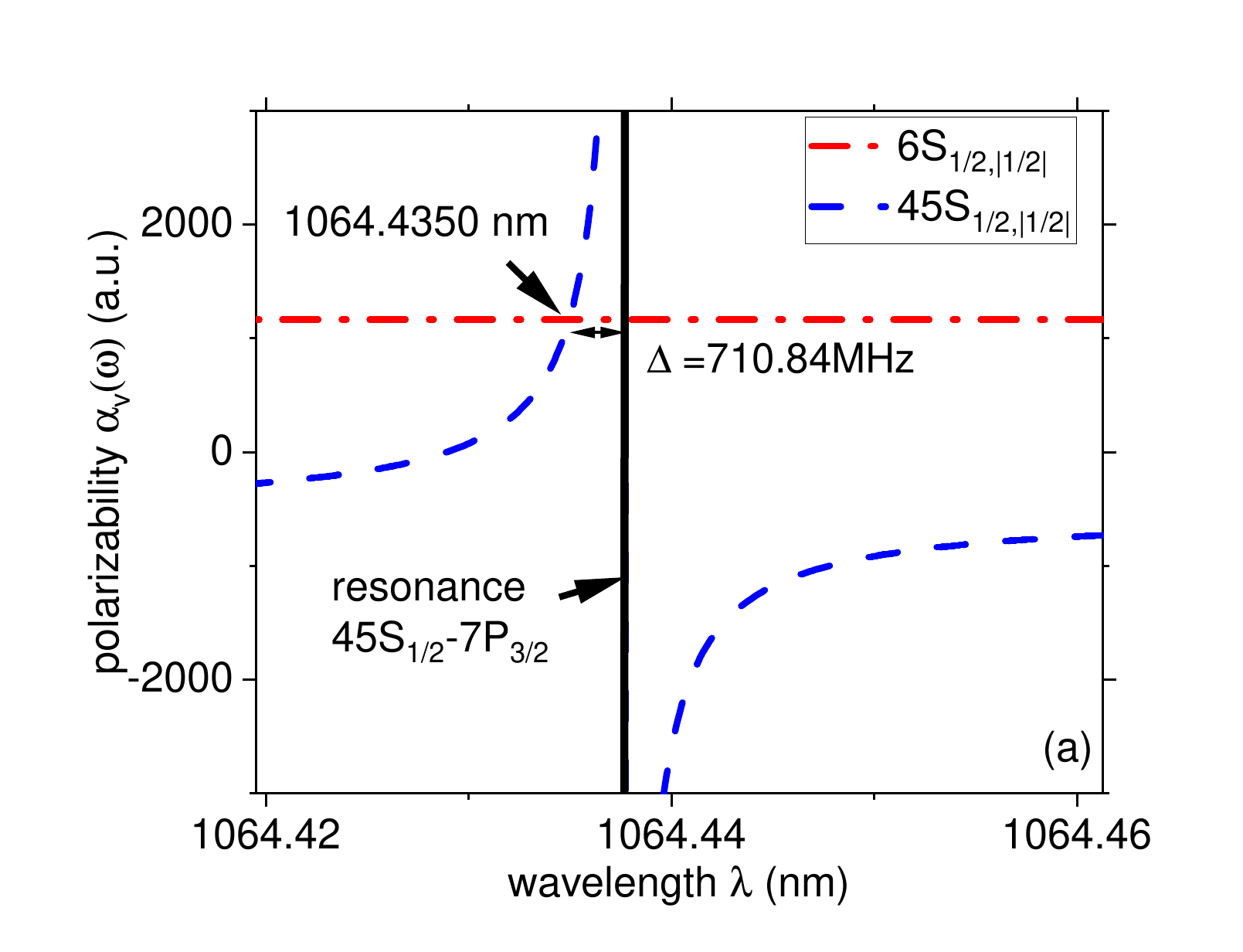}}\\
\vspace*{-0.25in}
{\includegraphics[ scale=.30]{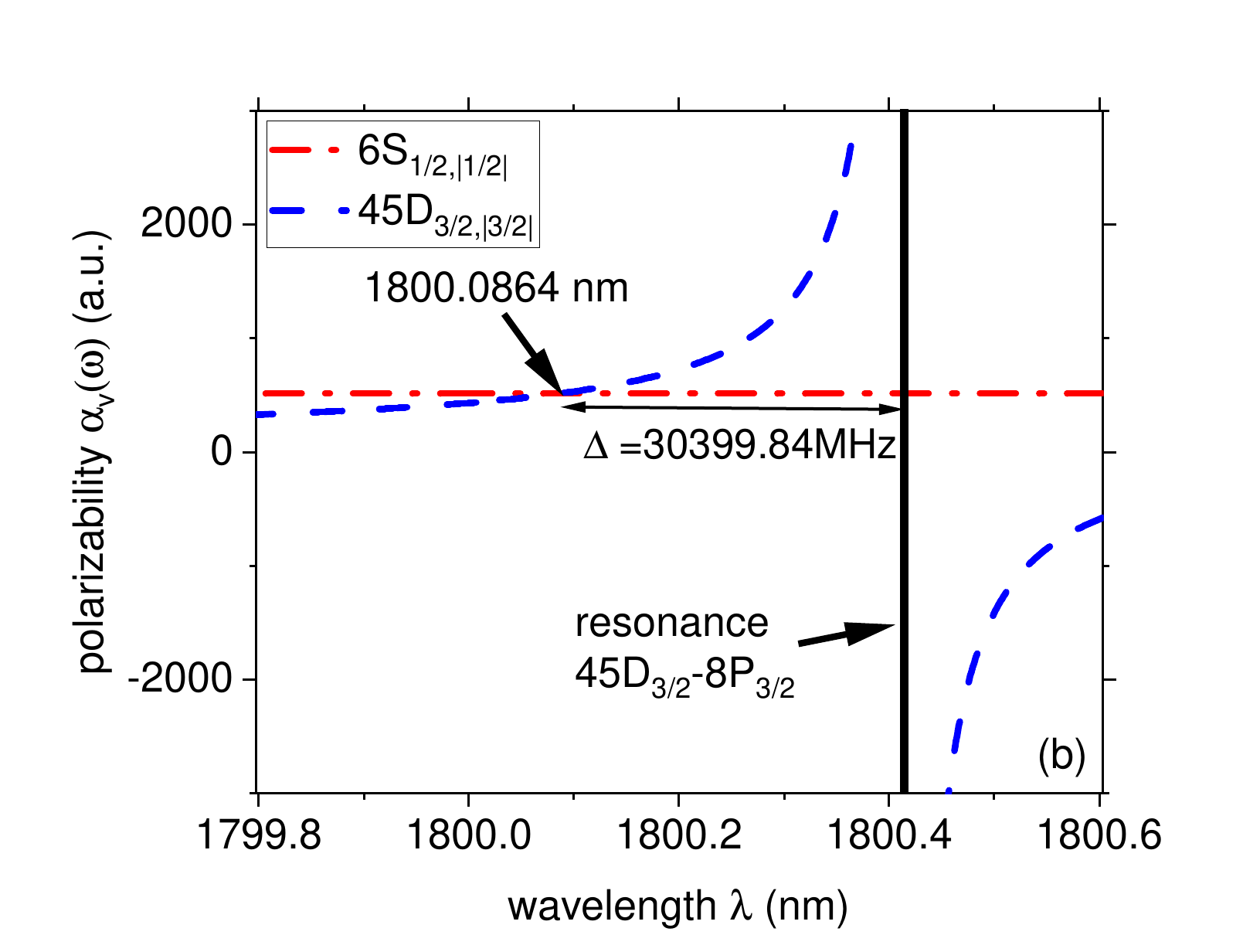}}\\
\vspace*{-0.05in}
\caption{Dynamic polarizability $\alpha_v(\omega)$ of the $6S_{1/2,|1/2|}$ ground state (red dash-dotted line) and the Rydberg state (blue dashed line) for (a) the $45S_{1/2, |1/2|}$ state with auxiliary state $7P_{3/2,|1/2|}$ and (b) the $45D_{3/2, |3/2|}$  state with auxiliary state $8P_{3/2,|3/2|}$.  The intersection of the polarizability curves occurs at the double magic wavelengths (marked by an arrow) of (a) $\lambda^{(d)}=1,064.4350$~nm 
and
(b) $\lambda^{(d)}=1,800.0864$~nm.  The resonance position is marked by the    black solid vertical lines.  The detuning of the double magic wavelength from the resonance transition is 
(a) $\Delta=710.84$~MHz
and
(b) $\Delta=30,399.84$~MHz.} 
\label{Fig4}
\end{figure}

To make the discussion concrete, Fig.~\ref{Fig3}  considers the $6S_{1/2,|1/2|}$ ground state and the $45S_{1/2,|1/2|}$  Rydberg state. Four auxiliary states, namely  $7P_{1/2,|1/2|}$,  $7P_{3/2,|1/2|}$, $8P_{1/2,|1/2|}$, and $8P_{3/2,|1/2|}$, have transition wavelengths between $1,000$~nm and $2,000$~nm
(see black numbers in Fig.~\ref{Fig3}).  To illustrate the emergence of double magic wavelengths, Fig.~\ref{Fig4}(a) shows the polarizability of the ground state and the  $45S_{1/2,|1/2|}$ Rydberg state in the vicinity of the $45S_{1/2,|1/2|}-7P_{3/2,|1/2|}$ resonance.
 While the polarizability of the ground state is essentially constant on the scale shown, the polarizability of the $45S_{1/2,|1/2|}$ Rydberg state changes sign at the resonance with the $7P_{3/2,|1/2|}$ auxiliary state [black vertical line in Fig.~\ref{Fig4}(a)]. The two polarizability curves cross at the double magic wavelength  $\lambda^{(d)}=1,064.4350$~nm, which is blue-detuned by $\Delta=710.84$~MHz from the resonance; here, the detuning $\Delta$ is calculated according to $\Delta=c/\lambda^{(d)}-c/\lambda^{\text{(res)}}$ ($\lambda^{\text{(res)}}$ denotes the resonance wavelength).  
 The  second column of Table~\ref{table_5}  contains   $\lambda^{(d)}$
for the ground state and the $nS_{1/2,|1/2|}$
Rydberg series  ($n$ is listed in the first column)  for the auxiliary state $7P_{3/2,|1/2|}$ [same auxiliary state as considered in Fig.~\ref{Fig4}(a)]. The third and fourth columns of Table~\ref{table_5} report   $\Delta$ and the corresponding polarizability $\alpha_v^{(d)}$.

\begin{figure}[!h]
{\includegraphics[ scale=.30]{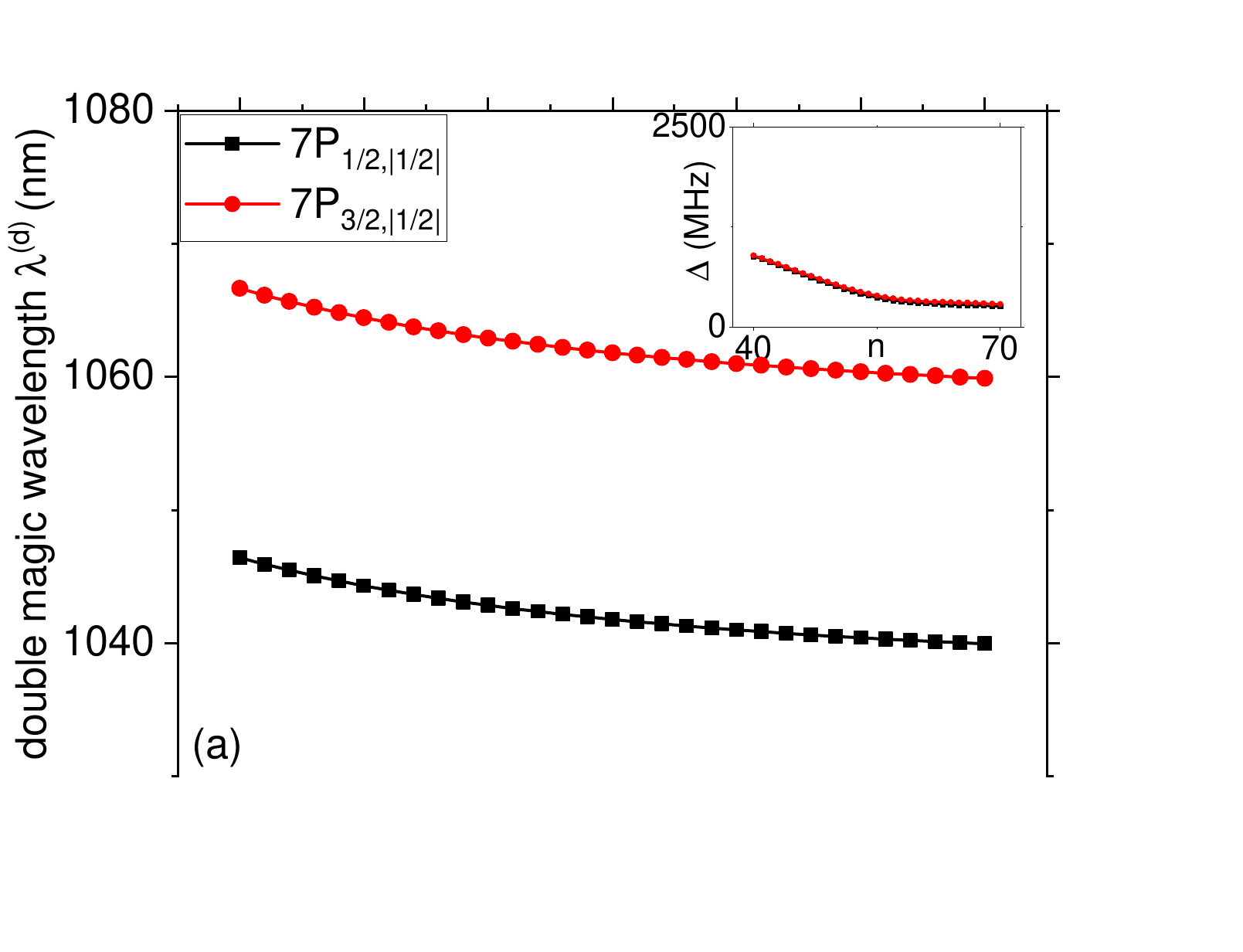}}\\
\vspace{-1.865cm}
{\includegraphics[ scale=.30]{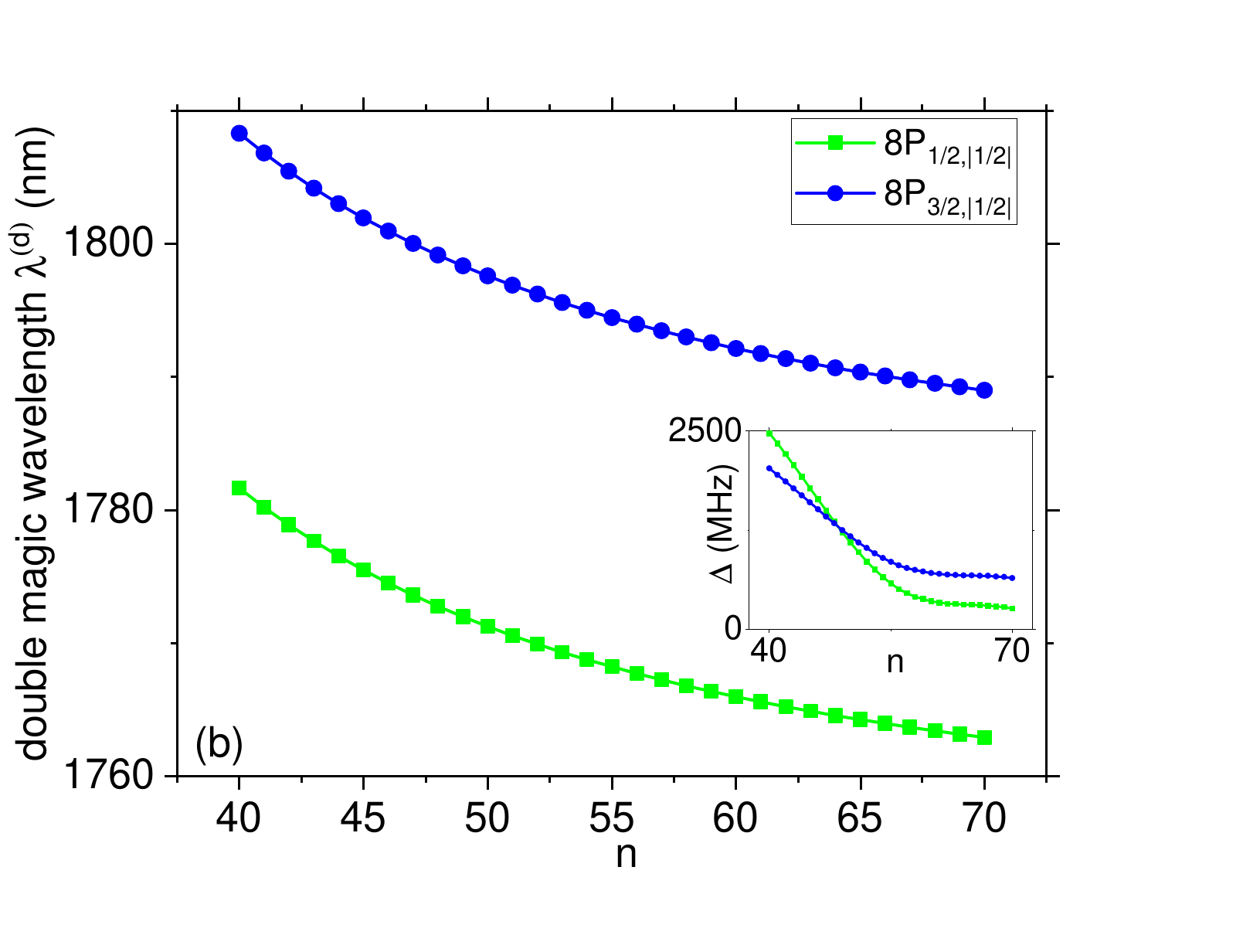}}\hspace{-0.14cm}
\vspace*{-0.2in}
\caption{Double magic wavelength $\lambda^{(d)}$ for  the ground state and the $nS_{1/2, |1/2|}$ Rydberg series   as a function of $n$ for the auxiliary states (a) $7P_{1/2,|1/2|}$ and $7P_{3/2,|1/2|}$, and (b) $8P_{1/2,|1/2|}$ and $8P_{3/2,|1/2|}$.  The insets show the corresponding detuning $\Delta$. In the inset of (a), the detunings for the $7P_{1/2,|1/2|}$ and $7P_{3/2,|1/2|}$ auxiliary states are nearly indistinguishable.}
\label{Fig5}
\end{figure}

Figures (not shown) similar to Fig.~\ref{Fig4} demonstrate that double magic wavelengths also exist for the other three auxiliary states enumerated in Fig.~\ref{Fig3}. Figure~\ref{Fig5} shows  $\lambda^{(d)}$ for the $nS_{1/2,|1/2|}$ series as a function of $n$ ($n \in [40,70]$) for the same auxiliary states as those considered in Fig.~\ref{Fig3}. Two double magic wavelengths lie around $1,050$~nm (auxiliary states with $n=7$ and $L=P$) and two lie around $1,790$~nm (auxiliary states with $n=8$ and $L=P$).   The corresponding detuning $\Delta$  from the resonance line  is shown in the insets. Figure~\ref{Fig5} shows that both the double magic wavelengths and detunings decrease monotonically  with increasing  $n$ for each of the  four auxiliary states considered.

Employing the same format as  used in Table~\ref{table_5}, Tables S.2-S.34 of the supplemental material~\cite{Supplement} contain the double magic wavelengths, detunings, and respective polarizabilities for    all other Rydberg series considered in this paper. Typically, one would like to work with large detunings to minimize light scattering~\cite{Saffman2010}. Inspection of Tables~\ref{table_5} and S.2-S.34 shows that $\Delta$ is particularly large  (greater than $5,000$~MHz for $n=40$) for the $nD_{3/2,|1/2|}$ series (auxiliary states $7P_{1/2,|1/2|}$, $8P_{1/2,|1/2|}$, and $4F_{5/2,|1/2|}$), the $nD_{3/2,|3/2|}$ series (auxiliary states $7P_{3/2,|3/2|}$, $8P_{3/2,|3/2|}$, and $4F_{5/2,|3/2|}$), the $nD_{5/2,|1/2|}$ series (auxiliary state $8P_{3/2,|1/2|}$), and the $nD_{5/2,|3/2|}$ series (auxiliary state  $8P_{3/2,|3/2|}$). As an example of these larger detuning cases, Fig.~\ref{Fig4}(b) shows the dynamic polarizability for the $45D_{3/2,|3/2|}$ state near the $45D_{3/2,|3/2|}-8P_{3/2,|3/2|}$ resonance line; for this case, $\Delta$ is equal to $30,399.84$~MHz. As discussed in Sec.~\ref{sec_4},  an assessment of the experimental feasibility of double magic wavelength trapping depends not only on the detuning value but also on the lifetime of the auxiliary state and on the coupling strength between the Rydberg state and the auxiliary state.

\begin{table}[h]
\caption{Rydberg states (column 1) and corresponding auxiliary states (columns 2) considered in this work (only auxiliary states that yield double magic wavelengths are listed). Auxiliary states for which triple magic wavelengths exist are in bold. Column~3 lists the  lifetime of the auxiliary state. Entries marked by the superscripts  $a$,  $b$, $c$,  $d$,  $e$, and $f$ are  experimental data taken from   Ref.~\cite{Ortiz1981}, Ref.~\cite{Marek1976}, Ref.~\cite{Marek1979}, Ref.~\cite{Marek1977}, Ref.~\cite{Neil1984}, and Ref.~\cite{Marek1977_2}, respectively. The lifetime of the auxiliary state $7P_{1/2,|1/2|}$, e.g., is $155$~ns, with an uncertainty of $4$~ns. Column 4 indicates the table number where the double and triple magic wavelengths are reported. For example, the double magic wavelengths $\lambda^{(d)}$ for simultaneous trapping of the ground state and a state from the $nS_{1/2,|1/2|}$ Rydberg series, using the auxiliary state $7P_{3/2,|1/2|}$, can be found in Table~\ref{table_5} while the triple magic wavelengths $\lambda^{(t)}$ can be found in Table S.35. } 
\centering 
\begin{tabular}{c |c|c | c} 
\hline\hline 
Rydberg series	&	auxiliary state	&	lifetime (ns)	&	table number\\
\hline
$nS_{1/2,|1/2|}$	&	$7P_{1/2,|1/2|}$   	&	155(4)$^a$& S.2/---   \\   
& \boldmath{$7P_{3/2,|1/2|}$} & 133(2)$^a$& \ref{table_5}/S.35\\

	&	\boldmath{$8P_{1/2,|1/2|}$}	& 	 307(14)$^b$	&S.3/S.36\\
 & \boldmath{$8P_{3/2,|1/2|}$}&	274(12)$^b$	&S.4/\ref{table_6} \\
\hline
$nP_{1/2,|1/2|}$	&	\boldmath{$6D_{3/2,|1/2|}$}   	&	60.0(2.5)$^c$ &
S.5/S.37\\
&\boldmath{$8S_{1/2,|1/2|}$}	&  87(9)$^d$&S.6/S.38 \\
&\boldmath{$7D_{3/2,|1/2|}$}	&		89(1)$^e$ &S.7/S.39\\
\hline
$nP_{3/2,|1/2|}$	&	\boldmath{$6D_{3/2,|1/2|}$}   	&	60.0(2.5)$^c$ &S.8/S.40 \\
&\boldmath{$6D_{5/2,|1/2|}$} & 60.7(2.5)$^c$&S.9/S.41\\
&	\boldmath{$8S_{1/2,|1/2|}$}&87(9)$^d$&S.10/S.42\\
&	\boldmath{$7D_{3/2,|1/2|}$}& 89(1)$^e$& S.11/S.43\\
	& \boldmath{$7D_{5/2,|1/2|}$}		&89(1)$^e$
 & S.12/S.44\\
\hline
$nP_{3/2,|3/2|}$	&	\boldmath{$6D_{3/2,|3/2|}$} 	&	60.0(2.5)$^c$& S.13/S.45\\
&	\boldmath{$6D_{5/2,|3/2|}$}& 60.7(2.5)$^c$ & S.14/S.46\\
&	\boldmath{$7D_{3/2,|3/2|}$}&89(1)$^e$& S.15/S.47\\
&	\boldmath{$7D_{5/2,|3/2|}$}		&	89(1)$^e$	 &S.16/S.48  \\
\hline
$nD_{3/2,|1/2|}$	&	$7P_{1/2,|1/2|}$  &	155(4)$^a$ &S.17/---\\
&	\boldmath{$7P_{3/2,|1/2|}$} &	133(2)$^a$& S.18/S.49\\
&	\boldmath{$8P_{1/2,|1/2|}$}&	307(14)$^b$&S.19/S.50\\
&	\boldmath{$8P_{3/2,|1/2|}$}&	274(12)$^b$	&S.20/S.51 \\
&\boldmath{$4F_{5/2,|1/2|}$} &40(6)$^f$  &S.21/S.52 \\
\hline
$nD_{3/2,|3/2|}$	&	\boldmath{$7P_{3/2,|3/2|}$}  &	133(2)$^a$   & S.22/S.53\\ 
&	\boldmath{$8P_{3/2,|3/2|}$}&	274(12)$^b$ & S.23/S.54\\
	&	\boldmath{$4F_{5/2,|3/2|}$} &			40(6)$^f$ &		S.24/S.55 \\
\hline
$nD_{5/2,|1/2|}$	&	\boldmath{$7P_{3/2,|1/2|}$}  &	133(2)$^a$   & S.25/S.56\\
&	\boldmath{$8P_{3/2,|1/2|}$}&	274(12)$^b$	& S.26/S.57\\
&	$4F_{5/2,|1/2|}$&	40(6)$^f$ &S.27/---\\
	&	$4F_{7/2,|1/2|}$		&	40(6)$^f$	
&S.28/--- \\
\hline
$nD_{5/2,|3/2|}$	&	\boldmath{$7P_{3/2,|3/2|}$}   	&	133(2)$^a$ & S.29/S.58\\
&	\boldmath{$8P_{3/2,|3/2|}$}&	274(12)$^b$	& S.30/S.59 \\
&	\boldmath{$4F_{5/2,|3/2|}$}&	40(6)$^f$&S.31/S.60\\
&	\boldmath{$4F_{7/2,|3/2|}$}&	40(6)$^f$	
&S.32/S.61  \\
\hline
$nD_{5/2,|5/2|}$	&	\boldmath{$4F_{5/2,|5/2|}$ }   	&	40(6)$^f$     & S.33/S.62\\
&	\boldmath{$4F_{7/2,|5/2|}$}	&		40(6)$^f$	
& S.34/S.63 \\
\hline
\hline 
\end{tabular}
\label{table_4} 
\end{table}

\begin{widetext}

\begin{table}[t]
\caption {Double magic wavelength $\lambda^{(d)}$ for the $6S_{1/2,|1/2|}$ ground state  and the $nS_{1/2,|1/2|}$ Rydberg series for $n=40-70$.  In all cases,   $\lambda^{(d)}$ is blue-detuned (positive $\Delta$) from the auxiliary  state $7P_{3/2,|1/2|}$.    The polarizability at the magic wavelength is $\alpha_v^{(d)}$.   For a trap  depth of $|U_{\text{trap}}|/k_B=1$~$\mu$K, the light intensity $I$, the Rabi frequency  $\Omega_R$ for the transition between the Rydberg state   and the auxiliary state,  and the associated maximal transition probability $P_{\text{max}}$ (see text)
  are presented. Following the same format as employed in this table, Tables S.2-S.34  of the supplemental material~\cite{Supplement} consider the other Rydberg series and auxiliary states considered in this work.} 
\centering 
\begin{tabular}{c |c|c | c| c|c|c} 
\hline\hline
&\multicolumn{6}{c}{\textbf{auxiliary state $7P_{3/2,|1/2|}$}}\\  
  \hline
state	&	$\lambda^{(d)}$	(nm) &	$\Delta$ (MHz)	&	$\alpha_v^{(d)}$	(a.u.) &	$I$  (Wcm$^{-2}$)	&	$\Omega_R$  (MHz)	&	$P_{\text{max}}$	\\
\hline
$40S_{1/2,|1/2|}$	&	1066.6447	&	892.33	&	1152.4154	&	386.01	&	205.58	&	0.050	\\
$41S_{1/2,|1/2|}$	&	1066.1293	&	858.13	&	1154.6188	&	385.27	&	196.94	&	0.050	\\
$42S_{1/2,|1/2|}$	&	1065.6545	&	822.53	&	1156.6592	&	384.59	&	188.91	&	0.050	\\
$43S_{1/2,|1/2|}$	&	1065.2162	&	785.89	&	1158.5519	&	383.96	&	181.41	&	0.051	\\
$44S_{1/2,|1/2|}$	&	1064.8108	&	748.54	&	1160.3102	&	383.38	&	174.40	&	0.051	\\
$45S_{1/2,|1/2|}$	&	1064.4350	&	710.84	&	1161.9475	&	382.84	&	167.84	&	0.053	\\
$46S_{1/2,|1/2|}$	&	1064.0860	&	673.14	&	1163.4732	&	382.34	&	161.68	&	0.055	\\
$47S_{1/2,|1/2|}$	&	1063.7613	&	635.79	&	1164.8982	&	381.87	&	156.02	&	0.057	\\
$48S_{1/2,|1/2|}$	&	1063.4586	&	599.15	&	1166.2306	&	381.43	&	150.45	&	0.059	\\
$49S_{1/2,|1/2|}$	&	1063.1762	&	563.55	&	1167.4782	&	381.03	&	145.31	&	0.062	\\
$50S_{1/2,|1/2|}$	&	1062.9121	&	529.35	&	1168.6479	&	380.65	&	140.47	&	0.066	\\
$51S_{1/2,|1/2|}$	&	1062.6648	&	496.90	&	1169.7463	&	380.29	&	135.89	&	0.070	\\
$52S_{1/2,|1/2|}$	&	1062.4329	&	466.55	&	1170.7788	&	379.95	&	131.56	&	0.074	\\
$53S_{1/2,|1/2|}$	&	1062.2152	&	438.66	&	1171.7506	&	379.64	&	127.45	&	0.078	\\
$54S_{1/2,|1/2|}$	&	1062.0105	&	413.56	&	1172.6664	&	379.34	&	123.56	&	0.082	\\
$55S_{1/2,|1/2|}$	&	1061.8178	&	391.61	&	1173.5301	&	379.06	&	119.86	&	0.086	\\
$56S_{1/2,|1/2|}$	&	1061.6362	&	373.04	&	1174.3461	&	378.80	&	116.34	&	0.089	\\
$57S_{1/2,|1/2|}$	&	1061.4648	&	357.63	&	1175.1176	&	378.55	&	112.99	&	0.091	\\
$58S_{1/2,|1/2|}$	&	1061.3029	&	345.02	&	1175.8474	&	378.31	&	109.80	&	0.092	\\
$59S_{1/2,|1/2|}$	&	1061.1498	&	334.85	&	1176.5390	&	378.09	&	106.76	&	0.092	\\
$60S_{1/2,|1/2|}$	&	1061.0049	&	326.79	&	1177.1947	&	377.88	&	103.86	&	0.092	\\
$61S_{1/2,|1/2|}$	&	1060.8676	&	320.47	&	1177.8167	&	377.68	&	101.09	&	0.091	\\
$62S_{1/2,|1/2|}$	&	1060.7373	&	315.56	&	1178.4080	&	377.49	&	98.44	&	0.089	\\
$63S_{1/2,|1/2|}$	&	1060.6137	&	311.69	&	1178.9695	&	377.31	&	95.91	&	0.086	\\
$64S_{1/2,|1/2|}$	&	1060.4962	&	308.53	&	1179.5044	&	377.14	&	93.48	&	0.084	\\
$65S_{1/2,|1/2|}$	&	1060.3845	&	305.71	&	1180.0132	&	376.98	&	91.15	&	0.082	\\
$66S_{1/2,|1/2|}$	&	1060.2781	&	302.90	&	1180.4984	&	376.82	&	88.92	&	0.079	\\
$67S_{1/2,|1/2|}$	&	1060.1768	&	299.73	&	1180.9608	&	376.68	&	86.78	&	0.077	\\
$68S_{1/2,|1/2|}$	&	1060.0803	&	295.87	&	1181.4020	&	376.54	&	84.72	&	0.076	\\
$69S_{1/2,|1/2|}$	&	1059.9882	&	290.95	&	1181.8237	&	376.40	&	82.75	&	0.075	\\
$70S_{1/2,|1/2|}$	&	1059.9002	&	284.64	&	1182.2263	&	376.27	&	80.85	&	0.075	\\
\hline 
\end{tabular}
\label{table_5} 
\end{table}

\end{widetext}

\section{Interpretation and experimental feasibility  of double magic wavelength trapping}\label{sec_4}
The polarizability for   the ground state and the  Rydberg state are equal at  $\lambda^{(d)}$ reported in the previous section.  For a given light intensity $I$, the depth $|U_{\text{trap}}|/k_B$ of the trapping potential is proportional to the polarizability, $U_{\text{trap}}=-I\alpha_v(\omega)/(2\epsilon_0 c)$,  where $k_B$ denotes the Boltzmann constant and $\epsilon_0$  the  vacuum permittivity~\cite{Saffman2005}.  Intuition  tells us that a magic wavelength for which the   value of the corresponding polarizability $\alpha_v^{(d)}$ is large  is a better candidate from an experimental perspective  than that for which $\alpha_v^{(d)}$ is small. For the double magic wavelengths reported, $\alpha_v^{(d)}$ is between about $500-1,300$~a.u. (these values are neither  particularly large nor particularly small). For a fixed polarizability,   $|U_{\text{trap}}|/k_B$ can be increased by    increasing $I$.  Unfortunately, though, there is a penalty associated with cranking up $I$  as this not only increases $|U_{\text{trap}}|/k_B$ but also enhances the coupling to the auxiliary state~\cite{Bai2020_1}.

Recall that the laser wavelength is, for the Rydberg state,  blue-detuned (positive detuning) relative to the auxiliary   state.    If the detuning $\Delta$ from the auxiliary state is very small, it inevitably leads to a mixing of the states, namely of the Rydberg state and the auxiliary state, resulting in potentially significant photon  scattering during the trapping process~\cite{Saffman2005}. The larger $I$,  the stronger the mixing of the states.  In experiment, one aims to minimize the mixing of the states by choosing $I$ such that the Rabi frequency $\Omega_R$ for the transition between the Rydberg state and the auxiliary state is much smaller than the corresponding  $\Delta$. 
Within a two-state model~\cite{Saffman2005}, 
the Rabi frequency $\Omega_R$ and the maximal transition probability $P_{\text{max}}$  from the Rydberg state $\psi_i$ to the auxiliary state $\psi_k$ are given by
\begin{eqnarray}  
\Omega_R=\sqrt{I}|\langle\psi_k||d||\psi_i\rangle|
\end{eqnarray} 
and
\begin{eqnarray}P_{\text{max}}=\dfrac{|\Omega_R|^2}{|\Omega_R|^2+\Delta^2},
\end{eqnarray}
respectively. 

\begin{widetext}

\begin{figure*}[!h]
{\includegraphics[ scale=.21]{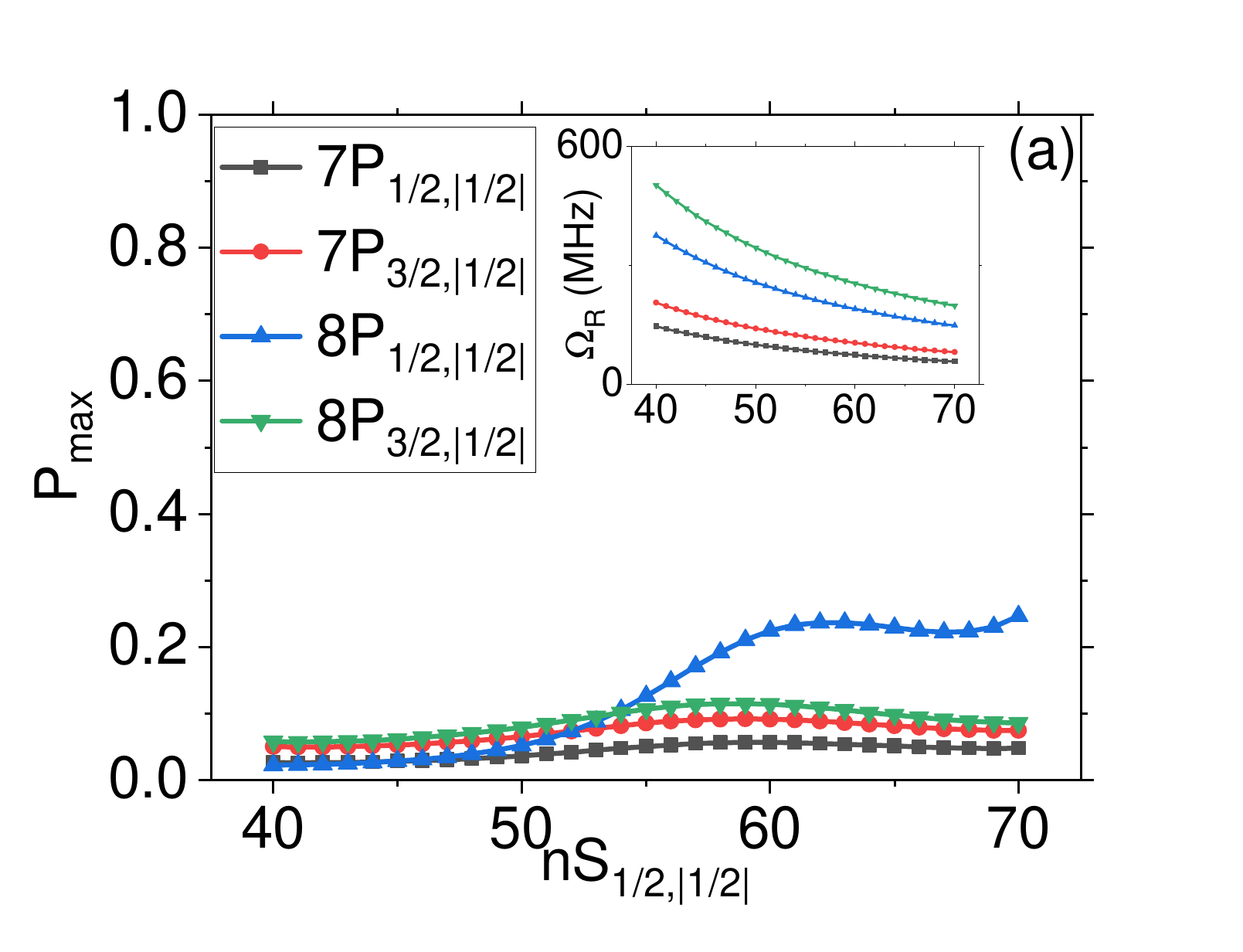}}\hspace{-1.78cm}
{\includegraphics[ scale=.21]{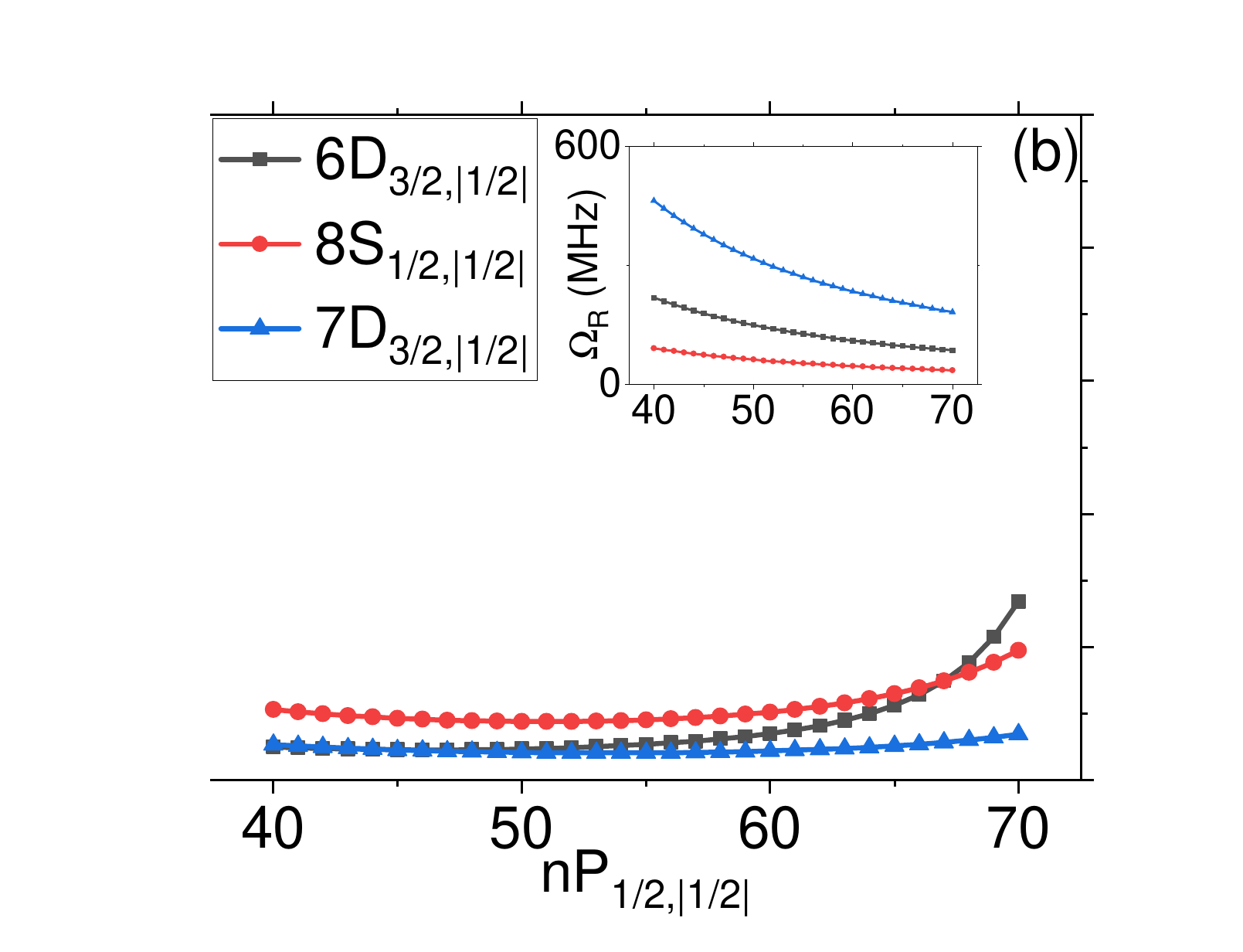}}\hspace{-1.78cm}
{\includegraphics[ scale=.21]{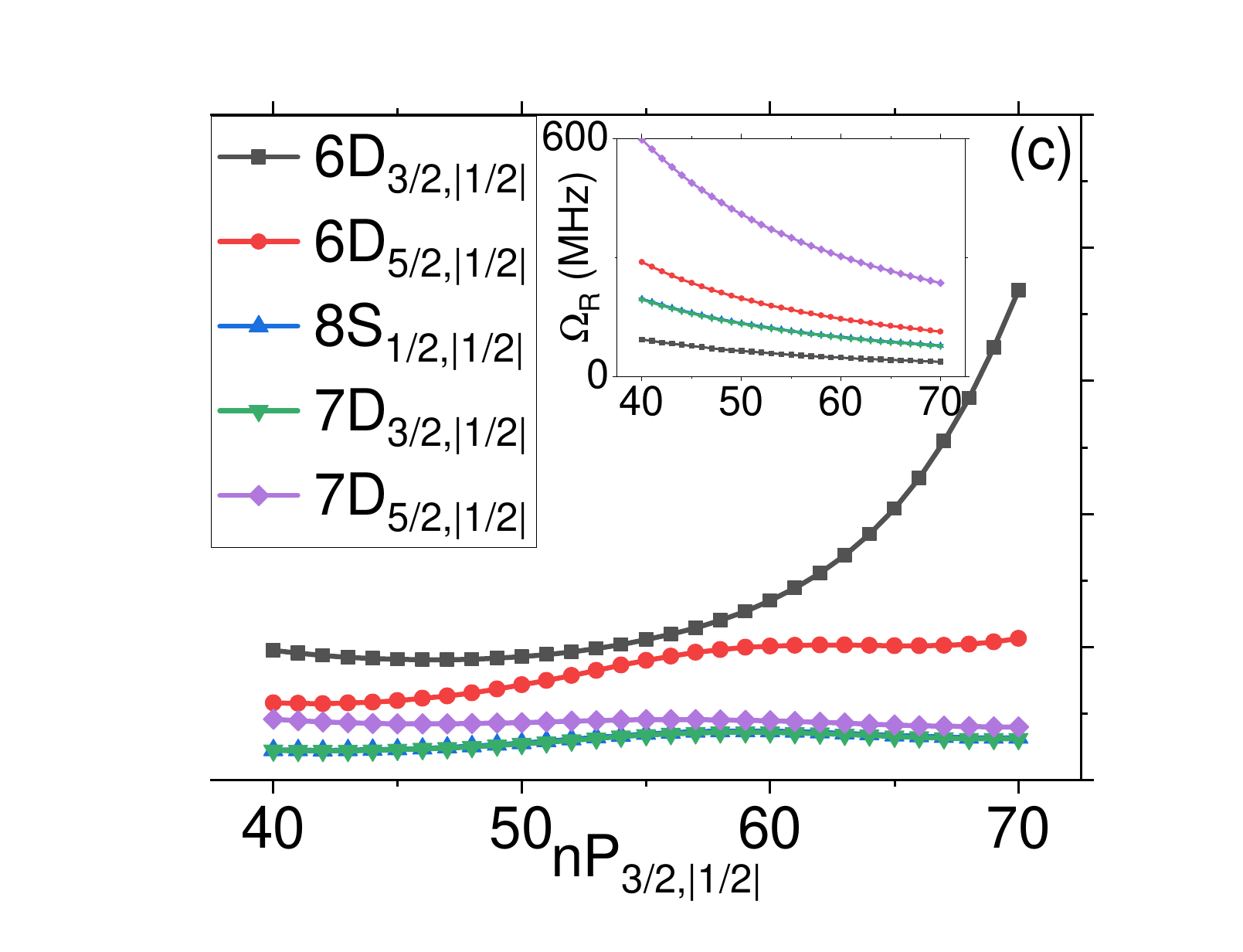}}\\
\vspace*{-0.15in}
{\includegraphics[ scale=.21]{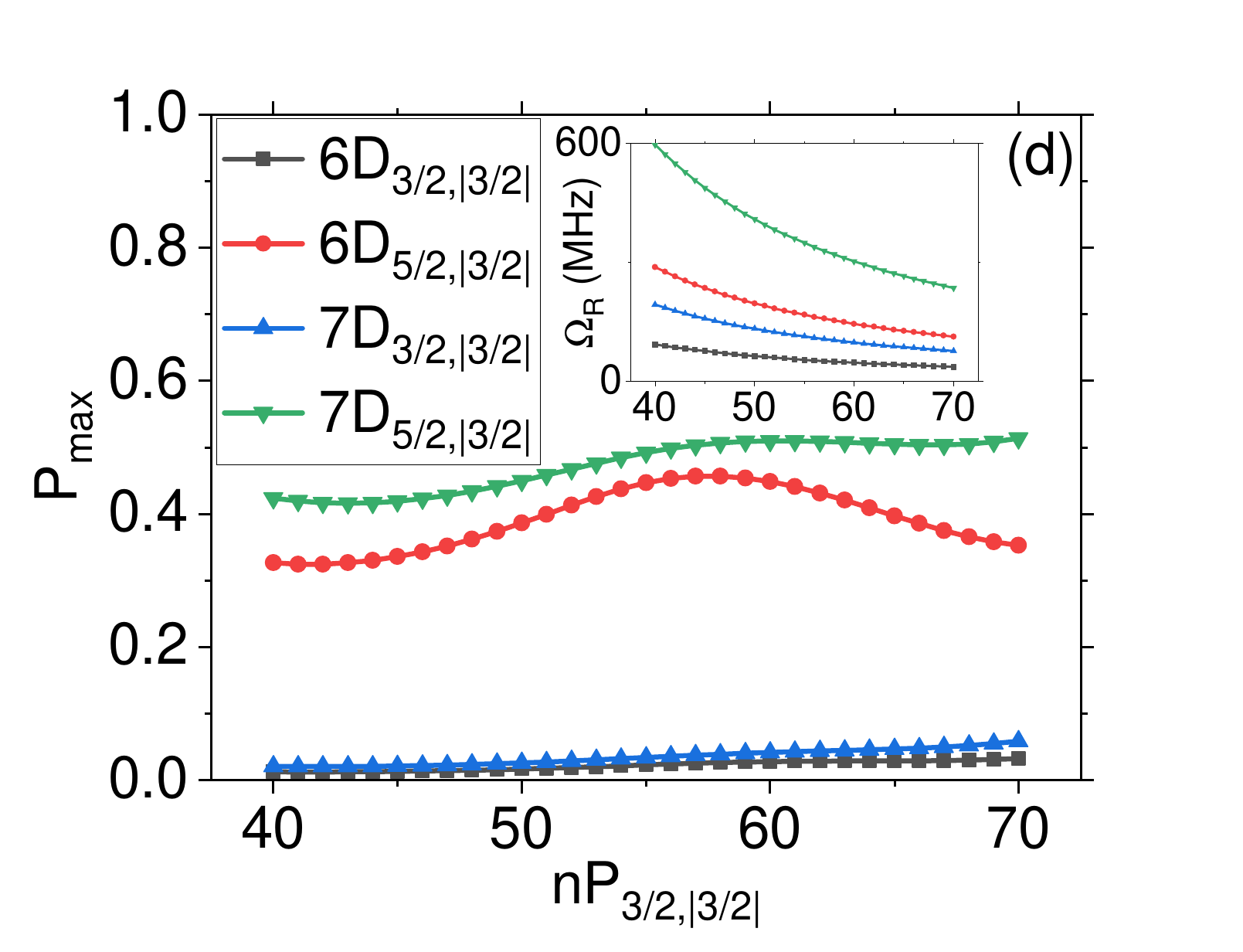}}\hspace{-1.78cm}
{\includegraphics[ scale=.21]{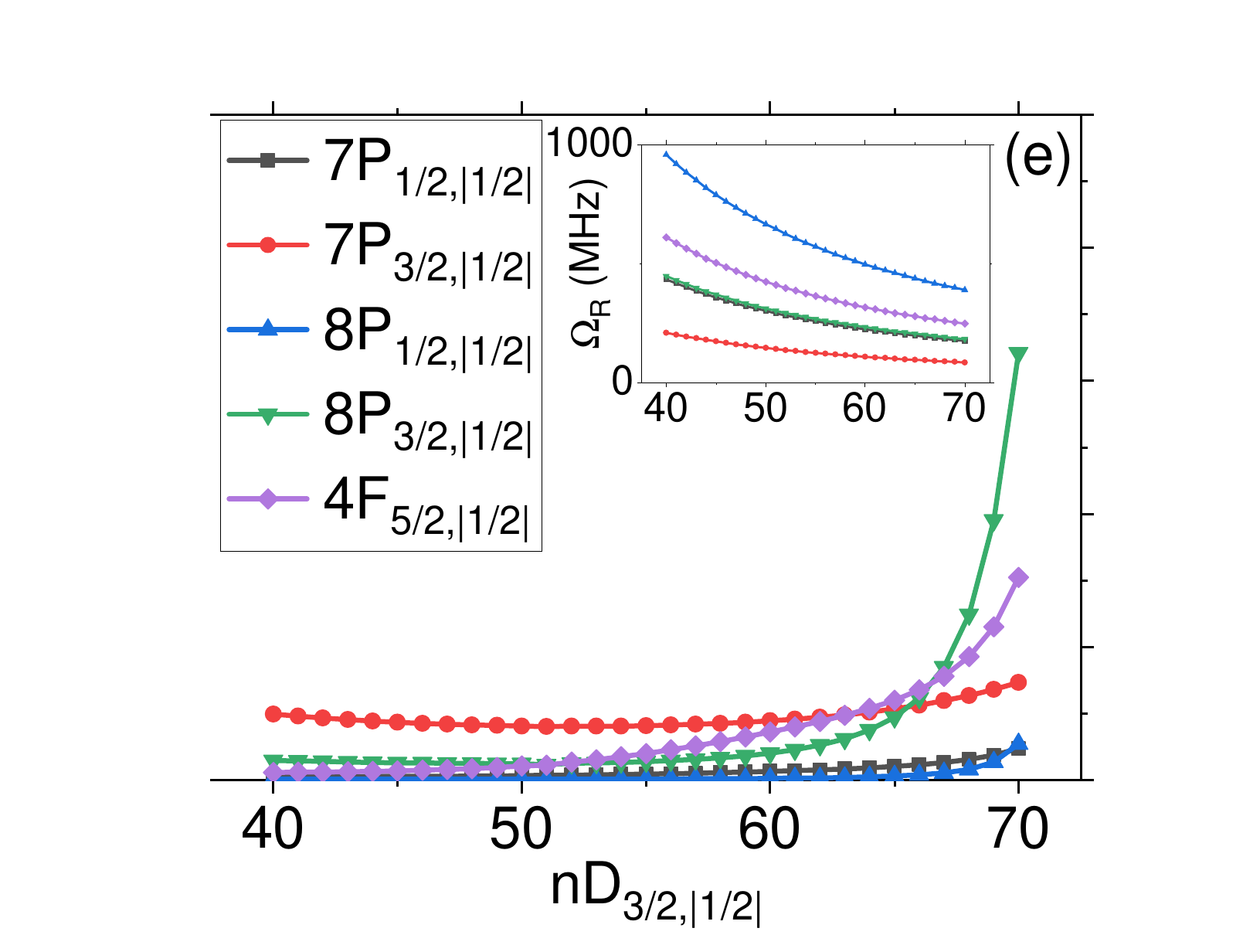}}\hspace{-1.78cm}
{\includegraphics[ scale=.21]{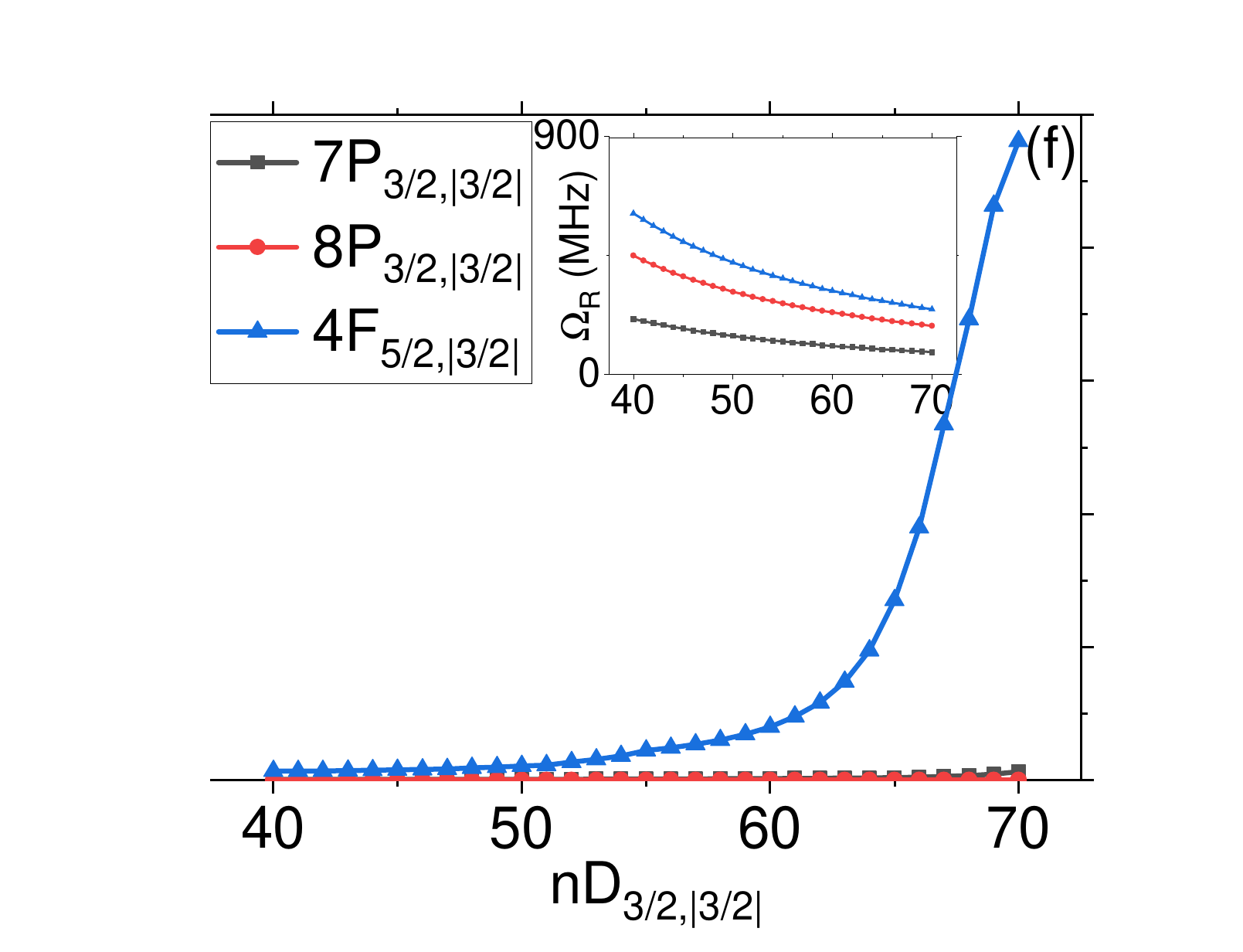}}\\
\vspace*{-0.15in}
{\includegraphics[ scale=.21]{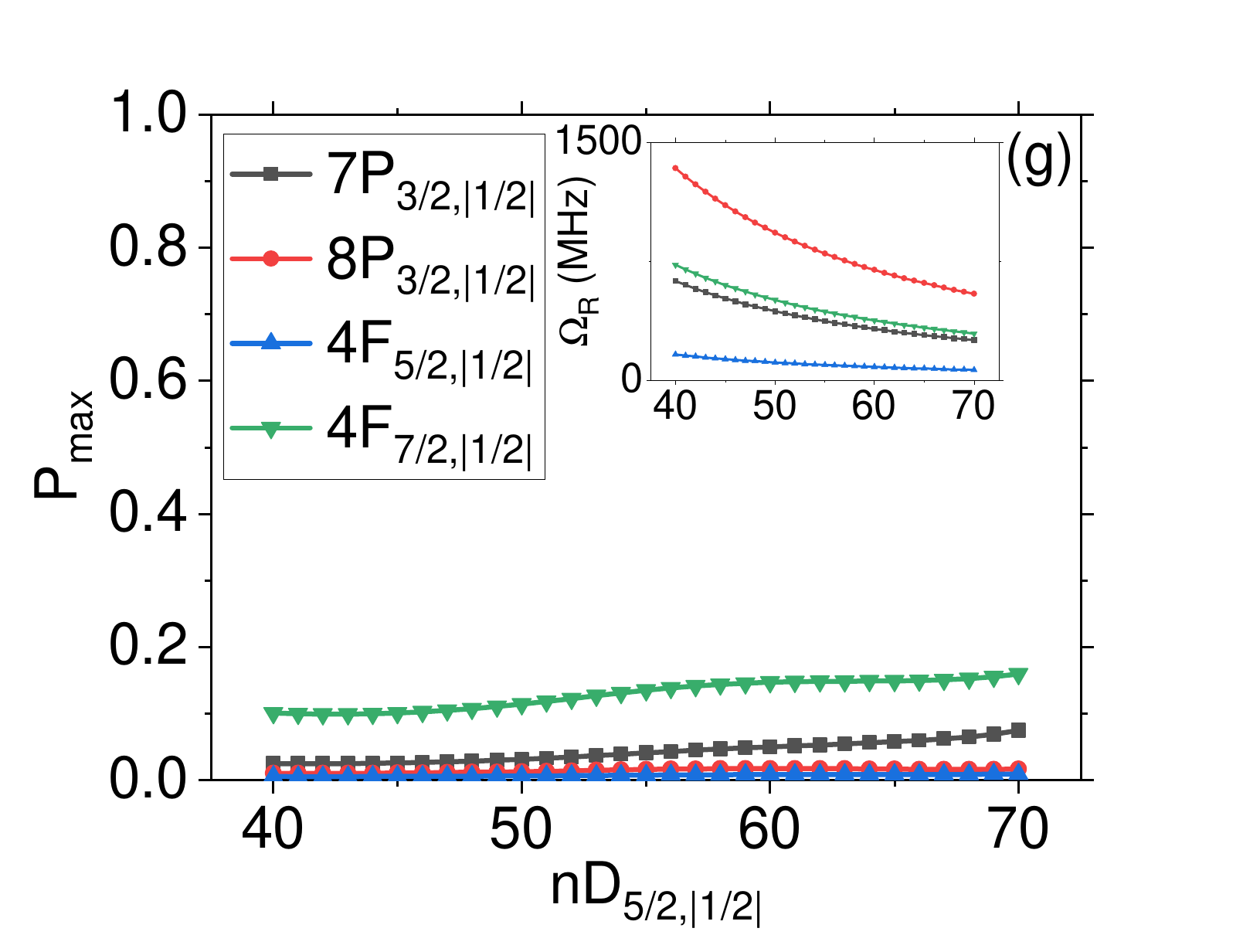}}\hspace{-1.78cm}
{\includegraphics[ scale=.21]{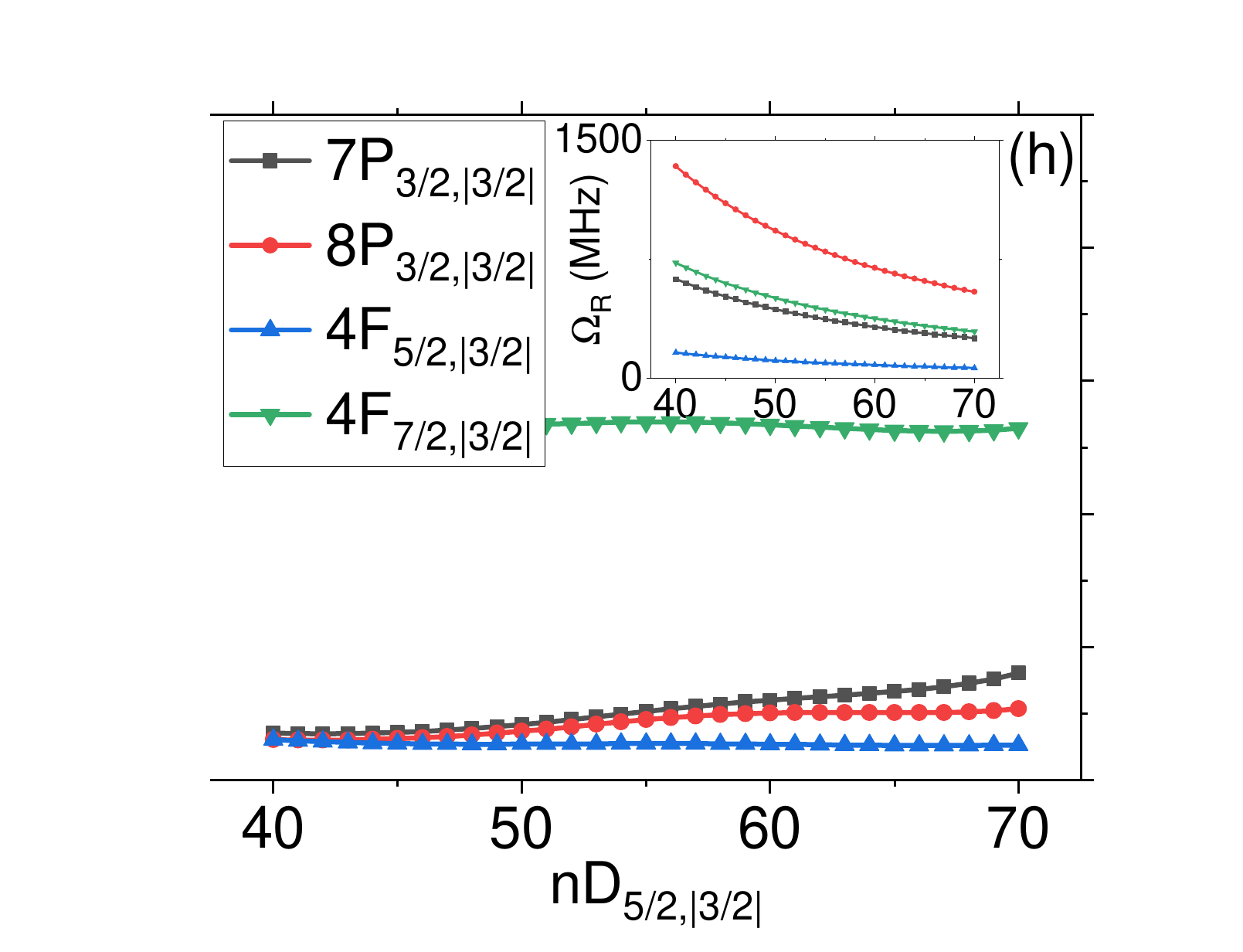}}\hspace{-1.78cm}
{\includegraphics[ scale=.21]{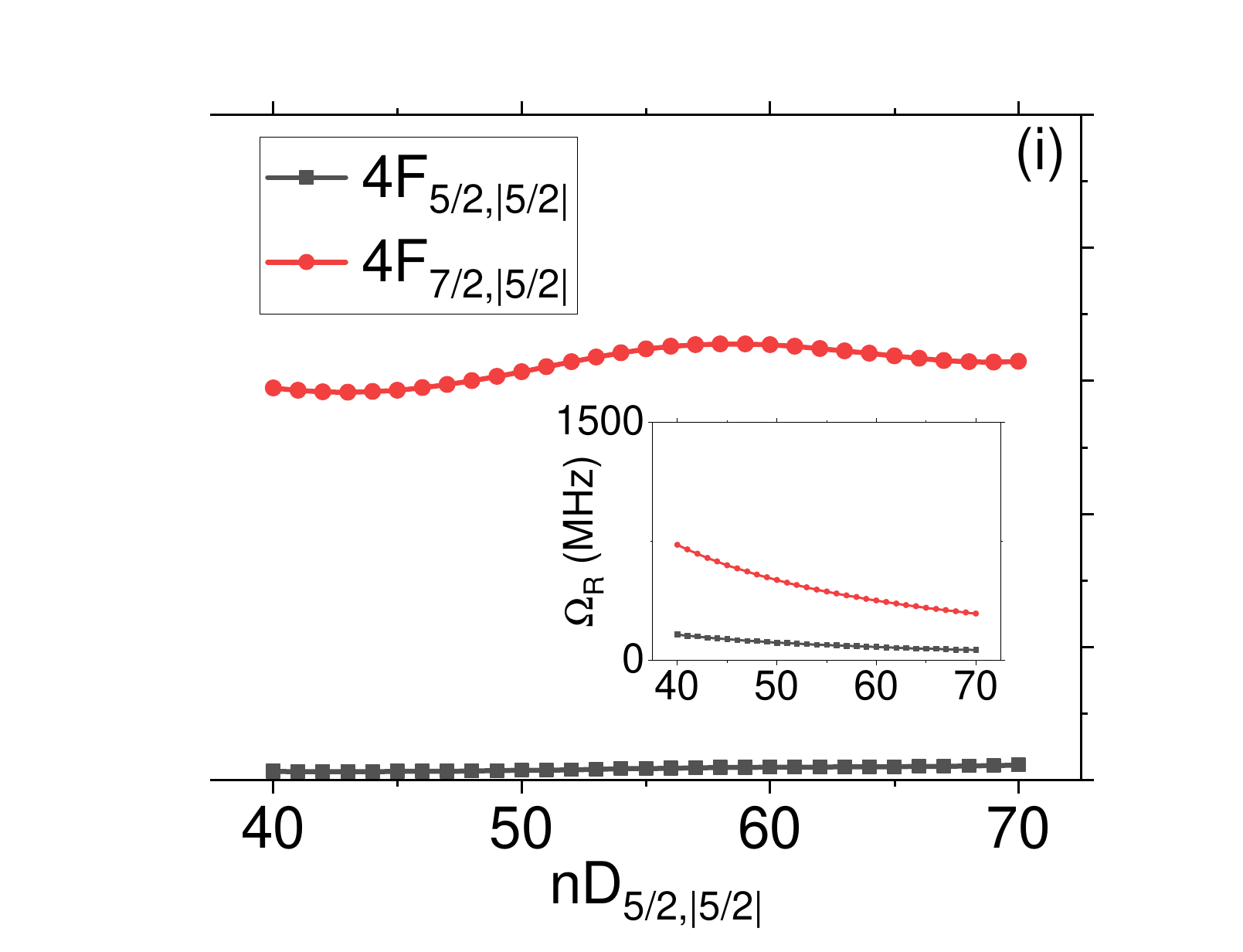}}\\
\vspace*{-0.05in}
\caption{ Maximal transition probability $P_{\text{max}}$, i.e., maximal probability  that the  Rydberg state transitions to the auxiliary state,  for  $|U_{\text{trap}}|/k_B=1 \; \mu$K as a function of the principal quantum number $n$; $P_{\text{max}}$ is estimated using a two-state model (see text for details).  The insets show the corresponding Rabi frequency $\Omega_R$.  Each panel corresponds to a different Rydberg series (see the $x$-axis). The auxiliary states considered are shown in the legend in each panel.  The values of $P_{\text{max}}$ and respective magic wavelengths, detunings, and polarizabilities are  tabulated in Table~\ref{table_5} and Tables S.2-S.34  of the supplemental material~\cite{Supplement}. } 
\label{Fig6}
\end{figure*}

\end{widetext}

Figure~\ref{Fig6}  presents the dependence of $P_{\text{max}}$ on $n$ for the nine  Rydberg series considered in this paper; as before, all the auxiliary states  listed in Table~\ref{table_4} are investigated. The corresponding Rabi frequencies are shown in the insets. Tables~\ref{table_5} and S.2-S.34 report  $\Omega_R$, $P_{\text{max}}$, and $I$ for simultaneously trapping the ground state and a Rydberg state. The calculations are performed for a trap depth of   $|U_{\text{trap}}|/k_B=1\;\mu$K.    As expected, for a fixed trap depth, $\Omega_R$ decreases monotonically  with  increasing  $n$. Even though $\Omega_R$ and $\Delta$ (see the insets of Fig.~\ref{Fig5})  decrease monotonically with $n$, $P_{\text{max}}$ displays a non-monotonic variation with $n$ due to the non-monotonic variation of the derivative of $\Delta$ with increasing $n$.

\begin{figure}[!h]
{\includegraphics[ scale=.30]{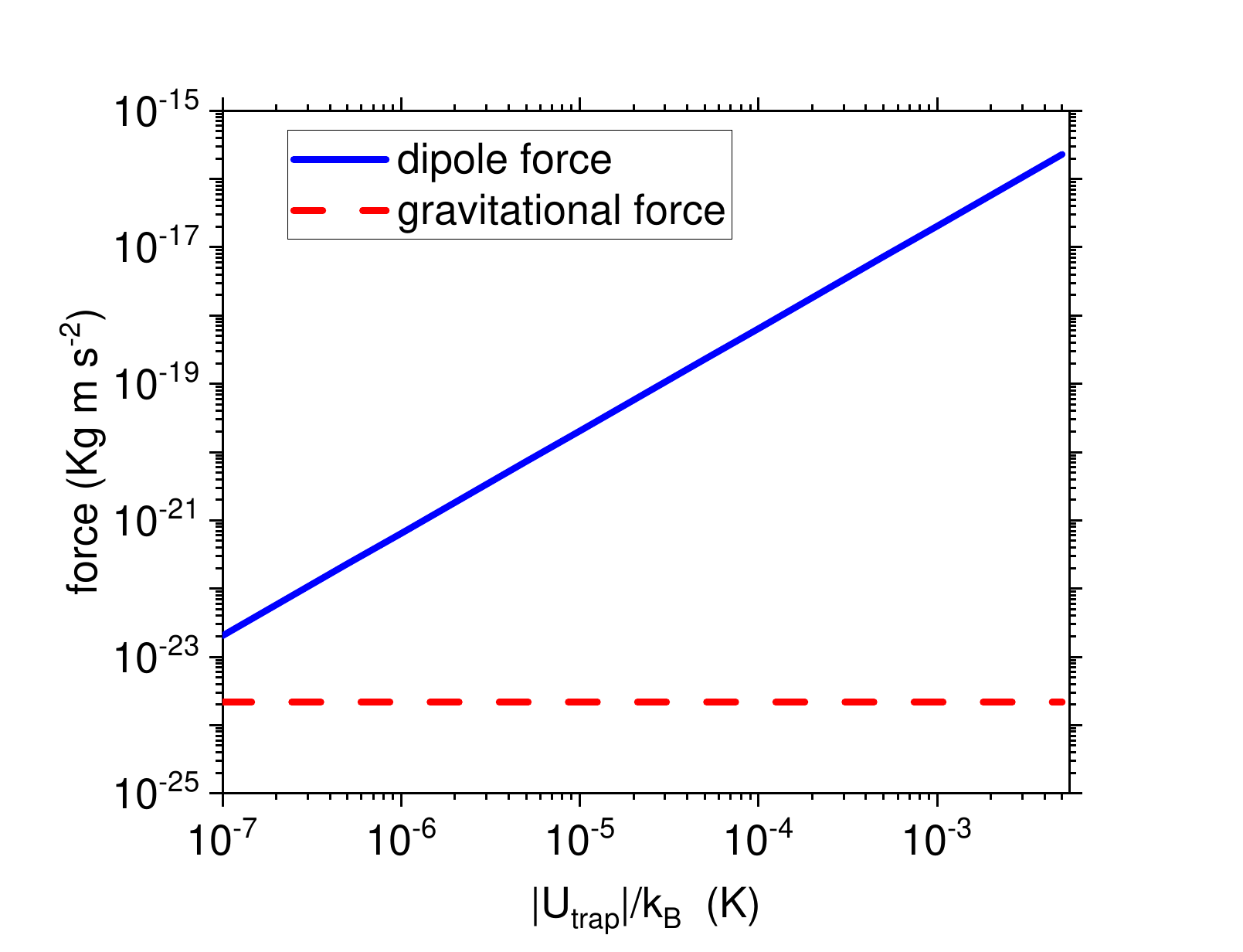}}
\vspace*{-0.05in}
\caption{Dipole and gravitational force for a cesium atom as a function of the trap depth $|U_{\text{trap}}|/k_B$ (note the log-log scale). The dipole force (blue solid line) is significantly larger than the gravitational force (red dashed line) for traps as shallow as $0.1$~$\mu$K.  } 
\label{Fig7}
\end{figure}

 Shallow traps with depths as small as $1$~$\mu$K can be realized experimentally, though reaching sufficiently low temperatures is, admittedly,  challenging~\cite{ Tuchendler2008}. We choose $1$~$\mu$K as a reference point, since the  dipole force $F_{\text{dipole}}$ for this trap depth is still significantly larger than the magnitude of the gravitational force $mg$, where $m$ denotes the mass of the $^{133}$Cs atom and  $g$ the gravitational acceleration.    To estimate $F_{\text{dipole}}$, we consider a one-dimensional harmonic potential with angular frequency $\omega_{\text{HO}}$ along the $z$-coordinate that changes to a constant for $|z|$ larger than the waist $z_R$ of the dipole trap beam.  Evaluating the force at $z_R$ and setting $\omega_{\text{HO}}=\sqrt{2|U_{\text{trap}}|/(mz_R^2)}$, we find $F_{\text{dipole}}=-2 U_{\text{trap}}/z_R$. To give a concrete example, 
a  beam waist of $0.5$~$\mu$m corresponds to  $\omega_{\text{HO}}=2 \pi \times 3.56$~kHz for $|U_{\text{trap}}|/k_B=1$~$\mu$K. 
Expressing $z_R$ are as $\gamma /\sqrt{2m|U_{\text{trap}}|}$, where $\gamma$ is a dimensionless scaling factor that is---for physical reasons---constrained to $\gamma \gtrsim 1$ (if $\gamma$ was smaller than $1$ the ground state wave function would not fit into the trap), 
we have  $F_{\text{dipole}}=\gamma^{-1}\sqrt{m}(2|U_{\text{trap}}|)^{3/2}$ (recall, $\hbar$ has been set to $1$).  Figure~\ref{Fig7} compares the dipole force and the gravitational force for cesium as a function of $|U_{\text{trap}}|/k_B$.   In plotting the dipole force, we set $\gamma=1$, i.e., we assumed the "best case scenario". If one wanted to trap multiple motional states, $\gamma$ would need to be larger than one, implying that $F_{\text{dipole}}$ would be smaller.   It can be seen that the dipole force for a trap  depth of $1 \; \mu$K  is three orders of magnitude larger than the gravitational  force. Figure~\ref{Fig7} thus suggests that one may even be able to achieve trapping for trap depths a bit shallower than $1$~$\mu$K. We note that the use of a more realistic functional form for the dipole trap potential yields to a dipole force that is four times smaller than the harmonic oscillator potential based estimate presented here. While the detailed functional form of the trap, including the confinement along the $x$- and $y$-directions, needs to be taken into account when analyzing a specific experimental set-up, the analysis presented in this paper should serve as a useful reference point.

    Figure~\ref{Fig6} as well as  Tables~\ref{table_5} and S.2-S.34  show  that $P_{\text{max}}$, calculated for $|U_{\text{trap}}|/k_B=1$~$\mu$K, varies greatly for the Rydberg series, auxiliary states, and principal quantum numbers considered.
   The smallest $P_{\text{max}}$ values are found for:
   \begin{itemize}
       \item $nD_{3/2,|1/2|}$ series with auxiliary state $7P_{1/2,|1/2|}$, 
       \item $nD_{3/2,|1/2|}$ series with auxiliary state $8P_{1/2,|1/2|}$, 
       \item $nD_{3/2,|3/2|}$ series with auxiliary state $7P_{3/2,|3/2|}$, 
       \item $nD_{3/2,|3/2|}$ series with auxiliary state $8P_{3/2,|3/2|}$, 
       and
       \item $nD_{5/2,|1/2|}$ series with auxiliary state $4F_{5/2,|1/2|}$. 
   \end{itemize}
Considering that the trapping of a Rydberg state does not only require a sufficiently small $P_{\text{max}}$ but also a sufficiently long lifetime of the auxiliary state (the lifetimes are listed in Table~\ref{table_4}), the $nD_{3/2,|1/2|}$ series with auxiliary state $8P_{1/2,|1/2|}$ [lifetime of $307(14)$~ns]
and 
the $nD_{3/2,|3/2|}$ series with auxiliary state $8P_{3/2,|3/2|}$
[lifetime of $274(12)$~ns]
are the most promising.
To estimate realistic trapping times, we solve a 12-state (13-state) master equation for the $45D_{3/2,|1/2|}$ ($45D_{3/2,|3/2|}$) cases~\cite{Breuer2007}.  Working in the rotating frame within the rotating wave approximation, the coupling matrix element for the Rydberg state, which is detuned by $\Delta$ and assumed to be infinitely long-lived, and the auxiliary state is given by $-\Omega_R/2$. The spontaneous lifetimes of the auxiliary state and the low-lying excited states that the auxiliary state can decay into, either  directly or through intermediate states such as $6P_{1/2}$, $6P_{3/2}$, etc.,   is treated through dissipators that depend on Lindbladian operators. 

\begin{figure}[!h]
{\includegraphics[ scale=.30]{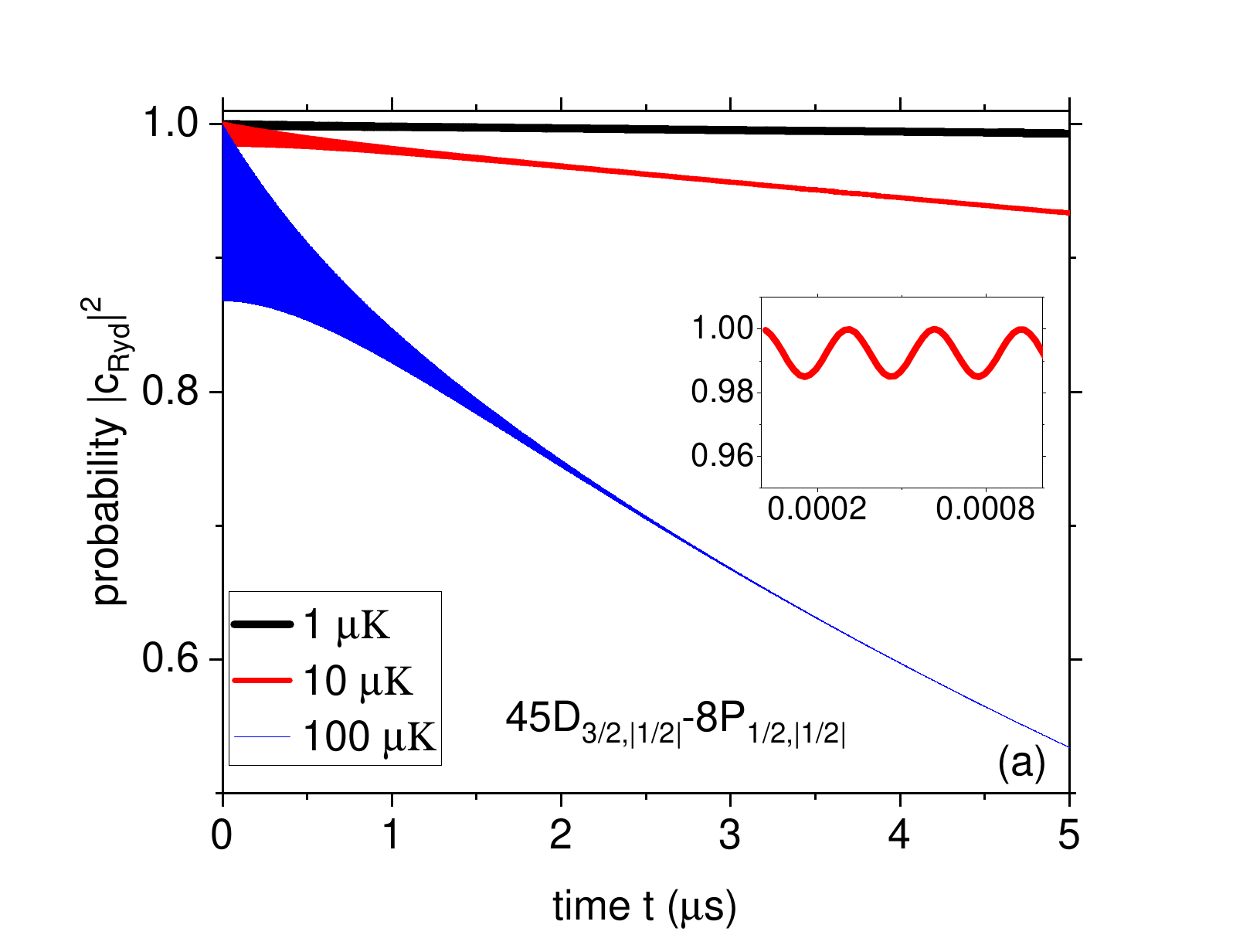}}\\
\vspace*{-0.2in}
{\includegraphics[ scale=.30]{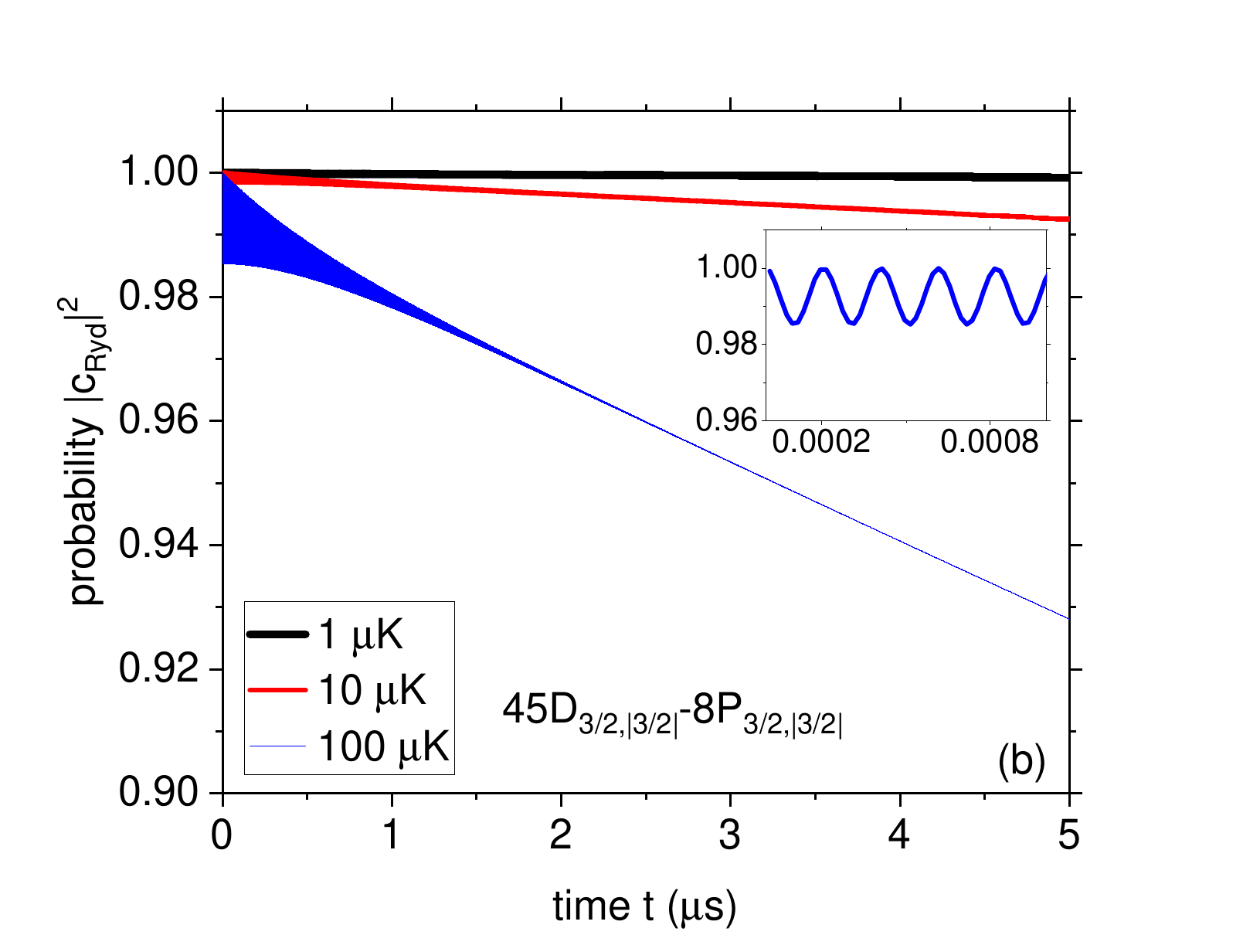}}\\
\caption{Probability $|c_{\text{Ryd}}|^2$ to be in the Rydberg state, determined by solving a master equation, for trap depths $|U_{\text{trap}}|/k_B= 1$~$\mu$K (black  line; top line at $t=1$~$\mu$s), $10$~$\mu$K (red  line; middle line at $t=1$~$\mu$s), and $100$~$\mu$K (blue  line; bottom line at $t=1$~$\mu$s). (a) and (b) consider the  Rydberg state $45D_{3/2, |1/2|}$ with  auxiliary state  $8P_{1/2, |1/2|}$ and the Rydberg state $45D_{3/2, |3/2|}$ with auxiliary state $8P_{3/2, |3/2|}$, respectively. Notes the different $y$ scales in (a) and (b). The insets show an enlargement of the short-time region for (a) the $10$~$\mu$K trap and (b) the $100$~$\mu$K trap. } 
\label{fig_newdecay}
\end{figure}

The black, red, and blue  lines in Fig.~\ref{fig_newdecay}
show the population $|c_{\text{Ryd}}|^2$
of the Rydberg state as a function of time for $|U_{\text{trap}}|/k_B=1$, $10$, and $100$~$\mu$K, respectively.
Figure~\ref{fig_newdecay}(a) considers the $45D_{3/2,|1/2|}$ state
while Fig.~\ref{fig_newdecay}(b) considers the $45D_{3/2,|3/2|}$ state.
At $t=0$, the system is initialized in
 the Rydberg state.
At short times (see the insets in Fig.~\ref{fig_newdecay}), the population oscillates rapidly at a time scale that is set by the effective Rabi coupling strength. The oscillations are damped at longer times by the incoherent decay processes.
Figure~\ref{fig_newdecay}(a) shows that the probability to be in the Rydberg state after $5$~$\mu$s is around $53.4$, $93.3$, and $99.3$~\% for the $45D_{3/2,|1/2|}$ state for trap depths of $100$, $10$, and $1$~$\mu$K, respectively.
For the $45D_{3/2,|3/2|}$ state, in contrast, 
the probability to be in the Rydberg state after $5$~$\mu$s is $92.8$, $99.2$, and $99.9$~\% for the same trap depths.
The probability $|c_{\text{Ryd}}|^2$ may, depending on the Rydberg series, vary notably with $n$. 
For the $nD_{3/2,|1/2|}$ series ($8P_{1/2,|1/2|}$ auxiliary state), e.g., 
$|c_{\text{Ryd}}|^2$ at $t=5$~$\mu$s is larger than $0.99$ for $n=40-57$ and then decreases to $0.777$ for $n=70$ for a trap depth of $1$~$\mu$K.
For the $nD_{3/2,|3/2|}$ series ($8P_{3/2,|3/2|}$ auxiliary state), in contrast, $|c_{\text{Ryd}}|^2$ at $t=5$~$\mu$s increases from $0.9990$ for $n=40$ to $0.9997$ for $n=70$ (these numbers are, again, for a trap depth of $1$~$\mu$K).
Our master equation simulations demonstrate that the proposed trapping scheme is feasible, opening the door for Rydberg state-based simulations on time scales up to tens of microseconds.

Our discussion up to now  assumed that the atomic energy levels and states are  labeled by the 
quantum numbers $n$, $L$, $J$, and the magnitude of $M_J$, i.e., we assumed that hyperfine state splittings are not resolved. Typical cold atom experiments are, however, sensitive to hyperfine splittings, dictating that one should consider that the total electron angular momentum (labeled by $J$) and the nuclear spin 
(labeled by $I_N$) combine to the total angular momentum (labeled by  $F$ with associated projection quantum number $M_F$). For $^{133}$Cs with  $I_N=7/2$, there exist two ground state manifolds ($F=3$ and $4$) that contain a total of 16 states. Since the scalar polarizability is unchanged when changing from the $(J,M_J,I_N,M_I)$ to the
$(J,I_N,F,M_F)$ basis, all the $F=3$ and $4$ ground states have the same polarizability   as the $6S_{1,2,|1/2|}$ state (the tensor polarizability does not enter). 
For the Rydberg states with $J=1/2$, the same argument applies. 
For Rydberg states with $J=3/2$ and $5/2$, in contrast, the tensor polarizability, which changes under the change of basis, enters. Appendix~\ref{Appendix_C} shows that the resulting change of the magic wavelengths is of the order of $0.01$~nm.
Since there is an uncertainty in the magic wavelength determination anyways due to
approximations made in the matrix element determination, etc., the  magic wavelengths predicted using the $(J,M_J,I_N,M_I) $ basis provide a faithful starting point for precisely pinpointing
 magic wavelength conditions experimentally.

\section{Multi-magic conditions: triple and quadruple magic wavelengths}\label{sec_5}
This section shows that there exist wavelengths that lead to trapping of not only  the ground state and a Rydberg state but also of low-lying excited states. Specifically, it is shown that there exist conditions for which the $6P_{3/2,|M_J|}$ state [radiative lifetime of $30.0(7)$~ns],
the $5D_{3/2,|M_J|}$ state [radiative lifetime of $966(34)$~ns], or 
the $5D_{5/2,|M_J|}$ state [radiative lifetime of $1,353(5)$~ns]
are trapped along with the ground state and a Rydberg state; the associated wavelengths are referred to as triple magic wavelengths $\lambda^{(t)}$. In addition, we identify seven quadruple magic wavelengths $\lambda^{(q)}$ at which four states (including the ground state and a Rydberg state) are trapped simultaneously (tens of approximate quadrupole magic wavelengths are also identified).
We refer to the low-lying excited states as intermediate states.

We do not find any multi-magic wavelengths in the  range $1,000-2,000$~nm   for the Rydberg series $nS_{1/2}$, $nP_{1/2}$, $nP_{3/2}$, $nD_{3/2}$, and $nD_{5/2}$ with $n=40-70$ if the quantization axis, set by $\vec{B}_{\text{ext}}$, and the polarization vector  are parallel [$\theta=0$; see Fig.~\ref{Fig_schematic}(a)].
 If, however, the angle $\theta$ between the quantization axis and the polarization vector, which enters through the  prefactor  ${(3 \cos^2\theta-1)}/{2}$ into the dynamic tensor polarizability [see Fig.~\ref{Fig_schematic}(b) and Eq.~(\ref{eq_A11})], is varied, we find---as discussed below---many multi-magic wavelengths. 
For what follows, it is important to keep in mind that the polarizability of states with $J=1/2$ does not depend on $\theta$ for linearly polarized Gaussian light. Moreover, the polarizability of the Rydberg states considered is essentially independent of $\theta$ for the $n$ values considered in this work. Thus, we aim to adjust the polarizability of candidate intermediate states that have $J>1/2$ by tuning $\theta$.

\begin{figure}[!h]
{\includegraphics[ scale=.30]{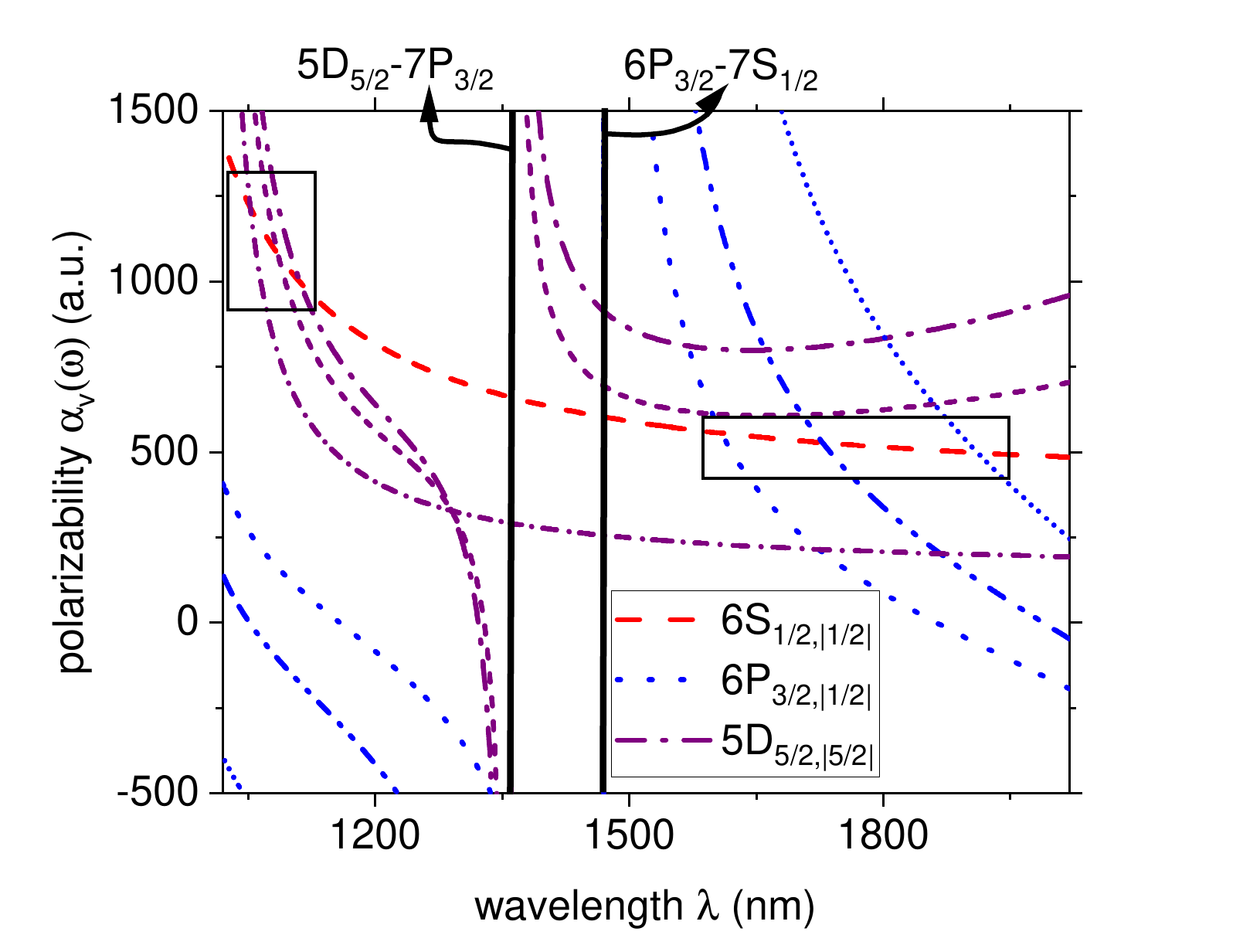}}
\vspace*{-0.05in}
\caption{Dynamic polarizability $\alpha_v(\omega)$ of the ground state $6S_{1/2, |1/2|}$ and two low-lying excited states, namely  $5D_{5/2, |5/2|}$ and $6P_{3/2, |1/2|}$, for three different angles $\theta$ between the quantization axis and the polarization vector for linearly polarized Gaussian light. The vertical black solid lines indicate  the resonance lines. The red dashed line shows the polarizability of the $6S_{1/2, |1/2|}$ state.  The purple   long-dash-dotted,  dashed, and short-dash-dotted lines show the polarizability for  the $5D_{5/2, |5/2|}$ state  for $\cos^2\theta=0, 1/3$, and 1, respectively. Similarly, blue wide-spaced-dotted, dash-dot-dotted, and dotted lines show the polarizability for  the  $6P_{3/2, |1/2|}$  state  for $\cos^2\theta=0, 1/3$, and 1, respectively. The black boxes draw the reader's attention to how the magic wavelength between the ground state and the low-lying excited state varies when $\theta$ is changed.}
\label{Fig8}
\end{figure}

To illustrate the tunability afforded by  $\theta$,   Fig.~\ref{Fig8} shows the dynamic polarizability of the ground state $6S_{1/2, |1/2|}$ (red dashed curve) and two low-lying excited states, namely  $5D_{5/2, |5/2|}$ (purple lines) and $6P_{3/2, |1/2|}$ (blue lines), for $\cos^2\theta=0,1/3$, and $1$. These angles correspond to $(3 \cos^2\theta-1)/2=-1/2, 0,$ and 1, respectively.  Interestingly, we find that the magic wavelength  for the  ground state and the  state $5D_{5/2,|5/2|}$ near the $5D_{5/2}-7P_{3/2}$ resonance line  decreases  with increasing $\cos^2\theta$  while that   for the ground state  and the state  $6P_{3/2, |1/2|}$  near the $6P_{3/2}-7S_{1/2}$ resonance line increases with increasing $\cos^2\theta$. It can be seen that changes in $\theta$ lead to magic wavelength shifts as large as $300$~nm.   This tunability of the dynamic polarizabilities of the low-lying states through the angle $\theta$ is leveraged in the discussion that follows.

\begin{widetext}

\begin{figure*}[!h]
{\includegraphics[ scale=.21]{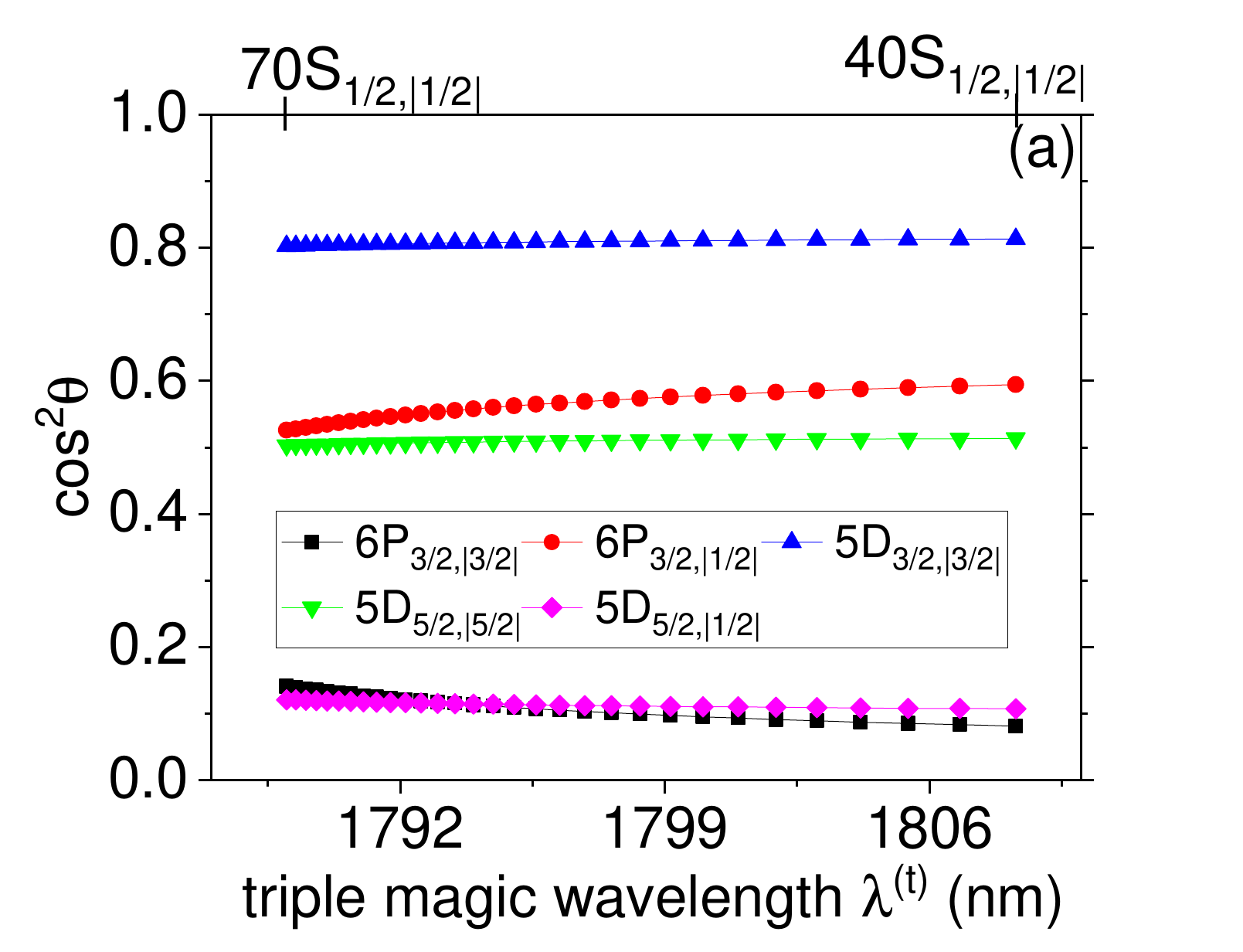}}\hspace{-1.8cm}
{\includegraphics[ scale=.21]{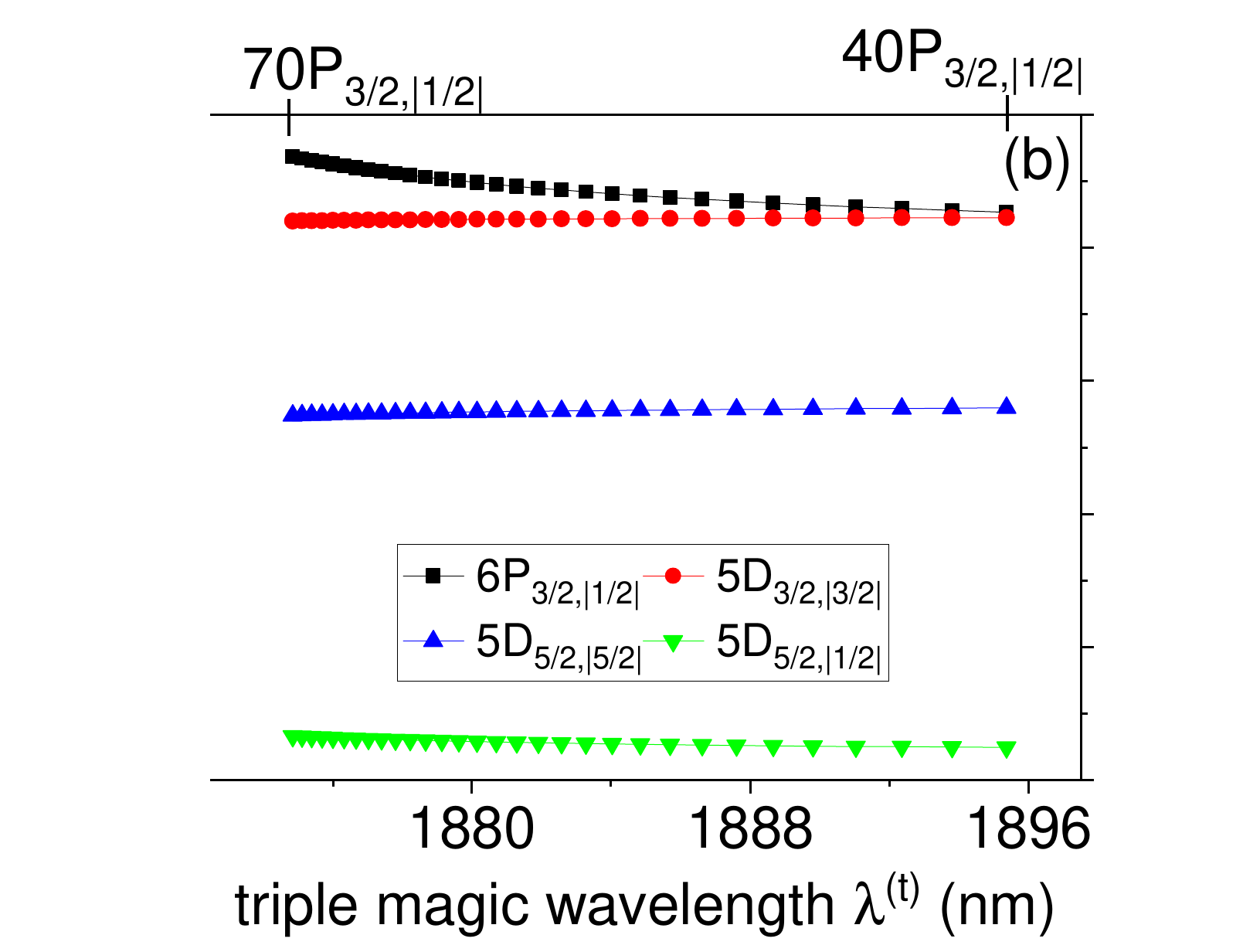}}\hspace{-1.8cm}
{\includegraphics[ scale=.21]{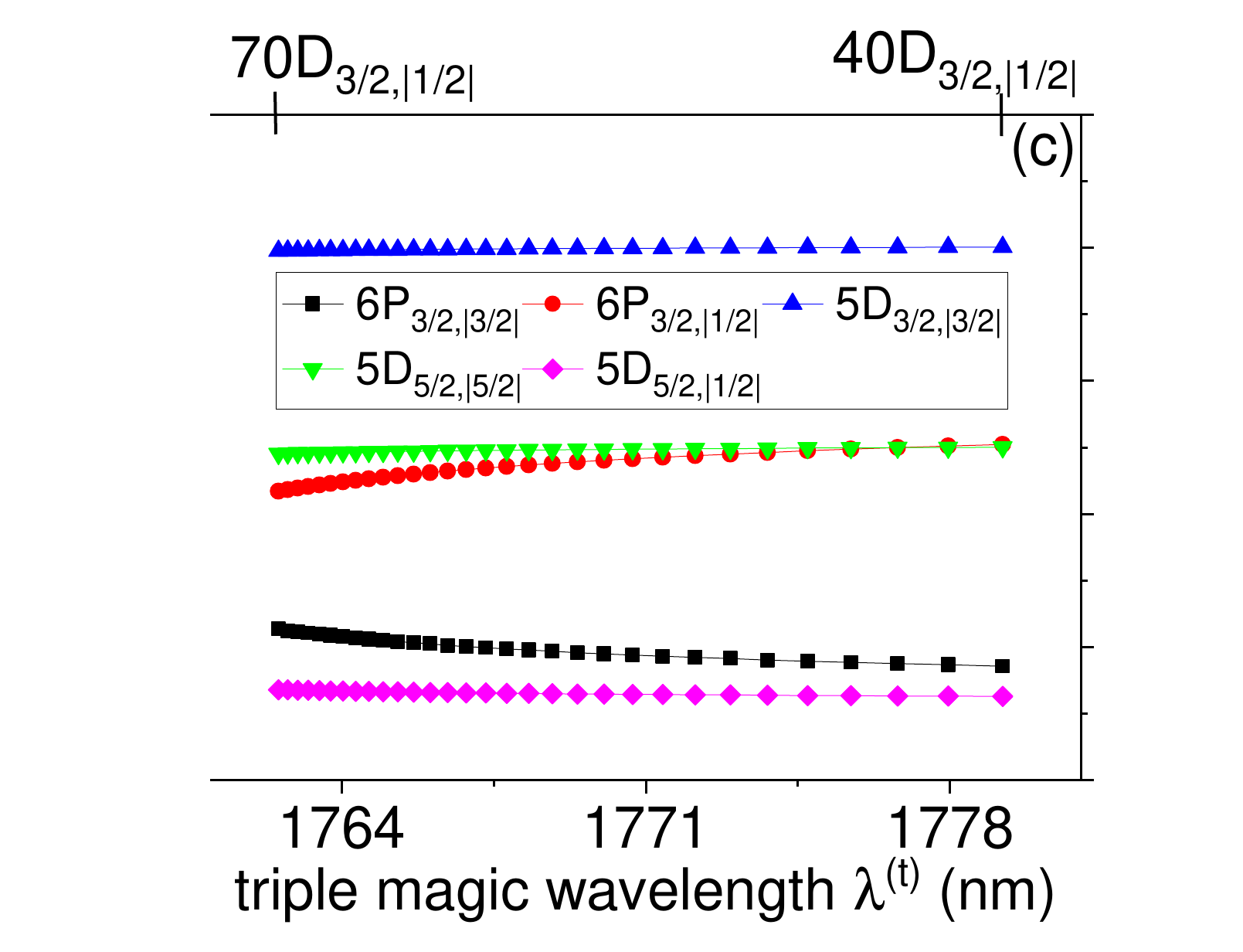}}\hspace{-1.8cm}
{\includegraphics[ scale=.21]{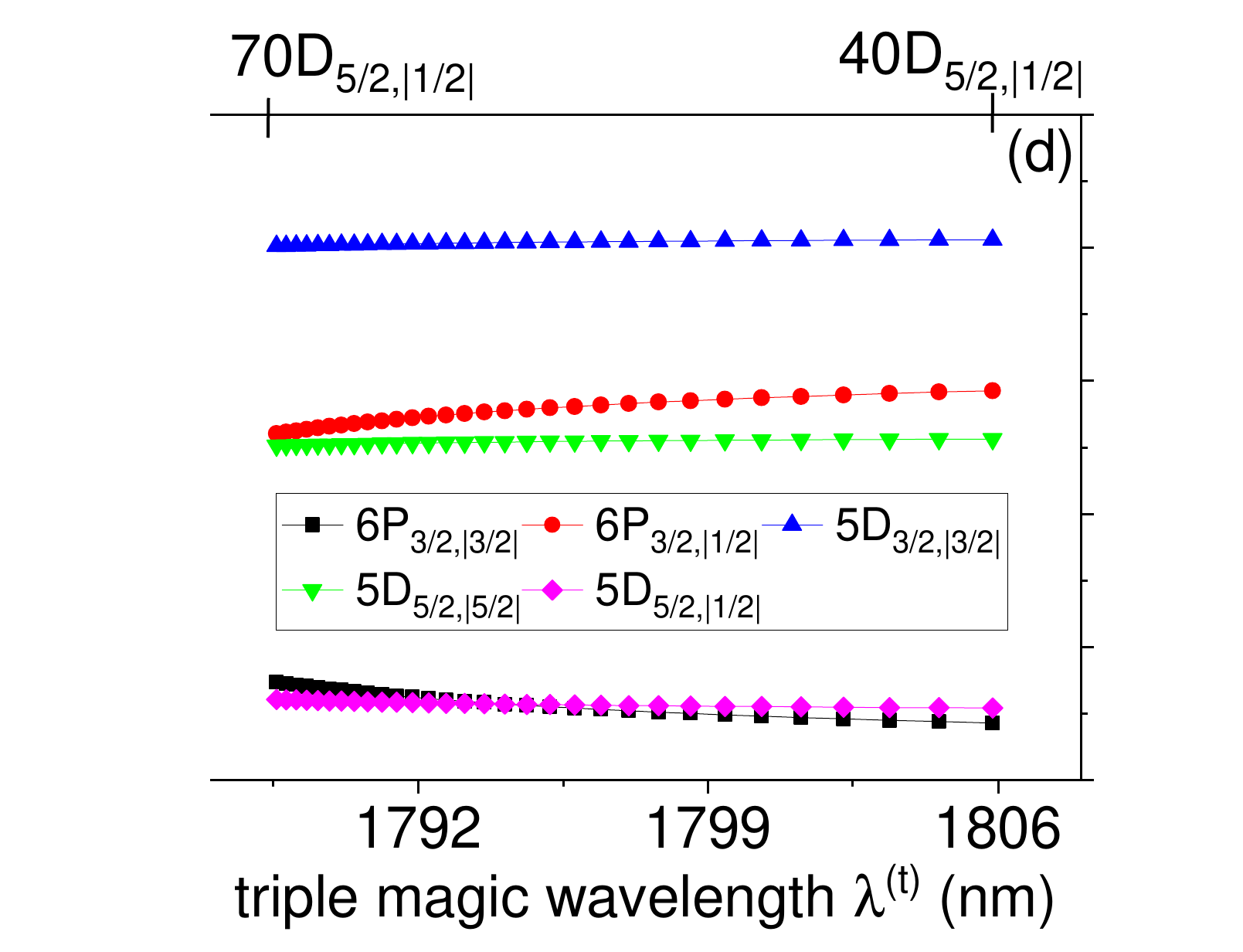}}\hspace{-1.8cm}\\
\caption{Triple magic conditions for the ground state, a Rydberg state, and an intermediate state. The value of $\cos^2 \theta$ for which the polarizability of the intermediate state coincides with the polarizability $\alpha_v^{(d)}$ for which the ground state and the Rydberg state are double magic, as a function of the triple magic wavelength $\lambda^{(t)}$ (adjusting $\theta$ turns the double magic wavelength into a triple magic wavelength). Each panel considers multiple intermediate states (see legend). The  Rydberg states are (a) $nS_{1/2,|1/2|}$ with auxiliary state $8P_{3/2,|1/2|}$ , (b) $nP_{3/2,|1/2|}$ with auxiliary state $7D_{3/2,|1/2|}$, (c) $nD_{3/2,|1/2|}$ with auxiliary state $8P_{1/2,|1/2|}$, and (d) $nD_{5/2,|1/2|}$ with auxiliary state $8P_{3/2,|1/2|}$. } 
\label{Fig9}
\end{figure*}

    \end{widetext}

Figure~\ref{Fig9} displays triple magic conditions for four Rydberg series, namely (a) the $nS_{1/2,|1/2|}$ series (auxiliary state $8P_{3/2,|1/2|}$),   (b) the $nP_{3/2,|1/2|}$ series (auxiliary state $7D_{3/2,|1/2|}$),   (c) the $nD_{3/2,|1/2|}$ series (auxiliary state $8P_{1/2,|1/2|}$), and (d) the $nD_{5/2,|1/2|}$ series (auxiliary state $8P_{3/2,|1/2|}$). To explain the plots, let us first focus on Fig.~\ref{Fig9}(a). The $x$-axis shows the previously identified  $\lambda^{(d)}$ for simultaneous trapping of the ground state and the $nS_{1/2,|1/2|}$ Rydberg state (even though the double magic wavelengths were previously discussed for $\theta=0$,  these double magic wavelengths are, as noted in the previous paragraph, independent of $\theta$). The top axis indicates the $n$ value: the state with $n=70$ corresponds to the smallest double magic wavelength and that with $n=40$ to the largest double magic wavelength. The blue up-triangles show the value of $\cos ^2 \theta$ for which the intermediate state $5D_{3/2,|3/2|}$ is trapped, i.e.,  the double magic wavelength becomes triple magic for this specific angle $\theta$. For the $nS_{1/2,|1/2|}$ series with auxiliary state $8P_{3/2,|1/2|}$, five intermediate states are identified that have triple magic wavelengths for specific $\theta$ for all $n$ considered.   Table~\ref{table_6} summarizes the triple magic wavelengths $\lambda^{(t)}$ as a function of $n$ for the auxiliary state $8P_{3/2,|1/2|}$. Tables S.35-S.63  show analogous results for different auxiliary states and different Rydberg series.

Interestingly, the curves for the $6P_{3/2,|3/2|}$ and $5D_{5/2,|1/2|}$ states in Fig.~\ref{Fig9}(a) cross at $n \approx 53$, signaling the existence of a quadruple magic wavelength $\lambda^{(q)}$.  Similarly, Fig.~\ref{Fig9}(c) indicates the existence of a quadruple magic wavelength that leads to trapping of the states $6S_{1/2,|1/2|}$,  $42D_{3/2,|1/2|}$, $6P_{3/2,|1/2|}$, and $5D_{5/2,|5/2|}$, 
and Fig.~\ref{Fig9}(d)  indicates the existence of a quadruple magic wavelength $\lambda^{(q)}$ that leads to trapping of the states  $6S_{1/2,|1/2|}$,  $54D_{5/2,|1/2|}$, $6P_{3/2,|3/2|}$, and $5D_{5/2,|1/2|}$.   Table~\ref{table_10} enumerates all $\lambda^{(q)}$ identified in this work, along with the corresponding detunings and polarizabilities.

Experimentally, one may also be able to use "approximate" quadruple magic wavelengths for which the difference between the $\theta$  values for the two intermediate states is rather small, say less than $1.5^\circ$ (this value is somewhat arbitrary---for some applications, a less stringent criterium might suffice). One such example is shown in Fig.~\ref{Fig9}(b) for the states $6S_{1/2,|1/2|}$,  $40P_{3/2,|1/2|}$, $6P_{3/2,|1/2|}$, and $5D_{3/2,|3/2|}$.  Tables S.64-S.65 of  the supplemental material enumerate  all the approximate quadruple magic wavelengths for which the difference between the $\theta$ values obtained for two intermediate states is less than $1.5^\circ$.

\begin{widetext}

\begin{table}[h]
\caption {Values of $\cos^2 \theta$ for which triple magic conditions are fulfilled  for the $6S_{1/2,|1/2|}$ ground state, 
the $nS_{1/2,|1/2|}$ Rydberg state  ($n=40-70$),
and several
intermediate states; see Fig.~\ref{Fig_schematic}(b) and Eq.~(\ref{eq_A11}) for the definition of $\theta$.
The last three columns report the triple magic wavelength $\lambda^{(t)}$, the detuning $\Delta$  between the triple magic wavelength and the Rydberg state--auxiliary state transition line, and the associated triple magic polarizability $\alpha_v^{(t)}$. The auxiliary state is $8P_{3/2,|1/2|}$. 
Following the same format as employed in this table, Tables S.35-S.63 of the supplemental material~\cite{Supplement} consider the other Rydberg series and auxiliary states considered in this work.  } 
\centering 
\begin{tabular}{c |c|c|c|c|c|c|c|c } 
\hline\hline
Rydberg 	&	$\cos^2\theta$ for  & $\cos^2\theta$ for  &  $\cos^2\theta$ for  & $\cos^2\theta$ for  & $\cos^2\theta$ for  & $\lambda^{(t)}$ (nm)& $\Delta$ (MHz) & $\alpha_v^{(t)}$ (a.u.)		\\
state &	$6P_{3/2,|3/2|}$  & $6P_{3/2,|1/2|}$  &  $5D_{3/2,|3/2|}$  & $5D_{5/2,|5/2|}$  & $5D_{5/2,|1/2|}$  & &   & 	\\
\hline
$40S_{1/2,|1/2|}$	&	0.0814	&	0.5942	&	0.8131	&	0.5136	&	0.1073	&	1808.2790	&	2027.01	&	512.7849	\\
$41S_{1/2,|1/2|}$	&	0.0834	&	0.5919	&	0.8128	&	0.5133	&	0.1077	&	1806.7987	&	1946.37	&	513.0281	\\
$42S_{1/2,|1/2|}$	&	0.0854	&	0.5896	&	0.8124	&	0.5129	&	0.1082	&	1805.4361	&	1862.27	&	513.2527	\\
$43S_{1/2,|1/2|}$	&	0.0873	&	0.5873	&	0.8120	&	0.5126	&	0.1086	&	1804.1790	&	1775.56	&	513.4605	\\
$44S_{1/2,|1/2|}$	&	0.0893	&	0.5850	&	0.8117	&	0.5122	&	0.1091	&	1803.0168	&	1687.11	&	513.6532	\\
$45S_{1/2,|1/2|}$	&	0.0913	&	0.5828	&	0.8113	&	0.5119	&	0.1095	&	1801.9401	&	1597.80	&	513.8322	\\
$46S_{1/2,|1/2|}$	&	0.0933	&	0.5805	&	0.8110	&	0.5115	&	0.1100	&	1800.9407	&	1508.48	&	513.9988	\\
$47S_{1/2,|1/2|}$	&	0.0953	&	0.5782	&	0.8106	&	0.5112	&	0.1105	&	1800.0113	&	1420.04	&	514.1540	\\
$48S_{1/2,|1/2|}$	&	0.0973	&	0.5759	&	0.8103	&	0.5108	&	0.1109	&	1799.1455	&	1333.33	&	514.2989	\\
$49S_{1/2,|1/2|}$	&	0.0993	&	0.5736	&	0.8099	&	0.5105	&	0.1114	&	1798.3377	&	1249.22	&	514.4344	\\
$50S_{1/2,|1/2|}$	&	0.1013	&	0.5713	&	0.8096	&	0.5101	&	0.1118	&	1797.5828	&	1168.59	&	514.5612	\\
$51S_{1/2,|1/2|}$	&	0.1033	&	0.5690	&	0.8092	&	0.5098	&	0.1123	&	1796.8761	&	1092.29	&	514.6801	\\
$52S_{1/2,|1/2|}$	&	0.1053	&	0.5667	&	0.8089	&	0.5094	&	0.1128	&	1796.2137	&	1021.20	&	514.7918	\\
$53S_{1/2,|1/2|}$	&	0.1073	&	0.5644	&	0.8085	&	0.5091	&	0.1132	&	1795.5919	&	956.19	&	514.8968	\\
$54S_{1/2,|1/2|}$	&	0.1093	&	0.5622	&	0.8082	&	0.5088	&	0.1137	&	1795.0074	&	898.13	&	514.9956	\\
$55S_{1/2,|1/2|}$	&	0.1113	&	0.5599	&	0.8078	&	0.5084	&	0.1141	&	1794.4573	&	847.87	&	515.0887	\\
$56S_{1/2,|1/2|}$	&	0.1133	&	0.5576	&	0.8075	&	0.5081	&	0.1146	&	1793.9388	&	806.01	&	515.1766	\\
$57S_{1/2,|1/2|}$	&	0.1150	&	0.5553	&	0.8071	&	0.5077	&	0.1150	&	1793.4498	&	771.96	&	515.2596	\\
$58S_{1/2,|1/2|}$	&	0.1173	&	0.5530	&	0.8067	&	0.5074	&	0.1155	&	1792.9878	&	744.85	&	515.3381	\\
$59S_{1/2,|1/2|}$	&	0.1192	&	0.5507	&	0.8064	&	0.5070	&	0.1160	&	1792.5510	&	723.81	&	515.4124	\\
$60S_{1/2,|1/2|}$	&	0.1212	&	0.5484	&	0.8060	&	0.5067	&	0.1164	&	1792.1376	&	707.99	&	515.4827	\\
$61S_{1/2,|1/2|}$	&	0.1232	&	0.5461	&	0.8057	&	0.5063	&	0.1169	&	1791.7459	&	696.50	&	515.5495	\\
$62S_{1/2,|1/2|}$	&	0.1252	&	0.5439	&	0.8053	&	0.5060	&	0.1173	&	1791.3744	&	688.49	&	515.6129	\\
$63S_{1/2,|1/2|}$	&	0.1272	&	0.5416	&	0.8050	&	0.5056	&	0.1178	&	1791.0218	&	683.08	&	515.6730	\\
$64S_{1/2,|1/2|}$	&	0.1292	&	0.5393	&	0.8046	&	0.5053	&	0.1183	&	1790.6868	&	679.41	&	515.7303	\\
$65S_{1/2,|1/2|}$	&	0.1312	&	0.5370	&	0.8043	&	0.5049	&	0.1187	&	1790.3682	&	676.60	&	515.7847	\\
$66S_{1/2,|1/2|}$	&	0.1332	&	0.5347	&	0.8039	&	0.5046	&	0.1192	&	1790.0651	&	673.80	&	515.8366	\\
$67S_{1/2,|1/2|}$	&	0.1352	&	0.5324	&	0.8036	&	0.5042	&	0.1196	&	1789.7764	&	670.13	&	515.8860	\\
$68S_{1/2,|1/2|}$	&	0.1372	&	0.5301	&	0.8032	&	0.5039	&	0.1201	&	1789.5013	&	664.72	&	515.9332	\\
$69S_{1/2,|1/2|}$	&	0.1392	&	0.5278	&	0.8029	&	0.5036	&	0.1205	&	1789.2388	&	656.70	&	515.9782	\\
$70S_{1/2,|1/2|}$	&	0.1412	&	0.5255	&	0.8025	&	0.5032	&	0.1210	&	1788.9884	&	645.22	&	516.0211	\\
\hline 
\end{tabular}
\label{table_6} 
\end{table}

\end{widetext}

\begin{widetext}

\begin{table}[h]
\caption{Values of $\cos^2 \theta$ for which quadruple magic conditions are fulfilled  for the $6S_{1/2,|1/2|}$ ground state, a Rydberg state, and two intermediate states (denoted as intermediate state-1 and intermediate state-2) for three different auxiliary states; see Fig.~\ref{Fig_schematic}(b) and Eq.~(\ref{eq_A11}) for the definition of $\theta$. A total of seven quadruple magic conditions were found. 
The last three columns report the quadruple magic wavelength $\lambda^{(q)}$, the detuning $\Delta$  between the quadruple magic wavelength and the Rydberg state--auxiliary state transition line, and the associated quadruple magic polarizability $\alpha_v^{(q)}$.} 
\centering 
\begin{tabular}{c|c | c| c|c|c|c} 
\hline\hline 
	Rydberg state 	&	intermediate state-1	&	intermediate state-2	& $\cos^2\theta$ &	$\lambda^{(q)}$ (nm) &  $\Delta$ (MHz)& $\alpha_v^{(q)}$  (a.u.) \\
\hline
\multicolumn{7}{c}{\textbf{auxiliary state $8P_{1/2,|1/2|}$}}\\
\hline
	$43S_{1/2,|1/2|}$	&	$6P_{3/2,|1/2|}$	&	$5D_{5/2,|5/2|}$	&	0.4996&	1777.6610	&	2067.04	&	517.9914
		\\

	$42D_{3/2,|1/2|}$	&	$6P_{3/2,|1/2|}$	&	$5D_{5/2,|5/2|}$	&	0.4995&	1776.7886 &	22356.16	&	518.6544
		\\
 \hline
\multicolumn{7}{c}{\textbf{auxiliary state $8P_{3/2,|1/2|}$}}\\
\hline
	$57S_{1/2,|1/2|}$	&	$6P_{3/2,|3/2|}$	&	$5D_{5/2,|1/2|}$	&	0.1150&	1793.4498	&	771.96	&	515.2596
 	\\

	$56D_{3/2,|1/2|}$	&	$6P_{3/2,|3/2|}$	&	$5D_{5/2,|1/2|}$	&	0.1150&	1793.1735&	1497.43	&	517.6532
		\\
	$54D_{5/2,|1/2|}$	&	$6P_{3/2,|3/2|}$	&	$5D_{5/2,|1/2|}$	&	0.1145&	1794.0898 &	6549.81	&	515.3139
		\\
	 \hline
\multicolumn{7}{c}{\textbf{auxiliary state $8P_{3/2,|3/2|}$}}\\
\hline
	$56D_{3/2,|3/2|}$	&	$6P_{3/2,|3/2|}$	&	$5D_{5/2,|1/2|}$	&	0.1153&	1792.8876 &	28171.61	&	511.6663
		\\
	
	$54D_{5/2,|3/2|}$	&	$6P_{3/2,|3/2|}$	&	$5D_{5/2,|1/2|}$	&	0.1138&	1794.1316 &	2652.96	&	515.3139
		\\
\hline 
\end{tabular}
\label{table_10} 
\end{table}

\end{widetext}

Since ultracold atom experiments are sensitive to the hyperfine structure, it is important to check how a change of the basis from $(I_N,M_I,J,M_J)$ to
$(J,I_N,F,M_F)$ impacts the triple magic wavelength conditions. As discussed earlier, the ground state is unaffected by the basis change while the Rydberg state is weakly affected. The low-lying excited states, in contrast, are comparatively strongly affected by the basis change, leading to shifts of the polarizability curves as well as changes of the functional form.    Appendix~\ref{Appendix_C} illustrates, by showing examples, that triple magic wavelengths also exist when accounting for the hyperfine structure. As such, the triple magic wavelength tables provided in this paper serve as a guide for the analysis that accounts for the hyperfine states.

\section{Conclusion}\label{sec_6}
This paper presented the lifetime and  dynamic polarizability  for a $^{133}$Cs atom in the  Rydberg state with  principal quantum number $40\leq n\leq 70$. Explicitly, we considered  all possible total angular momenta $J$ for the Rydberg series $nS$, $nP$, and $nD$.  We observed that, in general,  the room temperature ($T=300$~K) blackbody radiation decay  impacts the lifetime of a Rydberg state 
significantly. The effective lifetime, which is a combination of the spontaneous and blackbody radiation lifetimes,  does not scale as $n^{3}$ but with a lower power of $n$, namely  the scaling law for the states considered in this work is $n^{a}$ with $2.3 \le a \le 2.69$, where the value of $a$  depends on the orbital angular momentum and total electronic angular momentum of the Rydberg state.   

Moreover, we presented  static and dynamic polarizabilities for cesium Rydberg states for a linearly polarized Gaussian shaped laser beam, whose polarization vector is parallel to the quantization axis set by the external magnetic field vector $\vec{B}_{\text{ext}}$.  The polarizabilities were calculated using  the sum-over-state approach and  our  static scalar and tensor polarizabilities agree well with   available theoretical and experimental results.     

To confine the ground state and a Rydberg state simultaneously in the same potential well, we utilize a double magic wavelength based trapping scheme for which the wavelength of the laser is blue-detuned for the Rydberg state and red detuned for the ground state.
 A total of 34 sequences were identified (9 different Rydberg series and a varying number of auxiliary states for each Rydberg series), for which the differential ac-Stark shift between the ground state and the Rydberg state is zero.
  For  all the Rydberg series and respective auxiliary states considered,   with the exception of the Rydberg series $nP_{3/2,|1/2|}$ with auxiliary state $6D_{3/2,|1/2|}$, the detunings  are notably larger than $ 50$~MHz. 
The Rydberg series $nD_{3/2,|1/2|}$ with auxiliary state $8P_{1/2,|1/2|}$ and $nD_{3/2,|3/2|}$ with auxiliary state $8P_{3/2,|3/2|}$ were found to have particularly large detunings up to $20,000-30,000$~MHz.
  
  To further assess the applicability of  the   double magic wavelengths presented, we determined the Rabi frequency and discussed   the resulting maximal population transfer from the Rydberg state to the auxiliary state.
  While the reported $P_{\text{max}}$ are based on a trap depth of $1$~$\mu$K, the tables contain all the information needed to perform an analogous analysis for other trap depths.   We demonstrated, using a master equation formalism that accounts for the spontaneous lifetime of low-lying excited states and the auxiliary state, that an atom in the $nD_{3/2,|1/2|}$  or $nD_{3/2,|3/2|}$ states can be trapped up to several $\mu$s without significant losses, provided the trap is sufficiently shallow. For the $nD$ series, very encouraging results were found for trap depth as large as $10-100$~$\mu$K.
  
While only a subset of Rydberg and auxiliary states investigated promise to be suitable for long-time trapping, our work suggests other interesting applications. For example, the formalism developed can be used to determine tune-out wavelengths~\cite{LeBlanc2007, Ratkata2021}, for which the atom in the Rydberg state does neither feel an attractive nor a repulsive force. Operating at tune-out wavelengths might be viewed as a compromise, where the unwanted losses are reduced at the expense of loosing the trapping benefits.

Lastly, by tuning the angle $\theta$ between the quantization axis and the polarization vector, we  found   triple magic wavelengths for the ground state, a Rydberg state, and a low-lying excited state as well as several quadruple magic wavelengths for the ground state, a Rydberg state,  and  two low-lying excited states. The triple magic wavelength condition for a given Rydberg series and given intermediate state tends to have a small $\theta$ dependence, i.e., triple magic wavelengths were found for $n=40-70$. Quadruple magic wavelengths, in contrast, are---since they result from the crossing of two triple magic conditions---realized only for very specific $n$ and only in select incidences.

 The predicted double, triple, and quadruple magic wavelengths  are expected to be useful in, among other things, performing "spectroscopic studies" that can help pin down lifetimes and dipole matrix elements. Initializing the atom in a Rydberg state, the decay out of Rydberg states leads---on short time scales---to state populations that oscillate in time with an overall decay. Their analysis is expected to provide stringent constraints on lifetimes and dipole matrix elements. 
 In addition, the double magic wavelengths might find applications in implementations of different    quantum information and quantum metrology protocols. 

 A natural next step is to extend our study for linearly polarized light to   circularly polarized light. Preliminary explorations indicate that elliptically polarized light might offer intriguing advantages. Relatedly, it will be interesting to explore the influence of light sources with a non-Gaussian beam profile, such as vortex light, to confine cesium Rydberg atoms as well as other alkali Rydberg atoms, thereby opening up the possibility of    simultaneously trapping mixtures of Rydberg atoms. Such a setting would open the door for leveraging  long-range interactions between two Rydberg atoms.

\section*{Acknowledgement}
This work was supported by an award from the W. M. Keck Foundation. We thank Lindsay LeBlanc for discussions on tune-out wavelengths.

\appendix

\section{Energies, wavefunctions, and dipole matrix elements of  Rydberg states }\label{Appendix_A}

To determine the energies, wavefunctions, and dipole matrix elements of various Rydberg states, we apply  quantum defect theory and the Coulomb approximation method~\cite{Seaton1983,Wijngaarden1994}.   Within quantum defect theory,  the principal quantum  number $n$  and, correspondingly, the energy levels $\epsilon_{n,L,J}$ of single-valence Rydberg states are parameterized similar to those  of the hydrogen atom. Specifically, the energy $\epsilon_{n,L,J}$ of the state $nL_J$  is written in terms of the effective  fractional principal  quantum number $n_{\text{eff}}$,
\begin{equation}\label{eq_A1}
\epsilon_{n,L,J}=-\dfrac{m_r\alpha^2 c^2}{2(n_{\text{eff}})^2},
\end{equation}
where $m_r$ denotes the reduced mass
of the atom. 
The effective principal quantum number is parameterized in terms of the quantum defect parameters $\delta_{L,J,2i}$,  
\begin{equation}\label{eq_A2}
n_{\text{eff}}=n-\sum_{i=0}^\infty \dfrac{\delta_{L,J,2i}}{(n-\delta_{L,J,0})^{2i}}
\end{equation}
($i$ is a dummy index).
Note that $n_{\text{eff}}$ depends on the orbital angular momentum quantum number $L$ and the total angular momentum quantum number $J$ but that the subscripts are suppressed for notational convenience.
Table~\ref{table_A1} lists literature values of $\delta_{L,J,2i}$, extracted from experimental   data, for various Rydberg series.  
\setcounter{table}{0}
\renewcommand{\thetable}{A.\arabic{table}}

\begin{widetext}

\begin{table}[h]
\caption{Experimental quantum defect parameters for various Rydberg series of cesium. The entries with superscripts $a$, $b$, $c$, and $d$ are taken from Ref.~\cite{Deiglmayr2016},    Ref.~\cite{Lorenzen1983}, Ref.~\cite{Bai2023}, and Ref.~\cite{Weber1987}, respectively.  } 
\centering 
\begin{tabular}{c |c|c | c| c|c} 
\hline\hline 
state 	&	$\delta_{L,J,0}$ 	&	$\delta_{L,J,2}$	&	$\delta_{L,J,4}$	&	$\delta_{L,J,6}$ &   $\delta_{L,J,8}$\\
\hline
$nS_{1/2}$	&	$4.0493532(4)^a$   	&	$0.2391(5)^a$	&	$0.06(10)^a$	&	$11(7)^a$	& $-209(150)^a$\\
$nP_{1/2}$	&	$3.5915871(3)^a$   	&	$0.36273(16)^a$	&		&	\\
$nP_{3/2}$	&	$3.5590676(3)^a$   	&	$0.37469(14)^a$ 	&		&		&\\

$nD_{3/2}$	&	$2.47545(2)^b$    	&	$0.0099(40)^b$	&	$-0.43324^b$	&	$-0.96555^b$	&$-16.9464^b$\\

$nD_{5/2}$	&	$2.4663144(6)^a$    	&	$0.01381(15)^a$	&	$-0.392(12)^a$	&	$-1.9(3)^a$	& \\

$nF_{5/2}$	&	$0.03341537(70)^c$    	&	$-0.2014(16)^c$	&	$0.28953^d$	&	$-0.2601^d$	& \\

$nF_{7/2}$	&	$0.0335646(13)^c$    	&	$-0.2052(29)^c$	&		&		& \\
\hline 
\end{tabular}
\label{table_A1} 
\end{table}

\end{widetext}

Using $n_{\text{eff}}$,  the wavefunction of a Rydberg state can be efficiently approximated by the modified Coulomb solutions~\cite{Bates1949}. This approach has previously been   shown to yield accurate  polarizabilities for alkali atoms~\cite{Wijngaarden1997, Yerokhin2016}.  For Rydberg states, the  solution to the radial part of the Schr\"odinger equation that is regular in the $r \rightarrow \infty$ limit  ($r$ denotes the distance of the valence electron from the core) and accounts for the orbital angular momentum barrier and Coulomb potential reads~\cite{Bates1949}
\begin{eqnarray}\label{eq_A3}
R_{n_{\text{eff}},L}(r)=&&\dfrac{1}{[a_0 n_{\text{eff}}^2\Gamma(n_{\text{eff}}+L-1)\Gamma(n_{\text{eff}}-L)]^{1/2}}
\times \nonumber \\
&&W_{n_{\text{eff}}, L+1/2}(y),
\end{eqnarray}
where  $W_{n_{\text{eff}},L+1/2}(y)$
with $y= 2r/(n_{\text{eff}} a_0)$ denotes the Whittaker function and $a_0$ the Bohr radius.  One can check readily that the radial wavefunction $R_{n_{\text{eff}},L}(r)$ reduces to the well known non-relativistic bound state wavefunction for integer quantum numbers, i.e., for $n_{\text{eff}} = n$~\cite{Yerokhin2016}. The  radial transition dipole matrix element $R_{ki}$  is given by
\begin{eqnarray}\label{eq_A7}
R_{ki}=\int_{r_0}^\infty dr r R_{n_{\text{eff},k},L_k }(r) R_{n_{\text{eff},i},L_i}(r),
\end{eqnarray}
where the lower  integration limit $r_0$ needs to be chosen sufficiently small to ensure convergence of  the  results.
 We find that  
         \begin{eqnarray}
     r_0=\dfrac{sn_{\text{eff},k} n_{\text{eff},i} a_0}{n_{\text{eff},k}+n_{\text{eff},i}}\end{eqnarray} 
     with $s$ around $1/100$ yields converged results.

\section{Calculation of polarizability}\label{Appendix_B}

\setcounter{figure}{0}
\renewcommand{\thefigure}{B.\arabic{figure}}
The Stark shift $\Delta\xi_v $ for an  atom or ion 
with the valence electron in the  $v$-th state 
is given within second-order time-independent perturbation theory by 
\begin{eqnarray}\label{eq_A8}
\Delta\xi_v(\omega)=-\frac{1}{2}\alpha_v(\omega)\textit{E}^2,
\end{eqnarray}
where $E$ and $\omega$ are the amplitude and frequency of the external electric field, respectively. In Eq.~(\ref{eq_A8}), $\alpha_v(\omega)$ denotes---as already noted in the main text---the dynamic polarizability of the $v$-th state; 
it reduces to the static polarizability in the $\omega\rightarrow 0$ limit.
According to Eq.~(\ref{eq_4}),
$\alpha_v(\omega)$ is composed of the core polarizability $\alpha^C_v(\omega)$, the valence-core polarizability  $\alpha^{VC}_v(\omega)$, and the valence polarizability $\alpha^V_v(\omega)$.

For a Cs atom, the core is obtained by removing the valence electron, i.e., it is the singly-charged  Cs$^+$ ion.  The polarizability of Cs$^+$ can be estimated as~\cite{Mitroy2010,Bhowmik2018}
\begin{eqnarray}\label{eq_A10}
\alpha^C_v(\omega)= 
\frac{2}{3}e^2
\times 
\nonumber \\
\sum_{ap}\frac{|\langle \Phi_a||d_\textrm{DF}||\Phi_p\rangle \langle \Phi_a||d_\textrm{RMBPT(2)}||\Phi_p\rangle|(\epsilon_p-\epsilon_a)}{(\epsilon_p-\epsilon_a)^2-\omega^2},
\end{eqnarray}
where $a$  runs over all occupied core orbital configurations   
and $p$ represents all virtual
orbitals  with
appropriate symmetries with respect to the core orbitals. The matrix elements $\langle \Phi_a||d_\textrm{DF}||\Phi_p\rangle$ and   $\langle \Phi_a||d_\textrm{RMBPT(2)}||\Phi_p\rangle$ are reduced dipole matrix elements at the Dirac-Fock (DF)  and the second-order relativistic many-body perturbation theory (RMBPT(2)) levels, respectively~\cite{Mitroy2010,Bhowmik2018}. 
\begin{figure}[!h]
{\includegraphics[ scale=.30]{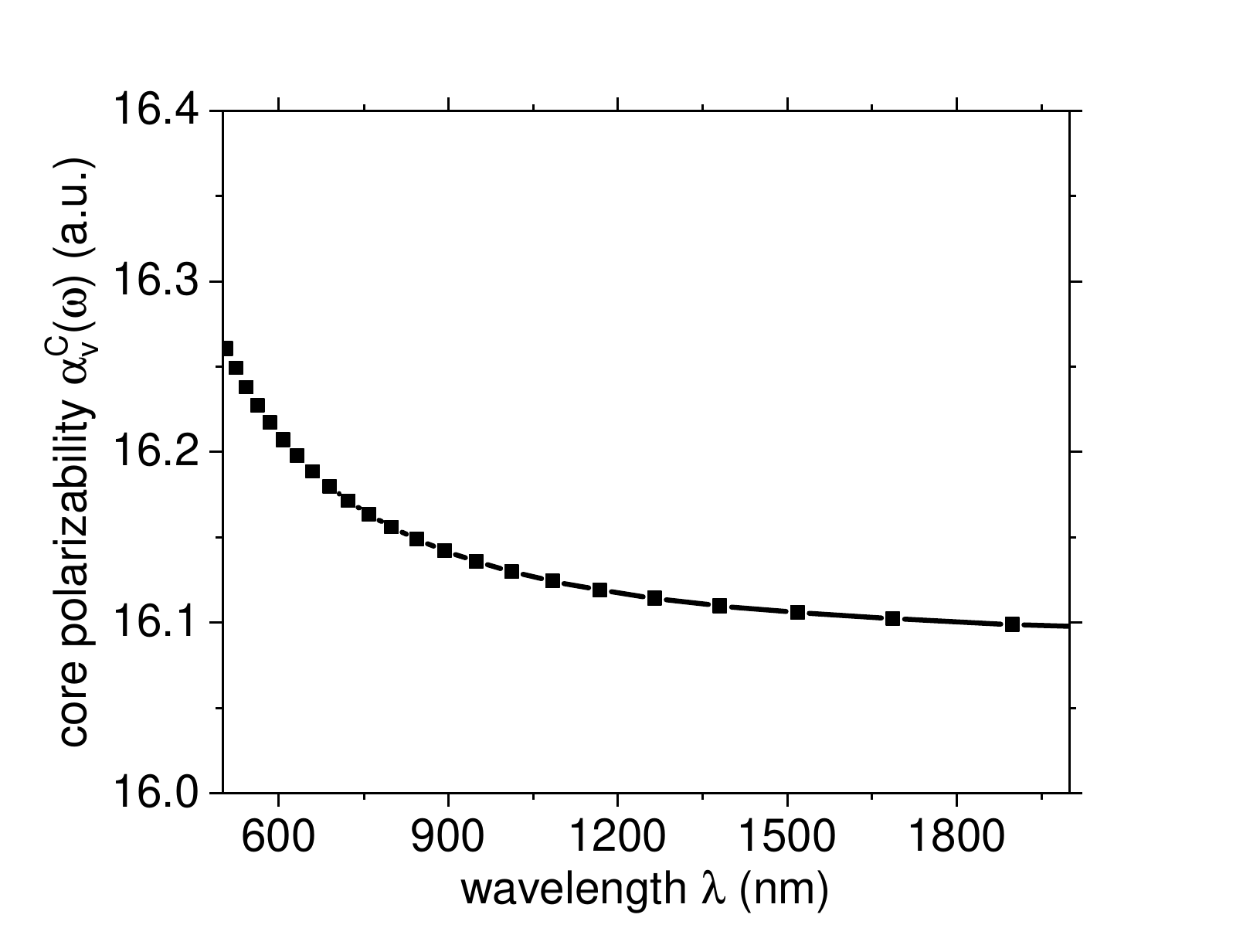}}
\vspace*{-0.05in}
\caption{Core polarizability $\alpha^C_v(\omega)$ 
of Cs$^+$ as a function of the wavelength $\lambda$.  It can be seen that the wavelength dependence or, equivalently, the frequency dependence of $\alpha^C_v(\omega)$ is negligible. }
\label{Fig_App_1}
\end{figure}
Figure~\ref{Fig_App_1} shows that the core polarizability varies slowly  with the wavelength of the light and approximately reaches its asymptotic value at the largest wavelength considered, i.e., around $\lambda=2,000$~nm. 

Since the core electrons are tightly bound to
the nucleus, the presence of the valence electron provides a very small change to  the core polarizability. Specifically,   $\alpha^{VC}_v(0)$  is  $-0.47$~a.u. for  the ground state of Cs \cite{Singh2016}. Since  $|\alpha^{VC}_v(0)|$ is very small and   essentially independent of $\omega$, we neglect the $\omega$ dependency of $\alpha^{VC}_v(\omega)$ ~\cite{Singh2016,Safronova2016}. Moreover, $\alpha^{VC}_v(0)$ is approximately zero for the low-lying excited states considered in this work~\cite{Singh2016}; correspondingly, we use $\alpha^{VC}_v(\omega)=0$ throughout for the low-lying excited  and  Rydberg states.

For linearly polarized light, the  valence polarizability $\alpha^V_v(\omega)$ is given by a weighted sum of the scalar polarizability $\alpha_{v,J}^{(0)}(\omega)$ and the tensor polarizability $\alpha_{v,J}^{(2)}(\omega)$~\cite{Manakov1986,Aora2012,Arora2007, Jiang2019, Bhowmik2022}, 
\begin{eqnarray}\label{eq_A11}
\alpha_{v}^{V}(\omega)= \alpha_{v,J}^{(0)}(\omega)+ \nonumber \\
\dfrac{3 \cos^2\theta-1}{2}\frac{3M_{J}^2-J(J+1)}{J(2J-1)}\alpha_{v,J}^{(2)}(\omega).
\end{eqnarray} 
The definition of the angle $\theta$ is illustrated in Fig.~\ref{Fig_schematic}.
The scalar polarizability has the form
\begin{eqnarray}\label{eq_A12}
\alpha_{v,J}^{(0)}(\omega)= \frac{2}{3(2J + 1)}\sum_n d_{nv},
\end{eqnarray}
where
$d_{nv}=
\frac{e^2|\langle\psi_v||d||\psi_n\rangle|^2 \omega_{nv}}{ \omega_{nv}^2-\omega^2}$, with $\omega_{nv}$ denoting the difference between the frequencies $\omega_n$ and $\omega_v$ of states $\psi_n$ and $\psi_v$, respectively, $\omega_{nv}=\omega_n-\omega_v$.  
 The  tensor polarizability is given by
 \begin{eqnarray}\label{eq_A13}
\alpha_{v,J}^{(2)}(\omega)=4\sqrt{\frac{5J(2J-1)}{6(J+1)(2J+1)(2J+3)}} \times \nonumber \\
\sum_n (-1)^{J_n+J} \left\{\begin{array}{ccc} J & 1 & J_n\\ 1 & J& 2 \end{array}\right \} d_{nv},
\end{eqnarray} 
where $\{\}$ denotes the $6j$ symbol~\cite{Manakov1986}.

 To obtain the triple and quadruple magic  wavelengths, we  consider the polarizabilities of  the low-lying states  $6S_{1/2,|1/2|}$, $6P_{1/2,|1/2|}$, $6P_{3/2,|1/2|}$,  $6P_{3/2,|3/2|}$,  $5D_{3/2,|1/2|}$, $5D_{3/2,|3/2|}$, $5D_{5/2,|1/2|}$, $5D_{5/2,|3/2|}$, and $5D_{5/2,|5/2|}$.  To calculate the polarizability of these low-lying states, we  collected 
 matrix elements from  Ref.~\cite{UDportal} and energies from Ref.~\cite{NIST_ASD}.

 \section{Changing   to $(J,I_N,F,M_F)$ basis}
 \label{Appendix_C}

The main text presents the  magic wavelengths for simultaneously trapping the ground state and  a Rydberg state. Throughout, the states were labeled by $n$, $L$, $J$, and $M_J$; the nuclear spin $I_N$ and associated projection quantum number $M_I$   were suppressed. 
Ultracold experiments, however, work with hyperfine states that are labeled by
$n$, $L$, $J$, $I_N$, $F$, and $M_F$.
In what follows, we account for the hyperfine splitting and estimate, by showing examples,  how the magic wavelengths change when one
goes from the $(J,M_J,I_N,M_I)$ to the
$(J,I_N,F,M_F)$ basis. At the hyperfine level, the dynamic polarizability of a valence state is given by \cite{Rosenbusch2009,Kaur2017,Das2020}
\begin{eqnarray}\label{eq_C1}
\alpha_{v}^{V}(\omega)= \alpha_{v,F}^{(0)}(\omega)+ \nonumber \\
\dfrac{3 \cos^2\theta-1}{2}\frac{3M_{F}^2-F(F+1)}{F(2F-1)}\alpha_{v,F}^{(2)}(\omega),
\end{eqnarray}
where $\alpha_{v,F}^{(0)}(\omega)$ and $\alpha_{v,F}^{(2)}(\omega)$  denote, respectively, the scalar and tensor polarizabilities  at the hyperfine level [note the similarity between Eqs.~(\ref{eq_C1}) and (\ref{eq_A12})].  The scalar polarizability does not change
under the basis change, i.e., $\alpha_{v,J}^{(0)}(\omega)=\alpha_{v,F}^{(0)}(\omega)$~\cite{Rosenbusch2009}. The tensor polarizability, in contrast, changes under the basis change~\cite{Kaur2017,Das2020}, 
 \begin{eqnarray}\label{eq_C2}
\alpha_{v,F}^{(2)}(\omega)=(-1)^{J+F+I_N}\left\{\begin{array}{ccc} F & J & I_N\\ J & F& 2 \end{array}\right \}\nonumber \\
\times \sqrt{\frac{F(2F-1)(2F+1)}{(2F+3)(F+1)}}\nonumber \\
\times \sqrt{\frac{(2J+3)(2J+1)(J+1)}{J(2J-1)}}
\alpha_{v,J}^{(2)}(\omega).
\end{eqnarray}

Since $I_N$ is equal to $7/2$ for $^{133}$Cs, the ground state splits into the $F=3$ and $F=4$ manifolds. To illustrate the effect of the basis change,
Fig.~\ref{Fig_App_2}(a) shows the 
 dynamic polarizabilities of the  $6S_{J=1/2,F=4,M_{F}}$ ground states
 ($M_F$ can take the values $-4,-3,\cdots,4$) and of the $6P_{J=3/2,F=5,M_{F}}$ low-lying excited states ($M_F$ can take the values $-5,-4,\cdots,5$) at $\theta=0$.
 As noted above, the polarizabilities 
  of the $6S_{J=1/2,F=4,M_{F}}$ states with different $M_F$ are identical.
 For the $6P_{J=3/2,F=5,M_{F}}$ states, the polarizabilities move further away from the resonance line as $|M_F|$ increases from $0$ to $5$.
 As a result,
  the  magic wavelengths for the low-lying state $6P_{J=3/2,F=5,M_{F}}$ and the ground state vary by around 220~nm  as the magnitude $|M_F|$ of the projection quantum number that characterizes the low-lying excited state changes [the black box in Fig.~\ref{Fig_App_2}(a) highlights the behavior].
  
 Figure~\ref{Fig_App_2}(b) considers the polarizabilities of the $F=4$ ground states and the  $45D_{J=3/2,F=5,M_{F}}$
  Rydberg states. Similar to the low-lying excited state in Fig.~\ref{Fig_App_2}(a), the polarizability of the Rydberg states is shown for  $|M_F|=0$ to $5$ for $\theta=0$. It can be seen that the magic wavelength changes comparatively little with $|M_F|$. Specifically, the magic wavelength increases by about $0.1$~nm as $|M_F|$ changes from $0$ to $5$ [the black box in Fig.~\ref{Fig_App_2}(b) highlights the behavior].
  The tiny shift suggests that the tabulated double magic wavelength predictions, calculated by using the $(J,M_J,I_N,M_I)$ basis provide a reliable starting point for locating double magic wavelengths experimentally. This conclusion does not only hold for the $45D_{J=3/2,F=5,M_{F}}$ Rydberg states but for the entire Rydberg series, other auxiliary states, as well as other Rydberg series.

\setcounter{figure}{0}
\renewcommand{\thefigure}{C.\arabic{figure}}
\begin{figure}[!h]
{\includegraphics[ scale=.29]{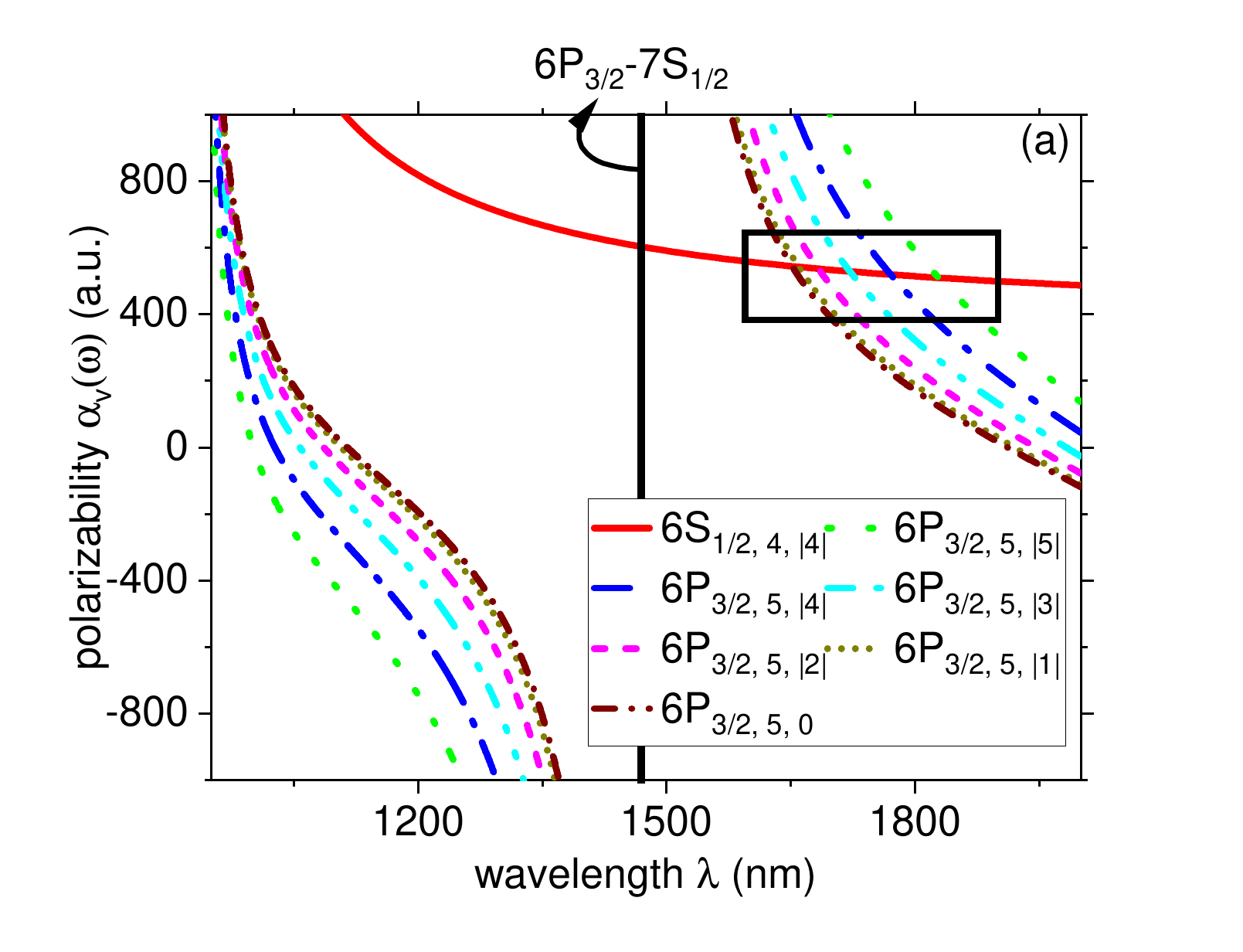}}
{\includegraphics[ scale=.29]{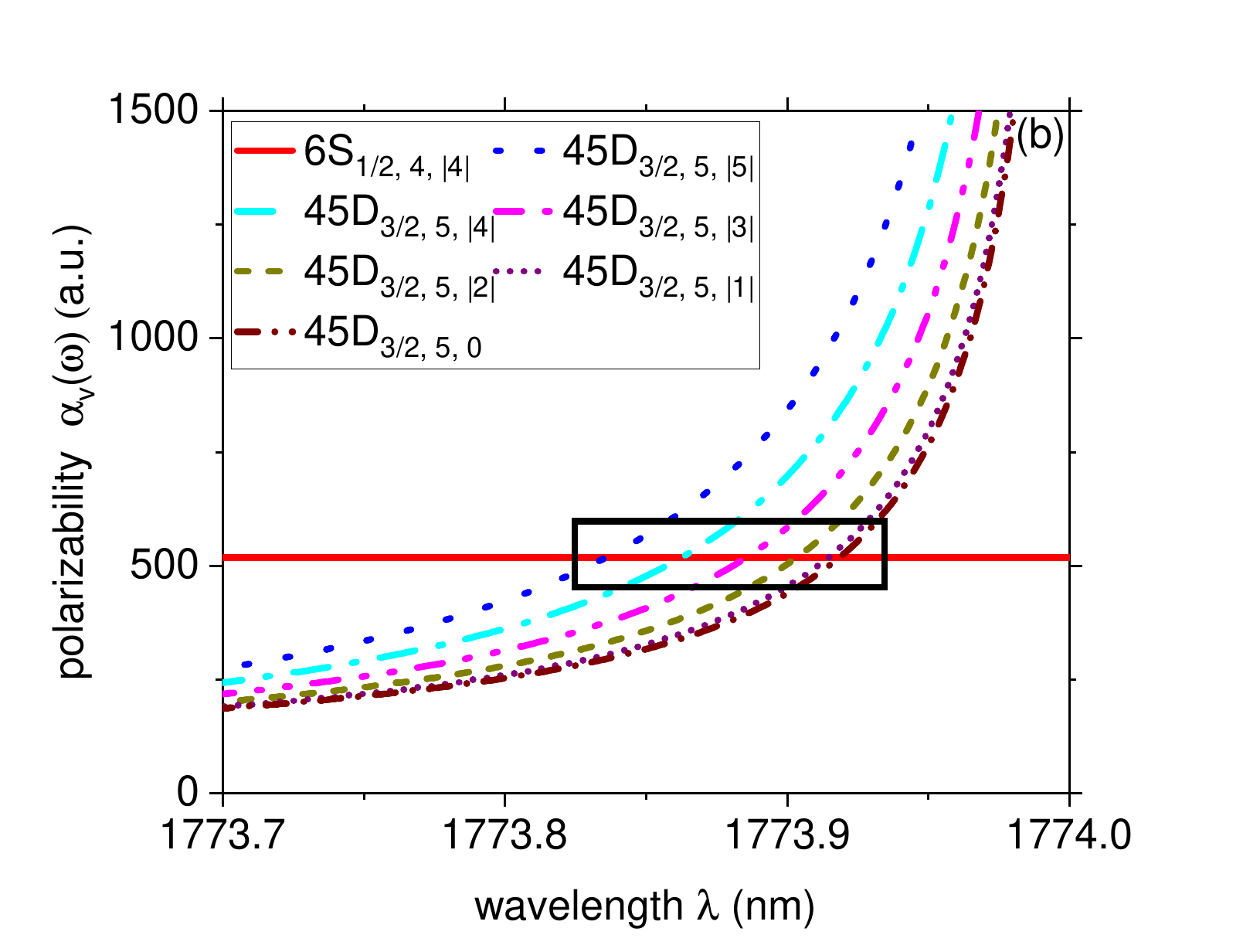}}
\vspace*{-0.05in}
\caption{Dynamic polarizability $\alpha_v(\omega)$ of (a) the $6S_{J=1/2,F=4,M_{F}=4}$ ground state  (red solid line) and the $6P_{J=3/2,F=5,M_{F}}$ low-lying excited state ("diving lines"),  and  (b) the $6S_{J=1/2,F=4,M_{F}=4}$ ground state (red solid line)  and the $45D_{J=3/2,F=5,M_{F}}$ Rydberg state ("diving lines") for $\theta=0$. The "diving lines" show the polarizability of the different $|M_F|$ states for $F=5$ (see legend). The black boxes draw the readers' attention to how the magic wavelengths between the ground state and the (a) low-lying excited states and (b) Rydberg states vary with $|M_{F}|$.  The solid black vertical line in (a)   indicates  the resonance line.   
} 
\label{Fig_App_2}
\end{figure}

\begin{figure}[!h]
{\includegraphics[ scale=.29]{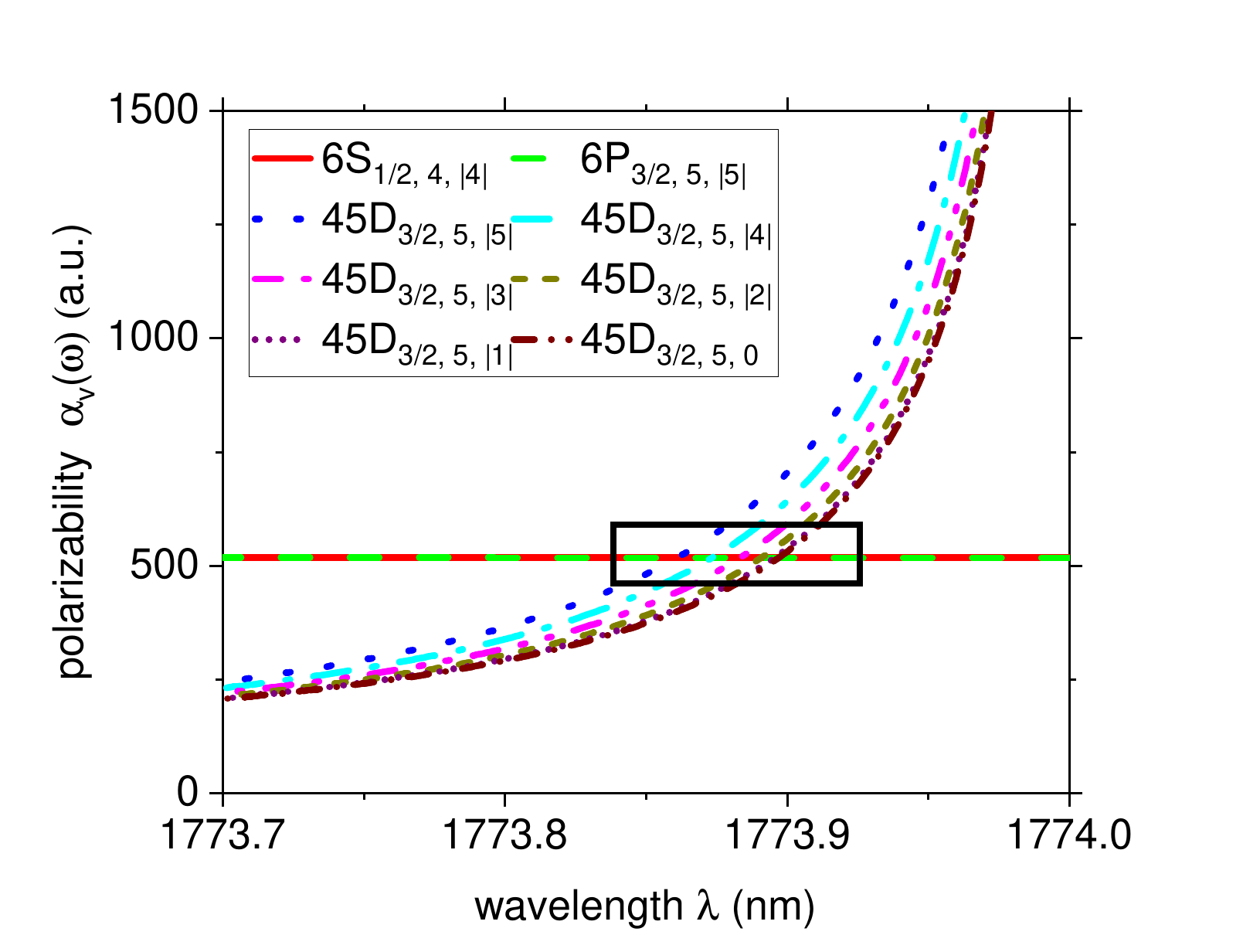}}
\caption{Dynamic polarizability $\alpha_v(\omega)$ of  the $6S_{J=1/2,F=4,M_{F}=4}$ ground state  (red solid line),  the $6P_{J=3/2,F=5,M_{F}=5}$ low-lying excited state (green dashed line),  and  the $45D_{J=3/2,F=5,M_{F}}$ Rydberg state ("diving lines") for $\cos^2\theta=0.62$. The "diving lines" show the polarizability of the different $|M_F|$ states for $F=5$ (see legend). The black boxes draw the readers' attention to how the triple magic wavelengths   vary with $|M_{F}|$.   
} 
\label{Fig_App_3}
\end{figure}

Now we tune  $\theta$, considering the intermediate state $6P_{J=3/2,F=5,M_{F}=5}$ as an example. We find that the ground state and the intermediate state $6P_{J=3/2,F=5,M_{F}=5}$ exhibit the essentially same ac-Stark shift for $\cos^2\theta=0.62$ over the shown wavelength range (the polarizability of the  intermediate state $6P_{J=3/2,F=5,M_{F}=5}$ varies appreciably with $\theta$). Using $\cos^2\theta=0.62$, Fig.~\ref{Fig_App_3} shows the dynamic polarizabilities of the ground state $6S_{J=1/2,F=4,M_{F}=4}$, the intermediate state $6P_{J=3/2,F=5,M_{F}=5}$, and the Rydberg states $45D_{J=3/2,F=5,M_{F}}$ with $|M_F|=0-5$.  As expected,  the polarizability of the $45D_{J=3/2,F=5,M_{F}}$ states depends weakly on $|M_F|$.  The black box in Fig.~\ref{Fig_App_3} highlights the  triple magic wavelengths for simultaneous trapping of the ground state, the intermediate state $6P_{J=3/2,F=5,M_{F}=5}$, and the $45D_{J=3/2,F=5,M_{F}}$ Rydberg states.

\bibliography{manuscript_Bhowmik.bib}

\begin{thebibliography}{75}%
\makeatletter
\providecommand \@ifxundefined [1]{%
 \@ifx{#1\undefined}
}%
\providecommand \@ifnum [1]{%
 \ifnum #1\expandafter \@firstoftwo
 \else \expandafter \@secondoftwo
 \fi
}%
\providecommand \@ifx [1]{%
 \ifx #1\expandafter \@firstoftwo
 \else \expandafter \@secondoftwo
 \fi
}%
\providecommand \natexlab [1]{#1}%
\providecommand \enquote  [1]{``#1''}%
\providecommand \bibnamefont  [1]{#1}%
\providecommand \bibfnamefont [1]{#1}%
\providecommand \citenamefont [1]{#1}%
\providecommand \href@noop [0]{\@secondoftwo}%
\providecommand \href [0]{\begingroup \@sanitize@url \@href}%
\providecommand \@href[1]{\@@startlink{#1}\@@href}%
\providecommand \@@href[1]{\endgroup#1\@@endlink}%
\providecommand \@sanitize@url [0]{\catcode `\\12\catcode `\$12\catcode
  `\&12\catcode `\#12\catcode `\^12\catcode `\_12\catcode `\%12\relax}%
\providecommand \@@startlink[1]{}%
\providecommand \@@endlink[0]{}%
\providecommand \url  [0]{\begingroup\@sanitize@url \@url }%
\providecommand \@url [1]{\endgroup\@href {#1}{\urlprefix }}%
\providecommand \urlprefix  [0]{URL }%
\providecommand \Eprint [0]{\href }%
\providecommand \doibase [0]{https://doi.org/}%
\providecommand \selectlanguage [0]{\@gobble}%
\providecommand \bibinfo  [0]{\@secondoftwo}%
\providecommand \bibfield  [0]{\@secondoftwo}%
\providecommand \translation [1]{[#1]}%
\providecommand \BibitemOpen [0]{}%
\providecommand \bibitemStop [0]{}%
\providecommand \bibitemNoStop [0]{.\EOS\space}%
\providecommand \EOS [0]{\spacefactor3000\relax}%
\providecommand \BibitemShut  [1]{\csname bibitem#1\endcsname}%
\let\auto@bib@innerbib\@empty
\bibitem [{\citenamefont {Robicheaux}\ \emph {et~al.}(2018)\citenamefont
  {Robicheaux}, \citenamefont {Booth},\ and\ \citenamefont
  {Saffman}}]{Robicheaux2018}%
  \BibitemOpen
  \bibfield  {author} {\bibinfo {author} {\bibfnamefont {F.}~\bibnamefont
  {Robicheaux}}, \bibinfo {author} {\bibfnamefont {D.~W.}\ \bibnamefont
  {Booth}},\ and\ \bibinfo {author} {\bibfnamefont {M.}~\bibnamefont
  {Saffman}},\ }\bibfield  {title} {\bibinfo {title} {Theory of long-range
  interactions for rydberg states attached to hyperfine-split cores},\ }\href
  {https://doi.org/10.1103/PhysRevA.97.022508} {\bibfield  {journal} {\bibinfo
  {journal} {Phys. Rev. A}\ }\textbf {\bibinfo {volume} {97}},\ \bibinfo
  {pages} {022508} (\bibinfo {year} {2018})}\BibitemShut {NoStop}%
\bibitem [{\citenamefont {Gallagher}(1994)}]{Gallagher1994}%
  \BibitemOpen
  \bibfield  {author} {\bibinfo {author} {\bibfnamefont {T.~F.}\ \bibnamefont
  {Gallagher}},\ }\href@noop {} {\emph {\bibinfo {title} {Rydberg Atoms}}},\
  Cambridge Monographs on Atomic, Molecular and Chemical Physics\ (\bibinfo
  {publisher} {Cambridge University Press},\ \bibinfo {year}
  {1994})\BibitemShut {NoStop}%
\bibitem [{\citenamefont {Saffman}\ \emph {et~al.}(2010)\citenamefont
  {Saffman}, \citenamefont {Walker},\ and\ \citenamefont
  {M\o{}lmer}}]{Saffman2010}%
  \BibitemOpen
  \bibfield  {author} {\bibinfo {author} {\bibfnamefont {M.}~\bibnamefont
  {Saffman}}, \bibinfo {author} {\bibfnamefont {T.~G.}\ \bibnamefont
  {Walker}},\ and\ \bibinfo {author} {\bibfnamefont {K.}~\bibnamefont
  {M\o{}lmer}},\ }\bibfield  {title} {\bibinfo {title} {Quantum information
  with rydberg atoms},\ }\href {https://doi.org/10.1103/RevModPhys.82.2313}
  {\bibfield  {journal} {\bibinfo  {journal} {Rev. Mod. Phys.}\ }\textbf
  {\bibinfo {volume} {82}},\ \bibinfo {pages} {2313} (\bibinfo {year}
  {2010})}\BibitemShut {NoStop}%
\bibitem [{\citenamefont {Sun}(2023)}]{Yuan2023}%
  \BibitemOpen
  \bibfield  {author} {\bibinfo {author} {\bibfnamefont {Y.}~\bibnamefont
  {Sun}},\ }\bibfield  {title} {\bibinfo {title} {Suppression of high-frequency
  components in off-resonant modulated driving protocols for rydberg-blockade
  gates},\ }\href {https://doi.org/10.1103/PhysRevApplied.20.L061002}
  {\bibfield  {journal} {\bibinfo  {journal} {Phys. Rev. Appl.}\ }\textbf
  {\bibinfo {volume} {20}},\ \bibinfo {pages} {L061002} (\bibinfo {year}
  {2023})}\BibitemShut {NoStop}%
\bibitem [{\citenamefont {Buchemmavari}\ \emph {et~al.}(2024)\citenamefont
  {Buchemmavari}, \citenamefont {Omanakuttan}, \citenamefont {Jau},\ and\
  \citenamefont {Deutsch}}]{Buchemmavari2024}%
  \BibitemOpen
  \bibfield  {author} {\bibinfo {author} {\bibfnamefont {V.}~\bibnamefont
  {Buchemmavari}}, \bibinfo {author} {\bibfnamefont {S.}~\bibnamefont
  {Omanakuttan}}, \bibinfo {author} {\bibfnamefont {Y.-Y.}\ \bibnamefont
  {Jau}},\ and\ \bibinfo {author} {\bibfnamefont {I.}~\bibnamefont {Deutsch}},\
  }\bibfield  {title} {\bibinfo {title} {Entangling quantum logic gates in
  neutral atoms via the microwave-driven spin-flip blockade},\ }\href
  {https://doi.org/10.1103/PhysRevA.109.012615} {\bibfield  {journal} {\bibinfo
   {journal} {Phys. Rev. A}\ }\textbf {\bibinfo {volume} {109}},\ \bibinfo
  {pages} {012615} (\bibinfo {year} {2024})}\BibitemShut {NoStop}%
\bibitem [{\citenamefont {Adams}\ \emph {et~al.}(2019)\citenamefont {Adams},
  \citenamefont {Pritchard},\ and\ \citenamefont {Shaffer}}]{Adams2020}%
  \BibitemOpen
  \bibfield  {author} {\bibinfo {author} {\bibfnamefont {C.~S.}\ \bibnamefont
  {Adams}}, \bibinfo {author} {\bibfnamefont {J.~D.}\ \bibnamefont
  {Pritchard}},\ and\ \bibinfo {author} {\bibfnamefont {J.~P.}\ \bibnamefont
  {Shaffer}},\ }\bibfield  {title} {\bibinfo {title} {Rydberg atom quantum
  technologies},\ }\href {https://doi.org/10.1088/1361-6455/ab52ef} {\bibfield
  {journal} {\bibinfo  {journal} {Journal of Physics B: Atomic, Molecular and
  Optical Physics}\ }\textbf {\bibinfo {volume} {53}},\ \bibinfo {pages}
  {012002} (\bibinfo {year} {2019})}\BibitemShut {NoStop}%
\bibitem [{\citenamefont {Saffman}(2016)}]{Saffman2016}%
  \BibitemOpen
  \bibfield  {author} {\bibinfo {author} {\bibfnamefont {M.}~\bibnamefont
  {Saffman}},\ }\bibfield  {title} {\bibinfo {title} {Quantum computing with
  atomic qubits and rydberg interactions: progress and challenges},\ }\href
  {https://doi.org/10.1088/0953-4075/49/20/202001} {\bibfield  {journal}
  {\bibinfo  {journal} {Journal of Physics B: Atomic, Molecular and Optical
  Physics}\ }\textbf {\bibinfo {volume} {49}},\ \bibinfo {pages} {202001}
  (\bibinfo {year} {2016})}\BibitemShut {NoStop}%
\bibitem [{\citenamefont {Zeytino\ifmmode~\breve{g}\else \u{g}\fi{}lu}\ and\
  \citenamefont {Sugiura}(2024)}]{Zeytinoglu2024}%
  \BibitemOpen
  \bibfield  {author} {\bibinfo {author} {\bibfnamefont {S.}~\bibnamefont
  {Zeytino\ifmmode~\breve{g}\else \u{g}\fi{}lu}}\ and\ \bibinfo {author}
  {\bibfnamefont {S.}~\bibnamefont {Sugiura}},\ }\bibfield  {title} {\bibinfo
  {title} {Error-robust quantum signal processing using rydberg atoms},\ }\href
  {https://doi.org/10.1103/PhysRevResearch.6.013003} {\bibfield  {journal}
  {\bibinfo  {journal} {Phys. Rev. Res.}\ }\textbf {\bibinfo {volume} {6}},\
  \bibinfo {pages} {013003} (\bibinfo {year} {2024})}\BibitemShut {NoStop}%
\bibitem [{\citenamefont {Malz}\ and\ \citenamefont {Cirac}(2023)}]{Malz2023}%
  \BibitemOpen
  \bibfield  {author} {\bibinfo {author} {\bibfnamefont {D.}~\bibnamefont
  {Malz}}\ and\ \bibinfo {author} {\bibfnamefont {J.~I.}\ \bibnamefont
  {Cirac}},\ }\bibfield  {title} {\bibinfo {title} {Few-body analog quantum
  simulation with rydberg-dressed atoms in optical lattices},\ }\href
  {https://doi.org/10.1103/PRXQuantum.4.020301} {\bibfield  {journal} {\bibinfo
   {journal} {PRX Quantum}\ }\textbf {\bibinfo {volume} {4}},\ \bibinfo {pages}
  {020301} (\bibinfo {year} {2023})}\BibitemShut {NoStop}%
\bibitem [{\citenamefont {Lee}\ \emph {et~al.}(2019)\citenamefont {Lee},
  \citenamefont {Kim}, \citenamefont {Jo}, \citenamefont {Song},\ and\
  \citenamefont {Ahn}}]{Lee2019}%
  \BibitemOpen
  \bibfield  {author} {\bibinfo {author} {\bibfnamefont {W.}~\bibnamefont
  {Lee}}, \bibinfo {author} {\bibfnamefont {M.}~\bibnamefont {Kim}}, \bibinfo
  {author} {\bibfnamefont {H.}~\bibnamefont {Jo}}, \bibinfo {author}
  {\bibfnamefont {Y.}~\bibnamefont {Song}},\ and\ \bibinfo {author}
  {\bibfnamefont {J.}~\bibnamefont {Ahn}},\ }\bibfield  {title} {\bibinfo
  {title} {Coherent and dissipative dynamics of entangled few-body systems of
  rydberg atoms},\ }\href {https://doi.org/10.1103/PhysRevA.99.043404}
  {\bibfield  {journal} {\bibinfo  {journal} {Phys. Rev. A}\ }\textbf {\bibinfo
  {volume} {99}},\ \bibinfo {pages} {043404} (\bibinfo {year}
  {2019})}\BibitemShut {NoStop}%
\bibitem [{\citenamefont {Zeiher}\ \emph {et~al.}(2016)\citenamefont {Zeiher},
  \citenamefont {van Bijnen}, \citenamefont {Schau{\ss}}, \citenamefont {Hild},
  \citenamefont {Choi}, \citenamefont {Pohl}, \citenamefont {Bloch},\ and\
  \citenamefont {Gross}}]{Zeiher2016}%
  \BibitemOpen
  \bibfield  {author} {\bibinfo {author} {\bibfnamefont {J.}~\bibnamefont
  {Zeiher}}, \bibinfo {author} {\bibfnamefont {R.}~\bibnamefont {van Bijnen}},
  \bibinfo {author} {\bibfnamefont {P.}~\bibnamefont {Schau{\ss}}}, \bibinfo
  {author} {\bibfnamefont {S.}~\bibnamefont {Hild}}, \bibinfo {author}
  {\bibfnamefont {J.-y.}\ \bibnamefont {Choi}}, \bibinfo {author}
  {\bibfnamefont {T.}~\bibnamefont {Pohl}}, \bibinfo {author} {\bibfnamefont
  {I.}~\bibnamefont {Bloch}},\ and\ \bibinfo {author} {\bibfnamefont
  {C.}~\bibnamefont {Gross}},\ }\bibfield  {title} {\bibinfo {title} {Many-body
  interferometry of a rydberg-dressed spin lattice},\ }\href
  {https://doi.org/10.1038/nphys3835} {\bibfield  {journal} {\bibinfo
  {journal} {Nature Physics}\ }\textbf {\bibinfo {volume} {12}},\ \bibinfo
  {pages} {1095} (\bibinfo {year} {2016})}\BibitemShut {NoStop}%
\bibitem [{\citenamefont {Bernien}\ \emph {et~al.}(2017)\citenamefont
  {Bernien}, \citenamefont {Schwartz}, \citenamefont {Keesling}, \citenamefont
  {Levine}, \citenamefont {Omran}, \citenamefont {Pichler}, \citenamefont
  {Choi}, \citenamefont {Zibrov}, \citenamefont {Endres}, \citenamefont
  {Greiner}, \citenamefont {Vuleti{\'{c}}},\ and\ \citenamefont
  {Lukin}}]{Bernien2017}%
  \BibitemOpen
  \bibfield  {author} {\bibinfo {author} {\bibfnamefont {H.}~\bibnamefont
  {Bernien}}, \bibinfo {author} {\bibfnamefont {S.}~\bibnamefont {Schwartz}},
  \bibinfo {author} {\bibfnamefont {A.}~\bibnamefont {Keesling}}, \bibinfo
  {author} {\bibfnamefont {H.}~\bibnamefont {Levine}}, \bibinfo {author}
  {\bibfnamefont {A.}~\bibnamefont {Omran}}, \bibinfo {author} {\bibfnamefont
  {H.}~\bibnamefont {Pichler}}, \bibinfo {author} {\bibfnamefont
  {S.}~\bibnamefont {Choi}}, \bibinfo {author} {\bibfnamefont {A.~S.}\
  \bibnamefont {Zibrov}}, \bibinfo {author} {\bibfnamefont {M.}~\bibnamefont
  {Endres}}, \bibinfo {author} {\bibfnamefont {M.}~\bibnamefont {Greiner}},
  \bibinfo {author} {\bibfnamefont {V.}~\bibnamefont {Vuleti{\'{c}}}},\ and\
  \bibinfo {author} {\bibfnamefont {M.~D.}\ \bibnamefont {Lukin}},\ }\bibfield
  {title} {\bibinfo {title} {Probing many-body dynamics on a 51-atom quantum
  simulator},\ }\href {https://doi.org/10.1038/nature24622} {\bibfield
  {journal} {\bibinfo  {journal} {Nature}\ }\textbf {\bibinfo {volume} {551}},\
  \bibinfo {pages} {579} (\bibinfo {year} {2017})}\BibitemShut {NoStop}%
\bibitem [{\citenamefont {Bharti}\ \emph {et~al.}(2023)\citenamefont {Bharti},
  \citenamefont {Sugawa}, \citenamefont {Mizoguchi}, \citenamefont {Kunimi},
  \citenamefont {Zhang}, \citenamefont {de~L\'es\'eleuc}, \citenamefont
  {Tomita}, \citenamefont {Franz}, \citenamefont {Weidem\"uller},\ and\
  \citenamefont {Ohmori}}]{Bharti2023}%
  \BibitemOpen
  \bibfield  {author} {\bibinfo {author} {\bibfnamefont {V.}~\bibnamefont
  {Bharti}}, \bibinfo {author} {\bibfnamefont {S.}~\bibnamefont {Sugawa}},
  \bibinfo {author} {\bibfnamefont {M.}~\bibnamefont {Mizoguchi}}, \bibinfo
  {author} {\bibfnamefont {M.}~\bibnamefont {Kunimi}}, \bibinfo {author}
  {\bibfnamefont {Y.}~\bibnamefont {Zhang}}, \bibinfo {author} {\bibfnamefont
  {S.}~\bibnamefont {de~L\'es\'eleuc}}, \bibinfo {author} {\bibfnamefont
  {T.}~\bibnamefont {Tomita}}, \bibinfo {author} {\bibfnamefont
  {T.}~\bibnamefont {Franz}}, \bibinfo {author} {\bibfnamefont
  {M.}~\bibnamefont {Weidem\"uller}},\ and\ \bibinfo {author} {\bibfnamefont
  {K.}~\bibnamefont {Ohmori}},\ }\bibfield  {title} {\bibinfo {title}
  {Picosecond-scale ultrafast many-body dynamics in an ultracold
  rydberg-excited atomic mott insulator},\ }\href
  {https://doi.org/10.1103/PhysRevLett.131.123201} {\bibfield  {journal}
  {\bibinfo  {journal} {Phys. Rev. Lett.}\ }\textbf {\bibinfo {volume} {131}},\
  \bibinfo {pages} {123201} (\bibinfo {year} {2023})}\BibitemShut {NoStop}%
\bibitem [{\citenamefont {Firstenberg}\ \emph {et~al.}(2016)\citenamefont
  {Firstenberg}, \citenamefont {Adams},\ and\ \citenamefont
  {Hofferberth}}]{Firstenberg2016}%
  \BibitemOpen
  \bibfield  {author} {\bibinfo {author} {\bibfnamefont {O.}~\bibnamefont
  {Firstenberg}}, \bibinfo {author} {\bibfnamefont {C.~S.}\ \bibnamefont
  {Adams}},\ and\ \bibinfo {author} {\bibfnamefont {S.}~\bibnamefont
  {Hofferberth}},\ }\bibfield  {title} {\bibinfo {title} {Nonlinear quantum
  optics mediated by rydberg interactions},\ }\href
  {https://doi.org/10.1088/0953-4075/49/15/152003} {\bibfield  {journal}
  {\bibinfo  {journal} {Journal of Physics B: Atomic, Molecular and Optical
  Physics}\ }\textbf {\bibinfo {volume} {49}},\ \bibinfo {pages} {152003}
  (\bibinfo {year} {2016})}\BibitemShut {NoStop}%
\bibitem [{\citenamefont {Huo}\ \emph {et~al.}(2022)\citenamefont {Huo},
  \citenamefont {Chen}, \citenamefont {Qian},\ and\ \citenamefont
  {Zhang}}]{Huo2022}%
  \BibitemOpen
  \bibfield  {author} {\bibinfo {author} {\bibfnamefont {X.}~\bibnamefont
  {Huo}}, \bibinfo {author} {\bibfnamefont {J.~F.}\ \bibnamefont {Chen}},
  \bibinfo {author} {\bibfnamefont {J.}~\bibnamefont {Qian}},\ and\ \bibinfo
  {author} {\bibfnamefont {W.}~\bibnamefont {Zhang}},\ }\bibfield  {title}
  {\bibinfo {title} {Interaction-enhanced transmission imaging with rydberg
  atoms},\ }\href {https://doi.org/10.1103/PhysRevA.105.012817} {\bibfield
  {journal} {\bibinfo  {journal} {Phys. Rev. A}\ }\textbf {\bibinfo {volume}
  {105}},\ \bibinfo {pages} {012817} (\bibinfo {year} {2022})}\BibitemShut
  {NoStop}%
\bibitem [{\citenamefont {Beterov}\ \emph {et~al.}(2009)\citenamefont
  {Beterov}, \citenamefont {Ryabtsev}, \citenamefont {Tretyakov},\ and\
  \citenamefont {Entin}}]{Beterov2009}%
  \BibitemOpen
  \bibfield  {author} {\bibinfo {author} {\bibfnamefont {I.~I.}\ \bibnamefont
  {Beterov}}, \bibinfo {author} {\bibfnamefont {I.~I.}\ \bibnamefont
  {Ryabtsev}}, \bibinfo {author} {\bibfnamefont {D.~B.}\ \bibnamefont
  {Tretyakov}},\ and\ \bibinfo {author} {\bibfnamefont {V.~M.}\ \bibnamefont
  {Entin}},\ }\bibfield  {title} {\bibinfo {title} {Quasiclassical calculations
  of blackbody-radiation-induced depopulation rates and effective lifetimes of
  rydberg $ns$, $np$, and $nd$ alkali-metal atoms with $n\ensuremath{\le}80$},\
  }\href {https://doi.org/10.1103/PhysRevA.79.052504} {\bibfield  {journal}
  {\bibinfo  {journal} {Phys. Rev. A}\ }\textbf {\bibinfo {volume} {79}},\
  \bibinfo {pages} {052504} (\bibinfo {year} {2009})}\BibitemShut {NoStop}%
\bibitem [{\citenamefont {Lukin}\ \emph {et~al.}(2001)\citenamefont {Lukin},
  \citenamefont {Fleischhauer}, \citenamefont {Cote}, \citenamefont {Duan},
  \citenamefont {Jaksch}, \citenamefont {Cirac},\ and\ \citenamefont
  {Zoller}}]{Lukin2001}%
  \BibitemOpen
  \bibfield  {author} {\bibinfo {author} {\bibfnamefont {M.~D.}\ \bibnamefont
  {Lukin}}, \bibinfo {author} {\bibfnamefont {M.}~\bibnamefont {Fleischhauer}},
  \bibinfo {author} {\bibfnamefont {R.}~\bibnamefont {Cote}}, \bibinfo {author}
  {\bibfnamefont {L.~M.}\ \bibnamefont {Duan}}, \bibinfo {author}
  {\bibfnamefont {D.}~\bibnamefont {Jaksch}}, \bibinfo {author} {\bibfnamefont
  {J.~I.}\ \bibnamefont {Cirac}},\ and\ \bibinfo {author} {\bibfnamefont
  {P.}~\bibnamefont {Zoller}},\ }\bibfield  {title} {\bibinfo {title} {Dipole
  blockade and quantum information processing in mesoscopic atomic ensembles},\
  }\href {https://doi.org/10.1103/PhysRevLett.87.037901} {\bibfield  {journal}
  {\bibinfo  {journal} {Phys. Rev. Lett.}\ }\textbf {\bibinfo {volume} {87}},\
  \bibinfo {pages} {037901} (\bibinfo {year} {2001})}\BibitemShut {NoStop}%
\bibitem [{\citenamefont {Xia}\ \emph {et~al.}(2015)\citenamefont {Xia},
  \citenamefont {Lichtman}, \citenamefont {Maller}, \citenamefont {Carr},
  \citenamefont {Piotrowicz}, \citenamefont {Isenhower},\ and\ \citenamefont
  {Saffman}}]{Xia2015}%
  \BibitemOpen
  \bibfield  {author} {\bibinfo {author} {\bibfnamefont {T.}~\bibnamefont
  {Xia}}, \bibinfo {author} {\bibfnamefont {M.}~\bibnamefont {Lichtman}},
  \bibinfo {author} {\bibfnamefont {K.}~\bibnamefont {Maller}}, \bibinfo
  {author} {\bibfnamefont {A.~W.}\ \bibnamefont {Carr}}, \bibinfo {author}
  {\bibfnamefont {M.~J.}\ \bibnamefont {Piotrowicz}}, \bibinfo {author}
  {\bibfnamefont {L.}~\bibnamefont {Isenhower}},\ and\ \bibinfo {author}
  {\bibfnamefont {M.}~\bibnamefont {Saffman}},\ }\bibfield  {title} {\bibinfo
  {title} {Randomized benchmarking of single-qubit gates in a 2d array of
  neutral-atom qubits},\ }\href
  {https://doi.org/10.1103/PhysRevLett.114.100503} {\bibfield  {journal}
  {\bibinfo  {journal} {Phys. Rev. Lett.}\ }\textbf {\bibinfo {volume} {114}},\
  \bibinfo {pages} {100503} (\bibinfo {year} {2015})}\BibitemShut {NoStop}%
\bibitem [{\citenamefont {Endres}\ \emph {et~al.}(2016)\citenamefont {Endres},
  \citenamefont {Bernien}, \citenamefont {Keesling}, \citenamefont {Levine},
  \citenamefont {Anschuetz}, \citenamefont {Krajenbrink}, \citenamefont
  {Senko}, \citenamefont {Vuletic}, \citenamefont {Greiner},\ and\
  \citenamefont {Lukin}}]{Endres2016}%
  \BibitemOpen
  \bibfield  {author} {\bibinfo {author} {\bibfnamefont {M.}~\bibnamefont
  {Endres}}, \bibinfo {author} {\bibfnamefont {H.}~\bibnamefont {Bernien}},
  \bibinfo {author} {\bibfnamefont {A.}~\bibnamefont {Keesling}}, \bibinfo
  {author} {\bibfnamefont {H.}~\bibnamefont {Levine}}, \bibinfo {author}
  {\bibfnamefont {E.~R.}\ \bibnamefont {Anschuetz}}, \bibinfo {author}
  {\bibfnamefont {A.}~\bibnamefont {Krajenbrink}}, \bibinfo {author}
  {\bibfnamefont {C.}~\bibnamefont {Senko}}, \bibinfo {author} {\bibfnamefont
  {V.}~\bibnamefont {Vuletic}}, \bibinfo {author} {\bibfnamefont
  {M.}~\bibnamefont {Greiner}},\ and\ \bibinfo {author} {\bibfnamefont {M.~D.}\
  \bibnamefont {Lukin}},\ }\bibfield  {title} {\bibinfo {title} {Atom-by-atom
  assembly of defect-free one-dimensional cold atom arrays},\ }\href
  {https://doi.org/10.1126/science.aah3752} {\bibfield  {journal} {\bibinfo
  {journal} {Science}\ }\textbf {\bibinfo {volume} {354}},\ \bibinfo {pages}
  {1024} (\bibinfo {year} {2016})},\ \Eprint
  {https://arxiv.org/abs/https://www.science.org/doi/pdf/10.1126/science.aah3752}
  {https://www.science.org/doi/pdf/10.1126/science.aah3752} \BibitemShut
  {NoStop}%
\bibitem [{\citenamefont {Laidig}\ and\ \citenamefont
  {Bader}(1990)}]{Laidig1990}%
  \BibitemOpen
  \bibfield  {author} {\bibinfo {author} {\bibfnamefont {K.~E.}\ \bibnamefont
  {Laidig}}\ and\ \bibinfo {author} {\bibfnamefont {R.~F.~W.}\ \bibnamefont
  {Bader}},\ }\bibfield  {title} {\bibinfo {title} {{Properties of atoms in
  molecules: Atomic polarizabilities}},\ }\href
  {https://doi.org/10.1063/1.459444} {\bibfield  {journal} {\bibinfo  {journal}
  {The Journal of Chemical Physics}\ }\textbf {\bibinfo {volume} {93}},\
  \bibinfo {pages} {7213} (\bibinfo {year} {1990})},\ \Eprint
  {https://arxiv.org/abs/https://pubs.aip.org/aip/jcp/article-pdf/93/10/7213/18989285/7213\_1\_online.pdf}
  {https://pubs.aip.org/aip/jcp/article-pdf/93/10/7213/18989285/7213\_1\_online.pdf}
  \BibitemShut {NoStop}%
\bibitem [{\citenamefont {Mitroy}\ \emph {et~al.}(2010)\citenamefont {Mitroy},
  \citenamefont {Safronova},\ and\ \citenamefont {Clark}}]{Mitroy2010}%
  \BibitemOpen
  \bibfield  {author} {\bibinfo {author} {\bibfnamefont {J.}~\bibnamefont
  {Mitroy}}, \bibinfo {author} {\bibfnamefont {M.~S.}\ \bibnamefont
  {Safronova}},\ and\ \bibinfo {author} {\bibfnamefont {C.~W.}\ \bibnamefont
  {Clark}},\ }\bibfield  {title} {\bibinfo {title} {Theory and applications of
  atomic and ionic polarizabilities},\ }\href
  {https://doi.org/10.1088/0953-4075/43/20/202001} {\bibfield  {journal}
  {\bibinfo  {journal} {Journal of Physics B: Atomic, Molecular and Optical
  Physics}\ }\textbf {\bibinfo {volume} {43}},\ \bibinfo {pages} {202001}
  (\bibinfo {year} {2010})}\BibitemShut {NoStop}%
\bibitem [{\citenamefont {Grimm}\ \emph {et~al.}(2000)\citenamefont {Grimm},
  \citenamefont {Weidemüller},\ and\ \citenamefont {Ovchinnikov}}]{GRIMM2000}%
  \BibitemOpen
  \bibfield  {author} {\bibinfo {author} {\bibfnamefont {R.}~\bibnamefont
  {Grimm}}, \bibinfo {author} {\bibfnamefont {M.}~\bibnamefont
  {Weidemüller}},\ and\ \bibinfo {author} {\bibfnamefont {Y.~B.}\ \bibnamefont
  {Ovchinnikov}},\ }\bibfield  {title} {\bibinfo {title} {Optical dipole traps
  for neutral atoms}\ }(\bibinfo  {publisher} {Academic Press},\ \bibinfo
  {year} {2000})\ pp.\ \bibinfo {pages} {95--170}\BibitemShut {NoStop}%
\bibitem [{\citenamefont {Younge}\ \emph {et~al.}(2010)\citenamefont {Younge},
  \citenamefont {Anderson},\ and\ \citenamefont {Raithel}}]{Younge2010}%
  \BibitemOpen
  \bibfield  {author} {\bibinfo {author} {\bibfnamefont {K.~C.}\ \bibnamefont
  {Younge}}, \bibinfo {author} {\bibfnamefont {S.~E.}\ \bibnamefont
  {Anderson}},\ and\ \bibinfo {author} {\bibfnamefont {G.}~\bibnamefont
  {Raithel}},\ }\bibfield  {title} {\bibinfo {title} {Adiabatic potentials for
  rydberg atoms in a ponderomotive optical lattice},\ }\href
  {https://doi.org/10.1088/1367-2630/12/2/023031} {\bibfield  {journal}
  {\bibinfo  {journal} {New Journal of Physics}\ }\textbf {\bibinfo {volume}
  {12}},\ \bibinfo {pages} {023031} (\bibinfo {year} {2010})}\BibitemShut
  {NoStop}%
\bibitem [{\citenamefont {de~Keijzer}\ \emph {et~al.}(2023)\citenamefont
  {de~Keijzer}, \citenamefont {Tse},\ and\ \citenamefont
  {Kokkelmans}}]{Keijzer2023}%
  \BibitemOpen
  \bibfield  {author} {\bibinfo {author} {\bibfnamefont {R.~J. P.~T.}\
  \bibnamefont {de~Keijzer}}, \bibinfo {author} {\bibfnamefont
  {O.}~\bibnamefont {Tse}},\ and\ \bibinfo {author} {\bibfnamefont {S.~J. J.
  M.~F.}\ \bibnamefont {Kokkelmans}},\ }\bibfield  {title} {\bibinfo {title}
  {Recapture probability for antitrapped rydberg states in optical tweezers},\
  }\href {https://doi.org/10.1103/PhysRevA.108.023122} {\bibfield  {journal}
  {\bibinfo  {journal} {Phys. Rev. A}\ }\textbf {\bibinfo {volume} {108}},\
  \bibinfo {pages} {023122} (\bibinfo {year} {2023})}\BibitemShut {NoStop}%
\bibitem [{\citenamefont {Anderson}\ \emph {et~al.}(2011)\citenamefont
  {Anderson}, \citenamefont {Younge},\ and\ \citenamefont
  {Raithel}}]{Anderson2011}%
  \BibitemOpen
  \bibfield  {author} {\bibinfo {author} {\bibfnamefont {S.~E.}\ \bibnamefont
  {Anderson}}, \bibinfo {author} {\bibfnamefont {K.~C.}\ \bibnamefont
  {Younge}},\ and\ \bibinfo {author} {\bibfnamefont {G.}~\bibnamefont
  {Raithel}},\ }\bibfield  {title} {\bibinfo {title} {Trapping rydberg atoms in
  an optical lattice},\ }\href {https://doi.org/10.1103/PhysRevLett.107.263001}
  {\bibfield  {journal} {\bibinfo  {journal} {Phys. Rev. Lett.}\ }\textbf
  {\bibinfo {volume} {107}},\ \bibinfo {pages} {263001} (\bibinfo {year}
  {2011})}\BibitemShut {NoStop}%
\bibitem [{\citenamefont {Li}\ \emph {et~al.}(2013)\citenamefont {Li},
  \citenamefont {Dudin},\ and\ \citenamefont {Kuzmich}}]{Li2013}%
  \BibitemOpen
  \bibfield  {author} {\bibinfo {author} {\bibfnamefont {L.}~\bibnamefont
  {Li}}, \bibinfo {author} {\bibfnamefont {Y.~O.}\ \bibnamefont {Dudin}},\ and\
  \bibinfo {author} {\bibfnamefont {A.}~\bibnamefont {Kuzmich}},\ }\bibfield
  {title} {\bibinfo {title} {Entanglement between light and an optical atomic
  excitation},\ }\href {https://doi.org/10.1038/nature12227} {\bibfield
  {journal} {\bibinfo  {journal} {Nature}\ }\textbf {\bibinfo {volume} {498}},\
  \bibinfo {pages} {466} (\bibinfo {year} {2013})}\BibitemShut {NoStop}%
\bibitem [{\citenamefont {Barredo}\ \emph {et~al.}(2020)\citenamefont
  {Barredo}, \citenamefont {Lienhard}, \citenamefont {Scholl}, \citenamefont
  {de~L\'es\'eleuc}, \citenamefont {Boulier}, \citenamefont {Browaeys},\ and\
  \citenamefont {Lahaye}}]{Barredo2020}%
  \BibitemOpen
  \bibfield  {author} {\bibinfo {author} {\bibfnamefont {D.}~\bibnamefont
  {Barredo}}, \bibinfo {author} {\bibfnamefont {V.}~\bibnamefont {Lienhard}},
  \bibinfo {author} {\bibfnamefont {P.}~\bibnamefont {Scholl}}, \bibinfo
  {author} {\bibfnamefont {S.}~\bibnamefont {de~L\'es\'eleuc}}, \bibinfo
  {author} {\bibfnamefont {T.}~\bibnamefont {Boulier}}, \bibinfo {author}
  {\bibfnamefont {A.}~\bibnamefont {Browaeys}},\ and\ \bibinfo {author}
  {\bibfnamefont {T.}~\bibnamefont {Lahaye}},\ }\bibfield  {title} {\bibinfo
  {title} {Three-dimensional trapping of individual rydberg atoms in
  ponderomotive bottle beam traps},\ }\href
  {https://doi.org/10.1103/PhysRevLett.124.023201} {\bibfield  {journal}
  {\bibinfo  {journal} {Phys. Rev. Lett.}\ }\textbf {\bibinfo {volume} {124}},\
  \bibinfo {pages} {023201} (\bibinfo {year} {2020})}\BibitemShut {NoStop}%
\bibitem [{\citenamefont {Xu}\ \emph {et~al.}(2010)\citenamefont {Xu},
  \citenamefont {He}, \citenamefont {Wang},\ and\ \citenamefont
  {Zhan}}]{Xu2010}%
  \BibitemOpen
  \bibfield  {author} {\bibinfo {author} {\bibfnamefont {P.}~\bibnamefont
  {Xu}}, \bibinfo {author} {\bibfnamefont {X.}~\bibnamefont {He}}, \bibinfo
  {author} {\bibfnamefont {J.}~\bibnamefont {Wang}},\ and\ \bibinfo {author}
  {\bibfnamefont {M.}~\bibnamefont {Zhan}},\ }\bibfield  {title} {\bibinfo
  {title} {Trapping a single atom in a blue detuned optical bottle beam trap},\
  }\href {https://doi.org/10.1364/OL.35.002164} {\bibfield  {journal} {\bibinfo
   {journal} {Opt. Lett.}\ }\textbf {\bibinfo {volume} {35}},\ \bibinfo {pages}
  {2164} (\bibinfo {year} {2010})}\BibitemShut {NoStop}%
\bibitem [{\citenamefont {Wilson}\ \emph {et~al.}(2022)\citenamefont {Wilson},
  \citenamefont {Saskin}, \citenamefont {Meng}, \citenamefont {Ma},
  \citenamefont {Dilip}, \citenamefont {Burgers},\ and\ \citenamefont
  {Thompson}}]{Wilson2022}%
  \BibitemOpen
  \bibfield  {author} {\bibinfo {author} {\bibfnamefont {J.~T.}\ \bibnamefont
  {Wilson}}, \bibinfo {author} {\bibfnamefont {S.}~\bibnamefont {Saskin}},
  \bibinfo {author} {\bibfnamefont {Y.}~\bibnamefont {Meng}}, \bibinfo {author}
  {\bibfnamefont {S.}~\bibnamefont {Ma}}, \bibinfo {author} {\bibfnamefont
  {R.}~\bibnamefont {Dilip}}, \bibinfo {author} {\bibfnamefont {A.~P.}\
  \bibnamefont {Burgers}},\ and\ \bibinfo {author} {\bibfnamefont {J.~D.}\
  \bibnamefont {Thompson}},\ }\bibfield  {title} {\bibinfo {title} {Trapping
  alkaline earth rydberg atoms optical tweezer arrays},\ }\href
  {https://doi.org/10.1103/PhysRevLett.128.033201} {\bibfield  {journal}
  {\bibinfo  {journal} {Phys. Rev. Lett.}\ }\textbf {\bibinfo {volume} {128}},\
  \bibinfo {pages} {033201} (\bibinfo {year} {2022})}\BibitemShut {NoStop}%
\bibitem [{\citenamefont {Katori}\ \emph {et~al.}(1999)\citenamefont {Katori},
  \citenamefont {Ido},\ and\ \citenamefont {Kuwata-Gonokami}}]{Katori1999}%
  \BibitemOpen
  \bibfield  {author} {\bibinfo {author} {\bibfnamefont {H.}~\bibnamefont
  {Katori}}, \bibinfo {author} {\bibfnamefont {T.}~\bibnamefont {Ido}},\ and\
  \bibinfo {author} {\bibfnamefont {M.}~\bibnamefont {Kuwata-Gonokami}},\
  }\bibfield  {title} {\bibinfo {title} {Optimal design of dipole potentials
  for efficient loading of sr atoms},\ }\href
  {https://doi.org/10.1143/JPSJ.68.2479} {\bibfield  {journal} {\bibinfo
  {journal} {Journal of the Physical Society of Japan}\ }\textbf {\bibinfo
  {volume} {68}},\ \bibinfo {pages} {2479} (\bibinfo {year} {1999})},\ \Eprint
  {https://arxiv.org/abs/https://doi.org/10.1143/JPSJ.68.2479}
  {https://doi.org/10.1143/JPSJ.68.2479} \BibitemShut {NoStop}%
\bibitem [{\citenamefont {Ye}\ \emph {et~al.}(2008)\citenamefont {Ye},
  \citenamefont {Kimble},\ and\ \citenamefont {Katori}}]{Ye2008}%
  \BibitemOpen
  \bibfield  {author} {\bibinfo {author} {\bibfnamefont {J.}~\bibnamefont
  {Ye}}, \bibinfo {author} {\bibfnamefont {H.~J.}\ \bibnamefont {Kimble}},\
  and\ \bibinfo {author} {\bibfnamefont {H.}~\bibnamefont {Katori}},\
  }\bibfield  {title} {\bibinfo {title} {Quantum state engineering and
  precision metrology using state-insensitive light traps},\ }\href
  {https://doi.org/10.1126/science.1148259} {\bibfield  {journal} {\bibinfo
  {journal} {Science}\ }\textbf {\bibinfo {volume} {320}},\ \bibinfo {pages}
  {1734} (\bibinfo {year} {2008})},\ \Eprint
  {https://arxiv.org/abs/https://www.science.org/doi/pdf/10.1126/science.1148259}
  {https://www.science.org/doi/pdf/10.1126/science.1148259} \BibitemShut
  {NoStop}%
\bibitem [{\citenamefont {Saffman}\ and\ \citenamefont
  {Walker}(2005)}]{Saffman2005}%
  \BibitemOpen
  \bibfield  {author} {\bibinfo {author} {\bibfnamefont {M.}~\bibnamefont
  {Saffman}}\ and\ \bibinfo {author} {\bibfnamefont {T.~G.}\ \bibnamefont
  {Walker}},\ }\bibfield  {title} {\bibinfo {title} {Analysis of a quantum
  logic device based on dipole-dipole interactions of optically trapped rydberg
  atoms},\ }\href {https://doi.org/10.1103/PhysRevA.72.022347} {\bibfield
  {journal} {\bibinfo  {journal} {Phys. Rev. A}\ }\textbf {\bibinfo {volume}
  {72}},\ \bibinfo {pages} {022347} (\bibinfo {year} {2005})}\BibitemShut
  {NoStop}%
\bibitem [{\citenamefont {Goldschmidt}\ \emph {et~al.}(2015)\citenamefont
  {Goldschmidt}, \citenamefont {Norris}, \citenamefont {Koller}, \citenamefont
  {Wyllie}, \citenamefont {Brown}, \citenamefont {Porto}, \citenamefont
  {Safronova},\ and\ \citenamefont {Safronova}}]{Goldschmidt2015}%
  \BibitemOpen
  \bibfield  {author} {\bibinfo {author} {\bibfnamefont {E.~A.}\ \bibnamefont
  {Goldschmidt}}, \bibinfo {author} {\bibfnamefont {D.~G.}\ \bibnamefont
  {Norris}}, \bibinfo {author} {\bibfnamefont {S.~B.}\ \bibnamefont {Koller}},
  \bibinfo {author} {\bibfnamefont {R.}~\bibnamefont {Wyllie}}, \bibinfo
  {author} {\bibfnamefont {R.~C.}\ \bibnamefont {Brown}}, \bibinfo {author}
  {\bibfnamefont {J.~V.}\ \bibnamefont {Porto}}, \bibinfo {author}
  {\bibfnamefont {U.~I.}\ \bibnamefont {Safronova}},\ and\ \bibinfo {author}
  {\bibfnamefont {M.~S.}\ \bibnamefont {Safronova}},\ }\bibfield  {title}
  {\bibinfo {title} {Magic wavelengths for the $5s\text{--}18s$ transition in
  rubidium},\ }\href {https://doi.org/10.1103/PhysRevA.91.032518} {\bibfield
  {journal} {\bibinfo  {journal} {Phys. Rev. A}\ }\textbf {\bibinfo {volume}
  {91}},\ \bibinfo {pages} {032518} (\bibinfo {year} {2015})}\BibitemShut
  {NoStop}%
\bibitem [{\citenamefont {Bai}\ \emph {et~al.}(2020{\natexlab{a}})\citenamefont
  {Bai}, \citenamefont {Bai}, \citenamefont {Han}, \citenamefont {Jiao},
  \citenamefont {Zhao},\ and\ \citenamefont {Jia}}]{Bai2020}%
  \BibitemOpen
  \bibfield  {author} {\bibinfo {author} {\bibfnamefont {J.}~\bibnamefont
  {Bai}}, \bibinfo {author} {\bibfnamefont {S.}~\bibnamefont {Bai}}, \bibinfo
  {author} {\bibfnamefont {X.}~\bibnamefont {Han}}, \bibinfo {author}
  {\bibfnamefont {Y.}~\bibnamefont {Jiao}}, \bibinfo {author} {\bibfnamefont
  {J.}~\bibnamefont {Zhao}},\ and\ \bibinfo {author} {\bibfnamefont
  {S.}~\bibnamefont {Jia}},\ }\bibfield  {title} {\bibinfo {title} {Precise
  measurements of polarizabilities of cesium ns rydberg states in an ultra-cold
  atomic ensemble},\ }\href {https://doi.org/10.1088/1367-2630/abaf30}
  {\bibfield  {journal} {\bibinfo  {journal} {New Journal of Physics}\ }\textbf
  {\bibinfo {volume} {22}},\ \bibinfo {pages} {093032} (\bibinfo {year}
  {2020}{\natexlab{a}})}\BibitemShut {NoStop}%
\bibitem [{\citenamefont {Yerokhin}\ \emph {et~al.}(2016)\citenamefont
  {Yerokhin}, \citenamefont {Buhmann}, \citenamefont {Fritzsche},\ and\
  \citenamefont {Surzhykov}}]{Yerokhin2016}%
  \BibitemOpen
  \bibfield  {author} {\bibinfo {author} {\bibfnamefont {V.~A.}\ \bibnamefont
  {Yerokhin}}, \bibinfo {author} {\bibfnamefont {S.~Y.}\ \bibnamefont
  {Buhmann}}, \bibinfo {author} {\bibfnamefont {S.}~\bibnamefont {Fritzsche}},\
  and\ \bibinfo {author} {\bibfnamefont {A.}~\bibnamefont {Surzhykov}},\
  }\bibfield  {title} {\bibinfo {title} {Electric dipole polarizabilities of
  rydberg states of alkali-metal atoms},\ }\href
  {https://doi.org/10.1103/PhysRevA.94.032503} {\bibfield  {journal} {\bibinfo
  {journal} {Phys. Rev. A}\ }\textbf {\bibinfo {volume} {94}},\ \bibinfo
  {pages} {032503} (\bibinfo {year} {2016})}\BibitemShut {NoStop}%
\bibitem [{\citenamefont {Bai}\ \emph {et~al.}(2020{\natexlab{b}})\citenamefont
  {Bai}, \citenamefont {Liu}, \citenamefont {He},\ and\ \citenamefont
  {Wang}}]{Bai2020_1}%
  \BibitemOpen
  \bibfield  {author} {\bibinfo {author} {\bibfnamefont {J.}~\bibnamefont
  {Bai}}, \bibinfo {author} {\bibfnamefont {S.}~\bibnamefont {Liu}}, \bibinfo
  {author} {\bibfnamefont {J.}~\bibnamefont {He}},\ and\ \bibinfo {author}
  {\bibfnamefont {J.}~\bibnamefont {Wang}},\ }\bibfield  {title} {\bibinfo
  {title} {Towards implementation of a magic optical-dipole trap for confining
  ground-state and rydberg-state cesium cold atoms},\ }\href
  {https://doi.org/10.1088/1361-6455/ab91de} {\bibfield  {journal} {\bibinfo
  {journal} {Journal of Physics B: Atomic, Molecular and Optical Physics}\
  }\textbf {\bibinfo {volume} {53}},\ \bibinfo {pages} {155302} (\bibinfo
  {year} {2020}{\natexlab{b}})}\BibitemShut {NoStop}%
\bibitem [{\citenamefont {Topcu}\ and\ \citenamefont
  {Derevianko}(2016)}]{Topcu_2016}%
  \BibitemOpen
  \bibfield  {author} {\bibinfo {author} {\bibfnamefont {T.}~\bibnamefont
  {Topcu}}\ and\ \bibinfo {author} {\bibfnamefont {A.}~\bibnamefont
  {Derevianko}},\ }\bibfield  {title} {\bibinfo {title} {Possibility of triple
  magic trapping of clock and rydberg states of divalent atoms in optical
  lattices},\ }\href {https://doi.org/10.1088/0953-4075/49/14/144004}
  {\bibfield  {journal} {\bibinfo  {journal} {Journal of Physics B: Atomic,
  Molecular and Optical Physics}\ }\textbf {\bibinfo {volume} {49}},\ \bibinfo
  {pages} {144004} (\bibinfo {year} {2016})}\BibitemShut {NoStop}%
\bibitem [{\citenamefont {Zhang}\ \emph {et~al.}(2024)\citenamefont {Zhang},
  \citenamefont {Jiang}, \citenamefont {Dong},\ and\ \citenamefont
  {Tang}}]{Zhang_2024}%
  \BibitemOpen
  \bibfield  {author} {\bibinfo {author} {\bibfnamefont {R.-K.}\ \bibnamefont
  {Zhang}}, \bibinfo {author} {\bibfnamefont {J.}~\bibnamefont {Jiang}},
  \bibinfo {author} {\bibfnamefont {C.-Z.}\ \bibnamefont {Dong}},\ and\
  \bibinfo {author} {\bibfnamefont {Y.-B.}\ \bibnamefont {Tang}},\ }\href@noop
  {} {\bibinfo {title} {Dynamic polarizabilities and triple magic trapping
  conditions for $5s^2~^1s_0\rightarrow 5s5p~^3p_{0,2}$ transitions of cd
  atoms}} (\bibinfo {year} {2024}),\ \Eprint {https://arxiv.org/abs/2403.05898}
  {arXiv:2403.05898 [physics.atom-ph]} \BibitemShut {NoStop}%
\bibitem [{\citenamefont {Johnson}(2007)}]{Johnson2007}%
  \BibitemOpen
  \bibfield  {author} {\bibinfo {author} {\bibfnamefont {W.~R.}\ \bibnamefont
  {Johnson}},\ }\href
  {https://doi.org/https://doi.org/10.1007/978-3-540-68013-0} {\emph {\bibinfo
  {title} {{Atomic Structure Theory}}}}\ (\bibinfo  {publisher} {Springer
  Berlin, Heidelberg},\ \bibinfo {year} {2007})\BibitemShut {NoStop}%
\bibitem [{\citenamefont {Gallagher}\ and\ \citenamefont
  {Cooke}(1979)}]{Gallagher1979}%
  \BibitemOpen
  \bibfield  {author} {\bibinfo {author} {\bibfnamefont {T.~F.}\ \bibnamefont
  {Gallagher}}\ and\ \bibinfo {author} {\bibfnamefont {W.~E.}\ \bibnamefont
  {Cooke}},\ }\bibfield  {title} {\bibinfo {title} {Interactions of blackbody
  radiation with atoms},\ }\href {https://doi.org/10.1103/PhysRevLett.42.835}
  {\bibfield  {journal} {\bibinfo  {journal} {Phys. Rev. Lett.}\ }\textbf
  {\bibinfo {volume} {42}},\ \bibinfo {pages} {835} (\bibinfo {year}
  {1979})}\BibitemShut {NoStop}%
\bibitem [{\citenamefont {Song}\ \emph {et~al.}(2022)\citenamefont {Song},
  \citenamefont {Bai}, \citenamefont {Jiao}, \citenamefont {Zhao},\ and\
  \citenamefont {Jia}}]{Song2022}%
  \BibitemOpen
  \bibfield  {author} {\bibinfo {author} {\bibfnamefont {R.}~\bibnamefont
  {Song}}, \bibinfo {author} {\bibfnamefont {J.}~\bibnamefont {Bai}}, \bibinfo
  {author} {\bibfnamefont {Y.}~\bibnamefont {Jiao}}, \bibinfo {author}
  {\bibfnamefont {J.}~\bibnamefont {Zhao}},\ and\ \bibinfo {author}
  {\bibfnamefont {S.}~\bibnamefont {Jia}},\ }\bibfield  {title} {\bibinfo
  {title} {Lifetime measurement of cesium atoms using a cold rydberg gas},\
  }\bibfield  {journal} {\bibinfo  {journal} {Applied Sciences}\ }\textbf
  {\bibinfo {volume} {12}},\ \href {https://doi.org/10.3390/app12052713}
  {10.3390/app12052713} (\bibinfo {year} {2022})\BibitemShut {NoStop}%
\bibitem [{\citenamefont {Lei}\ \emph {et~al.}(1995)\citenamefont {Lei},
  \citenamefont {Gu}, \citenamefont {Weng},\ and\ \citenamefont
  {Zeng}}]{Lei1995}%
  \BibitemOpen
  \bibfield  {author} {\bibinfo {author} {\bibfnamefont {T.}~\bibnamefont
  {Lei}}, \bibinfo {author} {\bibfnamefont {S.}~\bibnamefont {Gu}}, \bibinfo
  {author} {\bibfnamefont {Z.}~\bibnamefont {Weng}},\ and\ \bibinfo {author}
  {\bibfnamefont {X.}~\bibnamefont {Zeng}},\ }\bibfield  {title} {\bibinfo
  {title} {Measurement of the polarizabilities of n2d rydberg states of
  cesium},\ }\href {https://doi.org/10.1007/BF01437681} {\bibfield  {journal}
  {\bibinfo  {journal} {Zeitschrift für Physik D Atoms, Molecules and
  Clusters}\ }\textbf {\bibinfo {volume} {34}},\ \bibinfo {pages} {139}
  (\bibinfo {year} {1995})}\BibitemShut {NoStop}%
\bibitem [{\citenamefont {Zhao}\ \emph {et~al.}(2011)\citenamefont {Zhao},
  \citenamefont {Zhang}, \citenamefont {Feng}, \citenamefont {Zhu},
  \citenamefont {Zhang}, \citenamefont {Li},\ and\ \citenamefont
  {Jia}}]{Zhao2011}%
  \BibitemOpen
  \bibfield  {author} {\bibinfo {author} {\bibfnamefont {J.}~\bibnamefont
  {Zhao}}, \bibinfo {author} {\bibfnamefont {H.}~\bibnamefont {Zhang}},
  \bibinfo {author} {\bibfnamefont {Z.}~\bibnamefont {Feng}}, \bibinfo {author}
  {\bibfnamefont {X.}~\bibnamefont {Zhu}}, \bibinfo {author} {\bibfnamefont
  {L.}~\bibnamefont {Zhang}}, \bibinfo {author} {\bibfnamefont
  {C.}~\bibnamefont {Li}},\ and\ \bibinfo {author} {\bibfnamefont
  {S.}~\bibnamefont {Jia}},\ }\bibfield  {title} {\bibinfo {title} {Measurement
  of polarizability of cesium nd state in magneto-optical trap},\ }\href
  {https://doi.org/10.1143/JPSJ.80.034303} {\bibfield  {journal} {\bibinfo
  {journal} {Journal of the Physical Society of Japan}\ }\textbf {\bibinfo
  {volume} {80}},\ \bibinfo {pages} {034303} (\bibinfo {year} {2011})},\
  \Eprint {https://arxiv.org/abs/https://doi.org/10.1143/JPSJ.80.034303}
  {https://doi.org/10.1143/JPSJ.80.034303} \BibitemShut {NoStop}%
\bibitem [{Sup()}]{Supplement}%
  \BibitemOpen
  \bibfield  {title} {\bibinfo {title} {Supplemental material},\ }\href@noop {}
  {\ }\BibitemShut {NoStop}%
\bibitem [{\citenamefont {Ortiz}\ and\ \citenamefont
  {Campos}(1981)}]{Ortiz1981}%
  \BibitemOpen
  \bibfield  {author} {\bibinfo {author} {\bibfnamefont {M.}~\bibnamefont
  {Ortiz}}\ and\ \bibinfo {author} {\bibfnamefont {J.}~\bibnamefont {Campos}},\
  }\bibfield  {title} {\bibinfo {title} {Lifetime measurements of 7p levels of
  cs(i) by means of laser excitation},\ }\href
  {https://doi.org/https://doi.org/10.1016/0022-4073(81)90070-4} {\bibfield
  {journal} {\bibinfo  {journal} {Journal of Quantitative Spectroscopy and
  Radiative Transfer}\ }\textbf {\bibinfo {volume} {26}},\ \bibinfo {pages}
  {107} (\bibinfo {year} {1981})}\BibitemShut {NoStop}%
\bibitem [{\citenamefont {Marek}\ and\ \citenamefont
  {Niemax}(1976)}]{Marek1976}%
  \BibitemOpen
  \bibfield  {author} {\bibinfo {author} {\bibfnamefont {J.}~\bibnamefont
  {Marek}}\ and\ \bibinfo {author} {\bibfnamefont {K.}~\bibnamefont {Niemax}},\
  }\bibfield  {title} {\bibinfo {title} {The influence of collisions of xe
  atoms on the lifetime of atomic states of cs},\ }\href
  {https://doi.org/10.1088/0022-3700/9/16/005} {\bibfield  {journal} {\bibinfo
  {journal} {Journal of Physics B: Atomic and Molecular Physics}\ }\textbf
  {\bibinfo {volume} {9}},\ \bibinfo {pages} {L483} (\bibinfo {year}
  {1976})}\BibitemShut {NoStop}%
\bibitem [{\citenamefont {Marek}\ and\ \citenamefont
  {Ryschka}(1979)}]{Marek1979}%
  \BibitemOpen
  \bibfield  {author} {\bibinfo {author} {\bibfnamefont {J.}~\bibnamefont
  {Marek}}\ and\ \bibinfo {author} {\bibfnamefont {M.}~\bibnamefont
  {Ryschka}},\ }\bibfield  {title} {\bibinfo {title} {Lifetime measurements of
  f-levels of cs using partially superradiant population},\ }\href
  {https://doi.org/https://doi.org/10.1016/0375-9601(79)90580-2} {\bibfield
  {journal} {\bibinfo  {journal} {Physics Letters A}\ }\textbf {\bibinfo
  {volume} {74}},\ \bibinfo {pages} {51} (\bibinfo {year} {1979})}\BibitemShut
  {NoStop}%
\bibitem [{\citenamefont {Marek}(1977{\natexlab{a}})}]{Marek1977}%
  \BibitemOpen
  \bibfield  {author} {\bibinfo {author} {\bibfnamefont {J.}~\bibnamefont
  {Marek}},\ }\bibfield  {title} {\bibinfo {title} {Radiative lifetime of the
  8s, 9s and 7d levels of csi},\ }\href
  {https://doi.org/https://doi.org/10.1016/0375-9601(77)90810-6} {\bibfield
  {journal} {\bibinfo  {journal} {Physics Letters A}\ }\textbf {\bibinfo
  {volume} {60}},\ \bibinfo {pages} {190} (\bibinfo {year}
  {1977}{\natexlab{a}})}\BibitemShut {NoStop}%
\bibitem [{\citenamefont {Neil}\ and\ \citenamefont
  {Atkinson}(1984)}]{Neil1984}%
  \BibitemOpen
  \bibfield  {author} {\bibinfo {author} {\bibfnamefont {W.~S.}\ \bibnamefont
  {Neil}}\ and\ \bibinfo {author} {\bibfnamefont {J.~B.}\ \bibnamefont
  {Atkinson}},\ }\bibfield  {title} {\bibinfo {title} {Lifetimes of some
  excited s and d states of cs i},\ }\href
  {https://doi.org/10.1088/0022-3700/17/5/009} {\bibfield  {journal} {\bibinfo
  {journal} {Journal of Physics B: Atomic and Molecular Physics}\ }\textbf
  {\bibinfo {volume} {17}},\ \bibinfo {pages} {693} (\bibinfo {year}
  {1984})}\BibitemShut {NoStop}%
\bibitem [{\citenamefont {Marek}(1977{\natexlab{b}})}]{Marek1977_2}%
  \BibitemOpen
  \bibfield  {author} {\bibinfo {author} {\bibfnamefont {J.}~\bibnamefont
  {Marek}},\ }\bibfield  {title} {\bibinfo {title} {Study of the time-resolved
  fluorescence of the cs-xe molecular bands},\ }\href
  {https://doi.org/10.1088/0022-3700/10/9/001} {\bibfield  {journal} {\bibinfo
  {journal} {Journal of Physics B: Atomic and Molecular Physics}\ }\textbf
  {\bibinfo {volume} {10}},\ \bibinfo {pages} {L325} (\bibinfo {year}
  {1977}{\natexlab{b}})}\BibitemShut {NoStop}%
\bibitem [{\citenamefont {Tuchendler}\ \emph {et~al.}(2008)\citenamefont
  {Tuchendler}, \citenamefont {Lance}, \citenamefont {Browaeys}, \citenamefont
  {Sortais},\ and\ \citenamefont {Grangier}}]{Tuchendler2008}%
  \BibitemOpen
  \bibfield  {author} {\bibinfo {author} {\bibfnamefont {C.}~\bibnamefont
  {Tuchendler}}, \bibinfo {author} {\bibfnamefont {A.~M.}\ \bibnamefont
  {Lance}}, \bibinfo {author} {\bibfnamefont {A.}~\bibnamefont {Browaeys}},
  \bibinfo {author} {\bibfnamefont {Y.~R.~P.}\ \bibnamefont {Sortais}},\ and\
  \bibinfo {author} {\bibfnamefont {P.}~\bibnamefont {Grangier}},\ }\bibfield
  {title} {\bibinfo {title} {Energy distribution and cooling of a single atom
  in an optical tweezer},\ }\href {https://doi.org/10.1103/PhysRevA.78.033425}
  {\bibfield  {journal} {\bibinfo  {journal} {Phys. Rev. A}\ }\textbf {\bibinfo
  {volume} {78}},\ \bibinfo {pages} {033425} (\bibinfo {year}
  {2008})}\BibitemShut {NoStop}%
\bibitem [{\citenamefont {Breuer}\ and\ \citenamefont
  {Petruccione}(2007)}]{Breuer2007}%
  \BibitemOpen
  \bibfield  {author} {\bibinfo {author} {\bibfnamefont {H.-P.}\ \bibnamefont
  {Breuer}}\ and\ \bibinfo {author} {\bibfnamefont {F.}~\bibnamefont
  {Petruccione}},\ }\href
  {https://doi.org/10.1093/acprof:oso/9780199213900.001.0001} {\emph {\bibinfo
  {title} {{The Theory of Open Quantum Systems}}}}\ (\bibinfo  {publisher}
  {Oxford University Press},\ \bibinfo {year} {2007})\BibitemShut {NoStop}%
\bibitem [{\citenamefont {LeBlanc}\ and\ \citenamefont
  {Thywissen}(2007)}]{LeBlanc2007}%
  \BibitemOpen
  \bibfield  {author} {\bibinfo {author} {\bibfnamefont {L.~J.}\ \bibnamefont
  {LeBlanc}}\ and\ \bibinfo {author} {\bibfnamefont {J.~H.}\ \bibnamefont
  {Thywissen}},\ }\bibfield  {title} {\bibinfo {title} {Species-specific
  optical lattices},\ }\href {https://doi.org/10.1103/PhysRevA.75.053612}
  {\bibfield  {journal} {\bibinfo  {journal} {Phys. Rev. A}\ }\textbf {\bibinfo
  {volume} {75}},\ \bibinfo {pages} {053612} (\bibinfo {year}
  {2007})}\BibitemShut {NoStop}%
\bibitem [{\citenamefont {Ratkata}\ \emph {et~al.}(2021)\citenamefont
  {Ratkata}, \citenamefont {Gregory}, \citenamefont {Innes}, \citenamefont
  {Matthies}, \citenamefont {McArd}, \citenamefont {Mortlock}, \citenamefont
  {Safronova}, \citenamefont {Bromley},\ and\ \citenamefont
  {Cornish}}]{Ratkata2021}%
  \BibitemOpen
  \bibfield  {author} {\bibinfo {author} {\bibfnamefont {A.}~\bibnamefont
  {Ratkata}}, \bibinfo {author} {\bibfnamefont {P.~D.}\ \bibnamefont
  {Gregory}}, \bibinfo {author} {\bibfnamefont {A.~D.}\ \bibnamefont {Innes}},
  \bibinfo {author} {\bibfnamefont {A.~J.}\ \bibnamefont {Matthies}}, \bibinfo
  {author} {\bibfnamefont {L.~A.}\ \bibnamefont {McArd}}, \bibinfo {author}
  {\bibfnamefont {J.~M.}\ \bibnamefont {Mortlock}}, \bibinfo {author}
  {\bibfnamefont {M.~S.}\ \bibnamefont {Safronova}}, \bibinfo {author}
  {\bibfnamefont {S.~L.}\ \bibnamefont {Bromley}},\ and\ \bibinfo {author}
  {\bibfnamefont {S.~L.}\ \bibnamefont {Cornish}},\ }\bibfield  {title}
  {\bibinfo {title} {Measurement of the tune-out wavelength for
  $^{133}\mathrm{Cs}$ at 880 nm},\ }\href
  {https://doi.org/10.1103/PhysRevA.104.052813} {\bibfield  {journal} {\bibinfo
   {journal} {Phys. Rev. A}\ }\textbf {\bibinfo {volume} {104}},\ \bibinfo
  {pages} {052813} (\bibinfo {year} {2021})}\BibitemShut {NoStop}%
\bibitem [{\citenamefont {Seaton}(1983)}]{Seaton1983}%
  \BibitemOpen
  \bibfield  {author} {\bibinfo {author} {\bibfnamefont {M.~J.}\ \bibnamefont
  {Seaton}},\ }\bibfield  {title} {\bibinfo {title} {Quantum defect theory},\
  }\href {https://doi.org/10.1088/0034-4885/46/2/002} {\bibfield  {journal}
  {\bibinfo  {journal} {Reports on Progress in Physics}\ }\textbf {\bibinfo
  {volume} {46}},\ \bibinfo {pages} {167} (\bibinfo {year} {1983})}\BibitemShut
  {NoStop}%
\bibitem [{\citenamefont {{van Wijngaarden}}\ and\ \citenamefont
  {Li}(1994)}]{Wijngaarden1994}%
  \BibitemOpen
  \bibfield  {author} {\bibinfo {author} {\bibfnamefont {W.}~\bibnamefont {{van
  Wijngaarden}}}\ and\ \bibinfo {author} {\bibfnamefont {J.}~\bibnamefont
  {Li}},\ }\bibfield  {title} {\bibinfo {title} {Polarizabilities of cesium s,
  p, d, and f states},\ }\href
  {https://doi.org/https://doi.org/10.1016/0022-4073(94)90024-8} {\bibfield
  {journal} {\bibinfo  {journal} {Journal of Quantitative Spectroscopy and
  Radiative Transfer}\ }\textbf {\bibinfo {volume} {52}},\ \bibinfo {pages}
  {555} (\bibinfo {year} {1994})}\BibitemShut {NoStop}%
\bibitem [{\citenamefont {Deiglmayr}\ \emph {et~al.}(2016)\citenamefont
  {Deiglmayr}, \citenamefont {Herburger}, \citenamefont {Sa\ss{}mannshausen},
  \citenamefont {Jansen}, \citenamefont {Schmutz},\ and\ \citenamefont
  {Merkt}}]{Deiglmayr2016}%
  \BibitemOpen
  \bibfield  {author} {\bibinfo {author} {\bibfnamefont {J.}~\bibnamefont
  {Deiglmayr}}, \bibinfo {author} {\bibfnamefont {H.}~\bibnamefont
  {Herburger}}, \bibinfo {author} {\bibfnamefont {H.}~\bibnamefont
  {Sa\ss{}mannshausen}}, \bibinfo {author} {\bibfnamefont {P.}~\bibnamefont
  {Jansen}}, \bibinfo {author} {\bibfnamefont {H.}~\bibnamefont {Schmutz}},\
  and\ \bibinfo {author} {\bibfnamefont {F.}~\bibnamefont {Merkt}},\ }\bibfield
   {title} {\bibinfo {title} {Precision measurement of the ionization energy of
  cs i},\ }\href {https://doi.org/10.1103/PhysRevA.93.013424} {\bibfield
  {journal} {\bibinfo  {journal} {Phys. Rev. A}\ }\textbf {\bibinfo {volume}
  {93}},\ \bibinfo {pages} {013424} (\bibinfo {year} {2016})}\BibitemShut
  {NoStop}%
\bibitem [{\citenamefont {Lorenzen}\ and\ \citenamefont
  {Niemax}(1983)}]{Lorenzen1983}%
  \BibitemOpen
  \bibfield  {author} {\bibinfo {author} {\bibfnamefont {C.-J.}\ \bibnamefont
  {Lorenzen}}\ and\ \bibinfo {author} {\bibfnamefont {K.}~\bibnamefont
  {Niemax}},\ }\bibfield  {title} {\bibinfo {title} {Quantum defects of the
  n2p1/2,3/2 levels in 39k i and 85rb i},\ }\href
  {https://doi.org/10.1088/0031-8949/27/4/012} {\bibfield  {journal} {\bibinfo
  {journal} {Physica Scripta}\ }\textbf {\bibinfo {volume} {27}},\ \bibinfo
  {pages} {300} (\bibinfo {year} {1983})}\BibitemShut {NoStop}%
\bibitem [{\citenamefont {Bai}\ \emph {et~al.}(2023)\citenamefont {Bai},
  \citenamefont {Song}, \citenamefont {Fan}, \citenamefont {Jiao},
  \citenamefont {Zhao}, \citenamefont {Jia},\ and\ \citenamefont
  {Raithel}}]{Bai2023}%
  \BibitemOpen
  \bibfield  {author} {\bibinfo {author} {\bibfnamefont {J.}~\bibnamefont
  {Bai}}, \bibinfo {author} {\bibfnamefont {R.}~\bibnamefont {Song}}, \bibinfo
  {author} {\bibfnamefont {J.}~\bibnamefont {Fan}}, \bibinfo {author}
  {\bibfnamefont {Y.}~\bibnamefont {Jiao}}, \bibinfo {author} {\bibfnamefont
  {J.}~\bibnamefont {Zhao}}, \bibinfo {author} {\bibfnamefont {S.}~\bibnamefont
  {Jia}},\ and\ \bibinfo {author} {\bibfnamefont {G.}~\bibnamefont {Raithel}},\
  }\bibfield  {title} {\bibinfo {title} {Quantum defects of $n{F}_{J}$ levels
  of cs rydberg atoms},\ }\href {https://doi.org/10.1103/PhysRevA.108.022804}
  {\bibfield  {journal} {\bibinfo  {journal} {Phys. Rev. A}\ }\textbf {\bibinfo
  {volume} {108}},\ \bibinfo {pages} {022804} (\bibinfo {year}
  {2023})}\BibitemShut {NoStop}%
\bibitem [{\citenamefont {Weber}\ and\ \citenamefont
  {Sansonetti}(1987)}]{Weber1987}%
  \BibitemOpen
  \bibfield  {author} {\bibinfo {author} {\bibfnamefont {K.-H.}\ \bibnamefont
  {Weber}}\ and\ \bibinfo {author} {\bibfnamefont {C.~J.}\ \bibnamefont
  {Sansonetti}},\ }\bibfield  {title} {\bibinfo {title} {Accurate energies of
  ns, np, nd, nf, and ng levels of neutral cesium},\ }\href
  {https://doi.org/10.1103/PhysRevA.35.4650} {\bibfield  {journal} {\bibinfo
  {journal} {Phys. Rev. A}\ }\textbf {\bibinfo {volume} {35}},\ \bibinfo
  {pages} {4650} (\bibinfo {year} {1987})}\BibitemShut {NoStop}%
\bibitem [{\citenamefont {Bates}\ \emph {et~al.}(1949)\citenamefont {Bates},
  \citenamefont {Damgaard},\ and\ \citenamefont {Massey}}]{Bates1949}%
  \BibitemOpen
  \bibfield  {author} {\bibinfo {author} {\bibfnamefont {D.~R.}\ \bibnamefont
  {Bates}}, \bibinfo {author} {\bibfnamefont {A.}~\bibnamefont {Damgaard}},\
  and\ \bibinfo {author} {\bibfnamefont {H.~S.~W.}\ \bibnamefont {Massey}},\
  }\bibfield  {title} {\bibinfo {title} {The calculation of the absolute
  strengths of spectral lines},\ }\href
  {https://doi.org/10.1098/rsta.1949.0006} {\bibfield  {journal} {\bibinfo
  {journal} {Philosophical Transactions of the Royal Society of London. Series
  A, Mathematical and Physical Sciences}\ }\textbf {\bibinfo {volume} {242}},\
  \bibinfo {pages} {101} (\bibinfo {year} {1949})},\ \Eprint
  {https://arxiv.org/abs/https://royalsocietypublishing.org/doi/pdf/10.1098/rsta.1949.0006}
  {https://royalsocietypublishing.org/doi/pdf/10.1098/rsta.1949.0006}
  \BibitemShut {NoStop}%
\bibitem [{\citenamefont {{Van Wijngaarden}}(1997)}]{Wijngaarden1997}%
  \BibitemOpen
  \bibfield  {author} {\bibinfo {author} {\bibfnamefont {W.}~\bibnamefont {{Van
  Wijngaarden}}},\ }\bibfield  {title} {\bibinfo {title} {Scalar and tensor
  polarizabilities of low lying s, p, d, f and g states in rubidium},\ }\href
  {https://doi.org/https://doi.org/10.1016/S0022-4073(96)00111-2} {\bibfield
  {journal} {\bibinfo  {journal} {Journal of Quantitative Spectroscopy and
  Radiative Transfer}\ }\textbf {\bibinfo {volume} {57}},\ \bibinfo {pages}
  {275} (\bibinfo {year} {1997})}\BibitemShut {NoStop}%
\bibitem [{\citenamefont {Bhowmik}\ \emph {et~al.}(2018)\citenamefont
  {Bhowmik}, \citenamefont {Dutta},\ and\ \citenamefont
  {Majumder}}]{Bhowmik2018}%
  \BibitemOpen
  \bibfield  {author} {\bibinfo {author} {\bibfnamefont {A.}~\bibnamefont
  {Bhowmik}}, \bibinfo {author} {\bibfnamefont {N.~N.}\ \bibnamefont {Dutta}},\
  and\ \bibinfo {author} {\bibfnamefont {S.}~\bibnamefont {Majumder}},\
  }\bibfield  {title} {\bibinfo {title} {Tunable magic wavelengths for trapping
  with focused laguerre-gaussian beams},\ }\href
  {https://doi.org/10.1103/PhysRevA.97.022511} {\bibfield  {journal} {\bibinfo
  {journal} {Phys. Rev. A}\ }\textbf {\bibinfo {volume} {97}},\ \bibinfo
  {pages} {022511} (\bibinfo {year} {2018})}\BibitemShut {NoStop}%
\bibitem [{\citenamefont {Singh}\ \emph {et~al.}(2016)\citenamefont {Singh},
  \citenamefont {Kaur}, \citenamefont {Sahoo},\ and\ \citenamefont
  {Arora}}]{Singh2016}%
  \BibitemOpen
  \bibfield  {author} {\bibinfo {author} {\bibfnamefont {S.}~\bibnamefont
  {Singh}}, \bibinfo {author} {\bibfnamefont {K.}~\bibnamefont {Kaur}},
  \bibinfo {author} {\bibfnamefont {B.~K.}\ \bibnamefont {Sahoo}},\ and\
  \bibinfo {author} {\bibfnamefont {B.}~\bibnamefont {Arora}},\ }\bibfield
  {title} {\bibinfo {title} {Comparing magic wavelengths for the
  $6{s}^{2}{S}_{1/2}-6p{}^{2}{P}_{1/\mathrm{2,3}/2}$ transitions of cs using
  circularly and linearly polarized light},\ }\href
  {https://doi.org/10.1088/0953-4075/49/14/145005} {\bibfield  {journal}
  {\bibinfo  {journal} {Journal of Physics B: Atomic, Molecular and Optical
  Physics}\ }\textbf {\bibinfo {volume} {49}},\ \bibinfo {pages} {145005}
  (\bibinfo {year} {2016})}\BibitemShut {NoStop}%
\bibitem [{\citenamefont {Safronova}\ \emph {et~al.}(2016)\citenamefont
  {Safronova}, \citenamefont {Safronova},\ and\ \citenamefont
  {Clark}}]{Safronova2016}%
  \BibitemOpen
  \bibfield  {author} {\bibinfo {author} {\bibfnamefont {M.~S.}\ \bibnamefont
  {Safronova}}, \bibinfo {author} {\bibfnamefont {U.~I.}\ \bibnamefont
  {Safronova}},\ and\ \bibinfo {author} {\bibfnamefont {C.~W.}\ \bibnamefont
  {Clark}},\ }\bibfield  {title} {\bibinfo {title} {Magic wavelengths, matrix
  elements, polarizabilities, and lifetimes of cs},\ }\href
  {https://doi.org/10.1103/PhysRevA.94.012505} {\bibfield  {journal} {\bibinfo
  {journal} {Phys. Rev. A}\ }\textbf {\bibinfo {volume} {94}},\ \bibinfo
  {pages} {012505} (\bibinfo {year} {2016})}\BibitemShut {NoStop}%
\bibitem [{\citenamefont {Manakov}\ \emph {et~al.}(1986)\citenamefont
  {Manakov}, \citenamefont {Ovsiannikov},\ and\ \citenamefont
  {Rapoport}}]{Manakov1986}%
  \BibitemOpen
  \bibfield  {author} {\bibinfo {author} {\bibfnamefont {N.}~\bibnamefont
  {Manakov}}, \bibinfo {author} {\bibfnamefont {V.}~\bibnamefont
  {Ovsiannikov}},\ and\ \bibinfo {author} {\bibfnamefont {L.}~\bibnamefont
  {Rapoport}},\ }\bibfield  {title} {\bibinfo {title} {Atoms in a laser
  field},\ }\href
  {https://doi.org/https://doi.org/10.1016/S0370-1573(86)80001-1} {\bibfield
  {journal} {\bibinfo  {journal} {Physics Reports}\ }\textbf {\bibinfo {volume}
  {141}},\ \bibinfo {pages} {320} (\bibinfo {year} {1986})}\BibitemShut
  {NoStop}%
\bibitem [{\citenamefont {Arora}\ and\ \citenamefont {Sahoo}(2012)}]{Aora2012}%
  \BibitemOpen
  \bibfield  {author} {\bibinfo {author} {\bibfnamefont {B.}~\bibnamefont
  {Arora}}\ and\ \bibinfo {author} {\bibfnamefont {B.~K.}\ \bibnamefont
  {Sahoo}},\ }\bibfield  {title} {\bibinfo {title} {State-insensitive trapping
  of rb atoms: Linearly versus circularly polarized light},\ }\href
  {https://doi.org/10.1103/PhysRevA.86.033416} {\bibfield  {journal} {\bibinfo
  {journal} {Phys. Rev. A}\ }\textbf {\bibinfo {volume} {86}},\ \bibinfo
  {pages} {033416} (\bibinfo {year} {2012})}\BibitemShut {NoStop}%
\bibitem [{\citenamefont {Arora}\ \emph {et~al.}(2007)\citenamefont {Arora},
  \citenamefont {Safronova},\ and\ \citenamefont {Clark}}]{Arora2007}%
  \BibitemOpen
  \bibfield  {author} {\bibinfo {author} {\bibfnamefont {B.}~\bibnamefont
  {Arora}}, \bibinfo {author} {\bibfnamefont {M.~S.}\ \bibnamefont
  {Safronova}},\ and\ \bibinfo {author} {\bibfnamefont {C.~W.}\ \bibnamefont
  {Clark}},\ }\bibfield  {title} {\bibinfo {title} {Magic wavelengths for the
  $np\text{\ensuremath{-}}ns$ transitions in alkali-metal atoms},\ }\href
  {https://doi.org/10.1103/PhysRevA.76.052509} {\bibfield  {journal} {\bibinfo
  {journal} {Phys. Rev. A}\ }\textbf {\bibinfo {volume} {76}},\ \bibinfo
  {pages} {052509} (\bibinfo {year} {2007})}\BibitemShut {NoStop}%
\bibitem [{\citenamefont {Jiang}\ \emph {et~al.}(2019)\citenamefont {Jiang},
  \citenamefont {Jiang}, \citenamefont {Wu}, \citenamefont {Zhang},
  \citenamefont {Xie},\ and\ \citenamefont {Dong}}]{Jiang2019}%
  \BibitemOpen
  \bibfield  {author} {\bibinfo {author} {\bibfnamefont {J.}~\bibnamefont
  {Jiang}}, \bibinfo {author} {\bibfnamefont {L.}~\bibnamefont {Jiang}},
  \bibinfo {author} {\bibfnamefont {Z.~W.}\ \bibnamefont {Wu}}, \bibinfo
  {author} {\bibfnamefont {D.-H.}\ \bibnamefont {Zhang}}, \bibinfo {author}
  {\bibfnamefont {L.-Y.}\ \bibnamefont {Xie}},\ and\ \bibinfo {author}
  {\bibfnamefont {C.-Z.}\ \bibnamefont {Dong}},\ }\bibfield  {title} {\bibinfo
  {title} {Angle-dependent magic wavelengths for the
  $4{s}_{1/2}\ensuremath{\rightarrow}3{d}_{5/2,3/2}$ transitions of
  ${\mathrm{ca}}^{+}$ ions},\ }\href
  {https://doi.org/10.1103/PhysRevA.99.032510} {\bibfield  {journal} {\bibinfo
  {journal} {Phys. Rev. A}\ }\textbf {\bibinfo {volume} {99}},\ \bibinfo
  {pages} {032510} (\bibinfo {year} {2019})}\BibitemShut {NoStop}%
\bibitem [{\citenamefont {Bhowmik}\ \emph {et~al.}(2022)\citenamefont
  {Bhowmik}, \citenamefont {Dutta},\ and\ \citenamefont {Das}}]{Bhowmik2022}%
  \BibitemOpen
  \bibfield  {author} {\bibinfo {author} {\bibfnamefont {A.}~\bibnamefont
  {Bhowmik}}, \bibinfo {author} {\bibfnamefont {N.~N.}\ \bibnamefont {Dutta}},\
  and\ \bibinfo {author} {\bibfnamefont {S.}~\bibnamefont {Das}},\ }\bibfield
  {title} {\bibinfo {title} {Role of vector polarizability induced by a
  linearly polarized focused laguerre-gaussian light: applications in optical
  trapping and ultracold spinor mixture},\ }\href
  {https://doi.org/10.1140/epjd/s10053-022-00470-y} {\bibfield  {journal}
  {\bibinfo  {journal} {The European Physical Journal D}\ }\textbf {\bibinfo
  {volume} {76}},\ \bibinfo {pages} {139} (\bibinfo {year} {2022})}\BibitemShut
  {NoStop}%
\bibitem [{\citenamefont {Barakhshan}\ \emph {et~al.}(2022)\citenamefont
  {Barakhshan}, \citenamefont {Marrs}, \citenamefont {Bhosale}, \citenamefont
  {Arora}, \citenamefont {Eigenmann},\ and\ \citenamefont
  {Safronova}}]{UDportal}%
  \BibitemOpen
  \bibfield  {author} {\bibinfo {author} {\bibfnamefont {P.}~\bibnamefont
  {Barakhshan}}, \bibinfo {author} {\bibfnamefont {A.}~\bibnamefont {Marrs}},
  \bibinfo {author} {\bibfnamefont {A.}~\bibnamefont {Bhosale}}, \bibinfo
  {author} {\bibfnamefont {B.}~\bibnamefont {Arora}}, \bibinfo {author}
  {\bibfnamefont {R.}~\bibnamefont {Eigenmann}},\ and\ \bibinfo {author}
  {\bibfnamefont {M.~S.}\ \bibnamefont {Safronova}},\ }\href@noop {} {}\bibinfo
  {howpublished} {{\textit{Portal for High-Precision Atomic Data and
  Computation}} (version 2.0), University of Delaware, Newark, DE, USA,
  URL:{https://www.udel.edu/atom}} (\bibinfo {year} {February
  2022})\BibitemShut {NoStop}%
\bibitem [{\citenamefont {Kramida}\ \emph {et~al.}(2023)\citenamefont
  {Kramida}, \citenamefont {{Yu.~Ralchenko}}, \citenamefont {Reader},\ and\
  \citenamefont {{and NIST ASD Team}}}]{NIST_ASD}%
  \BibitemOpen
  \bibfield  {author} {\bibinfo {author} {\bibfnamefont {A.}~\bibnamefont
  {Kramida}}, \bibinfo {author} {\bibnamefont {{Yu.~Ralchenko}}}, \bibinfo
  {author} {\bibfnamefont {J.}~\bibnamefont {Reader}},\ and\ \bibinfo {author}
  {\bibnamefont {{and NIST ASD Team}}},\ }\href@noop {} {}\bibinfo
  {howpublished} {{NIST Atomic Spectra Database (ver. 5.11), [Online].
  Available: {\tt{https://physics.nist.gov/asd}} [2024, February 28]. National
  Institute of Standards and Technology, Gaithersburg, MD.}} (\bibinfo {year}
  {2023})\BibitemShut {NoStop}%
\bibitem [{\citenamefont {Rosenbusch}\ \emph {et~al.}(2009)\citenamefont
  {Rosenbusch}, \citenamefont {Ghezali}, \citenamefont {Dzuba}, \citenamefont
  {Flambaum}, \citenamefont {Beloy},\ and\ \citenamefont
  {Derevianko}}]{Rosenbusch2009}%
  \BibitemOpen
  \bibfield  {author} {\bibinfo {author} {\bibfnamefont {P.}~\bibnamefont
  {Rosenbusch}}, \bibinfo {author} {\bibfnamefont {S.}~\bibnamefont {Ghezali}},
  \bibinfo {author} {\bibfnamefont {V.~A.}\ \bibnamefont {Dzuba}}, \bibinfo
  {author} {\bibfnamefont {V.~V.}\ \bibnamefont {Flambaum}}, \bibinfo {author}
  {\bibfnamefont {K.}~\bibnamefont {Beloy}},\ and\ \bibinfo {author}
  {\bibfnamefont {A.}~\bibnamefont {Derevianko}},\ }\bibfield  {title}
  {\bibinfo {title} {ac stark shift of the cs microwave atomic clock
  transitions},\ }\href {https://doi.org/10.1103/PhysRevA.79.013404} {\bibfield
   {journal} {\bibinfo  {journal} {Phys. Rev. A}\ }\textbf {\bibinfo {volume}
  {79}},\ \bibinfo {pages} {013404} (\bibinfo {year} {2009})}\BibitemShut
  {NoStop}%
\bibitem [{\citenamefont {Kaur}\ \emph {et~al.}(2017)\citenamefont {Kaur},
  \citenamefont {Singh}, \citenamefont {Arora},\ and\ \citenamefont
  {Sahoo}}]{Kaur2017}%
  \BibitemOpen
  \bibfield  {author} {\bibinfo {author} {\bibfnamefont {J.}~\bibnamefont
  {Kaur}}, \bibinfo {author} {\bibfnamefont {S.}~\bibnamefont {Singh}},
  \bibinfo {author} {\bibfnamefont {B.}~\bibnamefont {Arora}},\ and\ \bibinfo
  {author} {\bibfnamefont {B.~K.}\ \bibnamefont {Sahoo}},\ }\bibfield  {title}
  {\bibinfo {title} {Annexing magic and tune-out wavelengths to the clock
  transitions of the alkaline-earth-metal ions},\ }\href
  {https://doi.org/10.1103/PhysRevA.95.042501} {\bibfield  {journal} {\bibinfo
  {journal} {Phys. Rev. A}\ }\textbf {\bibinfo {volume} {95}},\ \bibinfo
  {pages} {042501} (\bibinfo {year} {2017})}\BibitemShut {NoStop}%
\bibitem [{\citenamefont {Das}\ \emph {et~al.}(2020)\citenamefont {Das},
  \citenamefont {Bhowmik}, \citenamefont {Dutta},\ and\ \citenamefont
  {Majumder}}]{Das2020}%
  \BibitemOpen
  \bibfield  {author} {\bibinfo {author} {\bibfnamefont {A.}~\bibnamefont
  {Das}}, \bibinfo {author} {\bibfnamefont {A.}~\bibnamefont {Bhowmik}},
  \bibinfo {author} {\bibfnamefont {N.~N.}\ \bibnamefont {Dutta}},\ and\
  \bibinfo {author} {\bibfnamefont {S.}~\bibnamefont {Majumder}},\ }\bibfield
  {title} {\bibinfo {title} {Many-body calculations and hyperfine-interaction
  effect on dynamic polarizabilities at the low-lying energy levels of
  ${\mathrm{y}}^{2+}$},\ }\href {https://doi.org/10.1103/PhysRevA.102.012801}
  {\bibfield  {journal} {\bibinfo  {journal} {Phys. Rev. A}\ }\textbf {\bibinfo
  {volume} {102}},\ \bibinfo {pages} {012801} (\bibinfo {year}
  {2020})}\BibitemShut {NoStop}%
\end{thebibliography}%

\end{document}